\documentclass{aa}  
\usepackage{graphicx}
\usepackage{txfonts}
\usepackage{natbib}
\usepackage[group-minimum-digits=4]{siunitx}
\usepackage{gensymb}
\usepackage{subcaption}
\newcommand{\ri}{\ensuremath{r_\mathrm{i}}}
\newcommand{\ro}{\ensuremath{r_\mathrm{o}}}

\usepackage{amstext}

\begin{document} 

   \title{MRI-driven $\alpha\Omega$ dynamos in protoneutron stars}

   \subtitle{}

   \author{A. Reboul-Salze\thanks{\email{alexis.reboul-salze@aei.mpg.de}
}
          \inst{1,2}
          \and
          J. Guilet\inst{1}
          \and
          R. Raynaud\inst{3}
          \and
          M. Bugli\inst{1}
   }

   \institute{\inst{1}Laboratoire AIM, CEA/DRF-CNRS-Universit\'e Paris Cit\'e, IRFU/
              D\'epartement d'Astrophysique, CEA-Saclay, F-91191, France\\
              \inst{2}Max Planck Institute for Gravitational Physics (Albert Einstein Institute), D-14476 Potsdam, Germany\\
              \inst{3}Universit\'e Paris Cit\'e, Universit\'e Paris-Saclay, CNRS, CEA, Astrophysique, Instrumentation et Mod\'elisation, F-91191 Gif-sur-Yvette, France\\
             \email{alexis.reboul-salze@aei.mpg.de}
             }

   \date{Received 05 October, 2021; accepted 12 July, 2022}


  \abstract
  {Magnetars are highly magnetized neutron stars that can produce a wide diversity of X-ray and soft gamma-ray emissions that are powered by magnetic dissipation.
  Their magnetic dipole is constrained in the range of $10^{14}$ to \SI{e15}{G} by the measurement of their spin-down.
  In addition to fast rotation, these strong fields are also invoked to explain extreme stellar explosions, such as hypernovae, which are associated with long gamma-ray bursts and superluminous supernovae. A promising mechanism for explaining magnetar formation is the amplification of the magnetic field by the magnetorotational instability (MRI) in fast-rotating protoneutron stars (PNS). This scenario is supported by recent global incompressible models, which showed that a dipole field with magnetar-like intensity can be generated from small-scale turbulence.
  However, the impact of important physical ingredients, such as buoyancy and density stratification, on the efficiency of the MRI in generating a dipole field is still unknown.}
  {We assess the impact of the density and entropy profiles on the MRI dynamo in a global model of a fast-rotating PNS. The\,model focuses on the outer stratified region of the PNS that is stable to convection. }
  {Using the pseudo-spectral code MagIC, we performed 3D Boussinesq and anelastic magnetohydrodynamics simulations in spherical geometry with explicit diffusivities and with differential rotation forced at the outer boundary. The thermodynamic background of the anelastic models was retrieved from the data of 1D core-collapse supernova simulations from the Garching group. 
  We performed a parameter study in which we investigated the influence of different approximations and the effect of the thermal diffusion through the Prandtl number.}
  {We obtain a self-sustained turbulent MRI-driven dynamo. This confirms most of our previous incompressible results when they are rescaled for density. The MRI generates a strong turbulent magnetic field and a nondominant equatorial dipole, which represents about $4.3\%$ of the averaged magnetic field strength. 
  Interestingly, an axisymmetric magnetic field at large scales is observed to oscillate with time, which can be described as a mean-field $\alpha \Omega$ dynamo.
  By comparing these results with models without buoyancy or density stratification, we find that the key ingredient explaining the appearance of this mean-field behavior is the density gradient. Buoyancy due to the entropy gradient damps turbulence in the equatorial plane, but it has a relatively weak influence in the low Prandtl number regime overall, as expected from neutrino diffusion.
  However, the buoyancy starts to strongly impact the MRI dynamo for Prandtl numbers close to unity.}
  {Our results support the hypothesis that the MRI is able to generate magnetar-like large-scale magnetic fields. The results furthermore predict the presence of a $\alpha \Omega$ dynamo in the protoneutron star, which could be important to model in-situ magnetic field amplification in global models of core-collapse supernovae or binary neutron star mergers.}
\maketitle
%

\section{Introduction}

Magnetars are a special class of neutron stars that corresponds to isolated young neutron stars that are characterized by their variable high-energy emission, which is powered by the dissipation of enormous internal magnetic fields \citep[and references therein]{1998Kouveliotou,2015TurollaMagnetarObsReview, 2017Kaspi,2021EspositoMagnetarReview}. Observations of their pulsed X-ray activity constrain their rotation period $P$ and spin-down $\dot{P}$, two important timing parameters from which a magnetic field 
\begin{equation}
  B_\mathrm{dip} = 10^{14}\left( \frac{P}{5\ \mathrm{s}}\right)^{1/2} \left( \frac{\dot{P}}{\SI{e-11}{s.s^{-1}}}\right)^{1/2} \,\si{G}
  \label{e:Bdip}
\end{equation}
can be inferred under the assumption that only the dipolar field brakes the neutron star. Their activity can be explained by the decay of their ultrastrong magnetic field and includes short bursts \citep{2006Gotz}, large outbursts \citep{2018CotiZelati,2021CotiZelatiOutbursts}, and giant flares \citep{2005HurleyFlare,2021SvinkinGiantFlare} with quasi-periodic oscillations in their signal \citep{2005Israel,2021RobertsGiantFlare}. In addition, absorption lines that were interpreted as proton cyclotron lines have been detected in two outbursts \citep{2013Tiengo,2016RodriguezCastillo}, suggesting a nondipolar surface field stronger than the dipolar component derived from Eq.~\eqref{e:Bdip}.
Another class of transients that can be modeled using magnetars is represented by fast radio bursts (FRBs), that is, millisecond-duration bursts of radio waves discovered in 2007 \citep{2007LorimerfirstFRB}
that are observed from all directions in the sky. While the exact origin of these events is not yet well known, the observation of the signal FRB 200428 from the Galactic magnetar SGR 1935+2154 \citep{2020BochenekGalaticFRB,2020ChimeGalacticMagnetar} confirms that the magnetar model may be able to explain FRBs.

The formation of magnetars in the presence of fast rotation might also be important to explain the most extreme supernovae: hypernovae (also called broadline type Ic supernovae) associated with long gamma-ray bursts \citep{2011Drout} and superluminous supernovae \citep{2013NichollSLSN,2013InserraSLSN,2018MargalitSLSN}. These two classes of outstanding stellar explosions are rare events, representing 0.1$\%$ and 1$\%$ of all supernovae, respectively. The first are characterized by a kinetic energy that is ten times higher than that of standard supernovae, and the second by a luminosity that is one hundred times higher than that of standard supernovae. The central engine that can explain the kinetic energy of hypernovae is based on a strong large-scale magnetic field that can efficiently extract the high rotational energy of a fast-rotating protoneutron star (PNS). This large-scale magnetic field would then launch jets and lead to a magnetorotational explosion \citep[e.g.,][]{2006MoiseenkoMHDjet,2006ShibataHyper,2008DessartHyper,2012WintelerHyper,2014MostaHyper,2018Obergaulinger,2020Bugli,2020KurodaMRexplosion3D}. The jets that are launched by a millisecond protomagnetar could also lead to (ultra) long gamma-ray bursts \citep{1992DuncanLGRB,2011MetzgerLGRB,2018MetzgerLGRBs}. 
Millisecond magnetars have also been invoked to power superluminous supernovae with their spin-down luminosity, which corresponds to a delayed energy injection \citep{2013NichollSLSN,2013InserraSLSN,2018MargalitSLSN}. Other high-energy events could be explained by millisecond magnetars, such as short gamma-ray bursts \citep{2008MetzgerSGRB} and X-ray transients in the aftermath of binary neutron star mergers \citep{2019XueXBNS}. A millisecond magnetar might indeed form during neutron star mergers, thus providing an explanation for the plateau phase and the extended emission in X-ray sources associated with short gamma-ray bursts \citep{2008MetzgerSGRB,2012BucciantiniSGRB,2013RowlinsonSGRBR,2014GompertzSGRB}.

A scenario that might explain the magnetic field of some magnetars is magnetic flux conservation during the collapse of a highly magnetized progenitor. 
For example, the collapse of progenitors resulting from a stellar merger could lead to the strongest magnetic fields \citep{2019SchneiderStar}. However, this scenario predicts slowly rotating progenitors and therefore cannot explain the formation of millisecond magnetars. To have both fast rotation and a strong magnetic field, an in-situ magnetic field amplification  must be considered. Two mechanisms have been studied: convective dynamos \citep{1993ThompsonPNS,2020ScienceRaynaud,2020MasadaNSConvection,2021RaynaudGWPNS}, and the magnetorotational instability \citep[MRI; see][]{2003AkiyamaMRI,2009ObergaulingerMRI,2021ReboulSalze}.

The MRI has first been studied in Keplerian accretion disks, both analytically 
\citep{1991BalbusMRI} and numerically in local shearing box models \citep{1992HawleyMRI}. The first studies, which considered the incompressible ideal magnetohydrodynamics (MHD) framework, showed that in the presence of differential rotation, the turbulent velocity and magnetic field reach a statistically stationary state. 
Important physical ingredients, such as thermal stratification due to entropy and composition gradients and diffusivities (viscosity, resistivity, and thermal diffusion), were then taken into account by semi-analytical linear analysis \citep{1978Acheson,1995BalbusMRI,2004MenouMRI,2008PessahMRI} and numerical studies \citep{2007FromangMRIB,2007LesurMRIPm,2009SimonMRIPmeffectsonNetTor,2010LongarettiMRIPm,2015MeheutMRIPmlimit,2015GuiletBuoy}. 
An important aspect of disk models that include vertical density stratification is the appearance of oscillating dynamo cycles shaping the structure of the axisymmetric magnetic field \citep{2010DavisButterfly,2016Shi,2019DengMRI}.

In the context of core-collapse supernova (CCSN), numerical studies have shown that the MRI is a viable mechanism to efficiently amplify the small-scale magnetic field on timescales shorter than the successful explosion time (typically a few hundred milliseconds) \citep{2009ObergaulingerMRI,2015GuiletBuoy,2016RembiazMRIMaxB}. 
In protoneutron stars, the physical conditions differ from accretion disks in several ways. The strong differential rotation inside the PNS \citep{2003AkiyamaMRI,2006OttRotPNS,2020Bugli} is non-Keplerian because of the important role of the pressure gradient in the hydrostatic force balance \citep{2012MasadaRot}. 
Due to the high density and temperature, neutrinos play an important role in the dissipation of the kinetic energy, which can be modeled as a strong viscosity \citep{2007MasadaMRIneutrinolinear,2015GuiletVisc}.
Finally, the MRI turbulence can be reduced by the buoyancy forces driven by the gradients of entropy and lepton fraction in the case of stable stratification \citep{2015GuiletBuoy}.

The impact of spherical geometry, global entropy, and density gradients on MRI turbulence is still unknown.
The first attempts to address this question rely on semi-global models that include radial gradients of density and entropy \citep{2009ObergaulingerMRI,2015MasadaMRIdisk}. However, these models remain local at least vertically, and therefore only describe the small-scale turbulence. 
\citet{2015MostaMRI} showed the development of MRI turbulence in the first simulations by describing a quarter of the PNS whose resolution was high enough to resolve the MRI wavelength. These simulations started with a relatively strong initial large-scale magnetic field, and due to their high computational cost, they could not last long enough to show dynamo cycles.
\citet[hereafter referred to as paper I]{2021ReboulSalze} showed that the MRI is able to robustly generate a large-scale field with magnetar-like intensities under the incompressible approximation, and we extend these promising results to a more realistic setup including global gradients here. 

This article investigates the impact of a density gradient and stable stratification on the global properties of the MRI in a 3D spherical model. The anelastic approximation allows us to take the effects of density and entropy gradients into account while filtering sound waves, which drastically increases the time step and enables long simulation times. 
The paper is organized as follows: in Sect.~\ref{Numerical} we describe the physical and numerical setup. The results are then presented in Sect.~\ref{results} for the saturated nonlinear phase of the MRI and in Sect.~\ref{comp} for the comparison between different global models of the MRI. Finally, we discuss the validity of our assumptions in Sect.~\ref{discuss} and draw our conclusions in Sect.~\ref{conclusion}.
\section{Numerical setup}
\label{Numerical}

\subsection{1D protoneutron star model}

\begin{figure}[h]
\centering
\resizebox{\hsize}{!}
            {
     \includegraphics{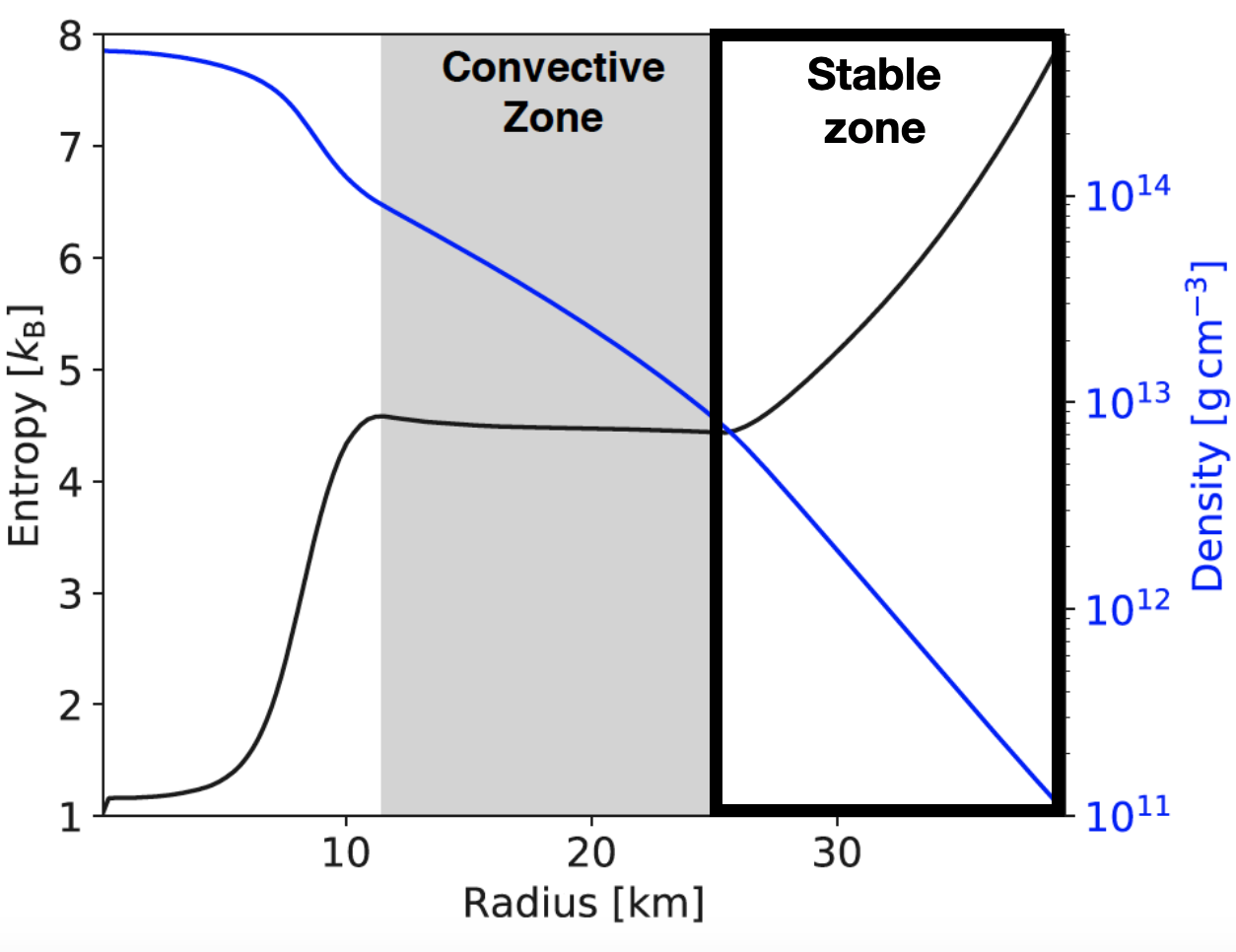}}
      \caption{1D radial profiles of the entropy per baryon (black line) and density (blue line) of our PNS model at  $t=\SI{0.2}{s}$ post-bounce. Our simulation domain corresponds to the outer stably stratified layer delimited by the thick black line.}
    \label{Background}
\end{figure}

The simulations performed in this article are designed to represent the outer region of a fast-rotating PNS, which is stable to the convection and unstable to the MRI. To model this stably stratified region, we used the same methods and internal structure as \citet{2020ScienceRaynaud}, but focused on a different part of the PNS. The PNS we considered has a baryonic (final gravitational) mass of 1.78 (1.59) M$_{\odot}$ and was taken from a 1D core-collapse supernova simulation from \citet{2014Hudepol}. This simulation used the high-density equation of state LS220 \citep{1991LattimerEOSdensematter} and the nonrotating $\SI{27}{M_\odot}$ progenitor s27.0 by \citet{2002WoosleyProgenitor}. The calculations were performed with the code Prometheus-Vertex, which combines the hydrodynamics solver Prometheus with the neutrino transport module Vertex \citep{2002RamppJankaVertex}. This module solves the energy-dependent moment equations of three species of neutrinos and antineutrinos using a variable Eddington factor closure and including a most recent set of neutrino interaction rates. 
The energy and lepton number transport by convection is modeled with a mixing length treatment.
We chose the physical parameters to represent the PNS at $t=\SI{0.2}{s}$ after bounce according to the 1D CCSN simulation, whose profiles of density and specific entropy are shown in Fig.~\ref{Background}.
While the models presented in \citet{2020ScienceRaynaud} focused on the convective zone, our simulation domain spans the outer stably stratified region, thus extending from $\ri=\SI{25.5}{km}$
to $\ro = \SI{39.25}{km}$, {corresponding to the PNS surface defined by the density $\rho_o = \SI{e11}{g.cm^{-3}}$.} 
All background profiles of density, temperature, entropy gradient, and gravitational acceleration $(\tilde{\rho}, \tilde{T}, \nabla \tilde{S}, \text{and } \tilde{g})$ were fit from the PNS model described above with fifth-order to eleventh-order polynomials, hence reproducing the profiles with a good accuracy. Their values at the outer boundary are noted with the letter `o' as a subscript. 
To have a self-consistent anelastic model, the thermal expansion coefficient at constant pressure $\tilde{\alpha}_T= -\left( \frac{\partial \rho}{\partial T}\right)_{P}$ was computed using the following thermodynamic relation: 
\begin{equation}
    \frac{d \ln \ \tilde{T}}{dr} = \frac{1}{c_p}\frac{d}{dr}{\tilde{S}} - \frac{\tilde{\alpha}_T \tilde{g}}{c_p}
    \,,
\end{equation}
where $c_p = \SI{3.6e8}{erg.K^{-1}.g^{-1}}$ is the specific heat capacity at constant pressure, assumed to be uniform. The profiles of the 1D CCSN model and the anelastic reference state agree well (see Appendix~\ref{comp_anel_state}).
For the sake of simplicity, several assumptions are made in this work. In order to describe the buoyancy associated with both entropy and lepton number gradients, we used an effective entropy gradient assuming  the thermal and lepton number diffusivities to be identical. 
In our model, we also assumed that the neutrinos are in a diffusive regime, so that the effects on the dynamics can be modeled by thermal and viscous diffusivities, $\kappa$ and~$\nu$, respectively.
The magnetic diffusivity $\eta$ was also explicitly included in our simulations.
Finally, we assumed all the diffusivities $\kappa$, $\nu,$ and $\eta$ to be constant inside the simulation domain.  

\subsection{Governing equations}
\label{Eq:intro}

We adopted the anelastic approximation to model the flow inside the PNS in order to take the density and effective entropy gradient into account while filtering out sound waves.
The MHD anelastic equations describing the dynamics of the PNS in a rotating frame at an angular frequency $\Omega_0 = \SI{958}{rad.s^{-1}}$ read
\citep{2011JonesAnelasticBench}
\begin{gather}
    \vec{\nabla}\cdot\tilde{\rho}\vec{u}=0\,, \label{eq:1}\\
    \begin{split}
    \dfrac{D \vec{u}}{D t}
     = -\vec{\nabla}\left({\dfrac{P}{\tilde{\rho}}}\right) - 2\Omega_0\vec{e}_z\times\vec{u} &
    -  \tilde{\alpha}_T \tilde{T} \,\frac{s'}{c_p}\, \tilde{g} \,\vec{e}_r \\
    & +\dfrac{1}{\mu_0\,\tilde{\rho}}\left(\vec{\nabla}\times \vec{B}
    \right)\times \vec{B}+ \mathbf{F}^{\nu}\,, \label{eq:2}
\end{split}\\
\tilde{\rho} \tilde{T}\left(\dfrac{D s'}{D t} + \vec{u} \cdot \nabla \tilde{S} \right) =
\kappa\vec{\nabla}\cdot\left(\tilde{\rho}\tilde{T}\vec{\nabla} s'\right) +
\Phi_\nu +
\dfrac{\eta}{\mu_0}\left(\vec{\nabla}
\times\vec{B}\right)^2\,, \label{eq:3}\\
\dfrac{\partial \vec{B}}{\partial t} = \vec{\nabla} \times \left( \vec{u}\times\vec{B}\right)
+\eta \nabla^2 \vec{B}\,, \label{eq:4}\\
\vec{\nabla}\cdot\vec{B}=0\,, \label{eq:5}
\end{gather}
where $\vec{u}$ is the flow velocity, $\vec{B}$ is the magnetic field, $P$ is the pressure, $s'$ is the entropy perturbation, and $\mu_0$ is the vacuum permeability.  
In this system, viscous force and viscous heating are given by $F^{\nu}_i= \tilde{\rho}^{-1} \partial_j \sigma_{ij}$ and $\Phi_\nu= \partial_j u_i \sigma_{ij}$, where $\sigma_{ij} = 2 \tilde{\rho} \nu (e_{ij} - e_{kk} \delta_{ij}/3)$ is the rate of strain tensor and $e_{ij}= (\partial_j u_i + \partial_i u_j)/2$ is the deformation tensor. Tensors are expressed with the Einstein summation convention and the Kronecker symbol $\delta_{ij}$.

\subsection{Numerical methods}

In order to integrate the system of Eqs.~\eqref{eq:1}--\eqref{eq:5} in time, we used the benchmarked pseudo-spectral code MagIC\footnote{\url{https://magic-sph.github.io}} \citep{2002WichtMagic,2012GastineCode,2013SHTNS}.
MagIC solves the 3D MHD equations in a spherical shell using a poloidal-toroidal decomposition for the velocity and the magnetic field,
\begin{align}
\tilde{\rho} \vec{u} &= \vec{\nabla} \times \vec{\nabla} \times \left(W \ \vec{e_r}\right) + \vec{\nabla} \times \left(Z \ \vec{e_r}\right), \\
\vec{B} &= \vec{\nabla} \times \vec{\nabla} \times \left(b \ \vec{e_r}\right) + \vec{\nabla} \times \left(a_j \ \vec{e_r}\right),
\end{align}
where $W$ and $Z$ are the poloidal and toroidal kinetic scalar potentials, respectively, while $b$ and $a_j$ are the magnetic potentials.
The scalar potentials and the pressure $P$ are decomposed on spherical harmonics for the colatitude $\theta$ and the longitude $\phi$ angles, together with Chebyshev polynomials in the radial direction.
The linear terms are computed in the spectral space, while the nonlinear terms and the Coriolis force are computed in the physical space and transformed back to the spectral space. For more detailed descriptions of the numerical method and the associated spectral transforms, we refer to \citet{1981GilmanAnel}, \citet{1997TilgnerJFM}, and \citet{2015CHRISTENSEN245}.

The simulations presented in this paper were performed either using a standard grid resolution of $(n_r,n_{\theta},n_{\phi})=(257,512,1024)$ or a higher resolution of $(n_r,n_{\theta},n_{\phi})=(385,768,1536)$.
The resolution was chosen to ensure that the dissipation scales were resolved. For model \texttt{Standard}, about nine cells resolve the resistive scale as the maxima of viscous and resistive dissipation are at the spherical harmonic orders $l_{\nu}\simeq70$ and $l_{\eta}\simeq100$, respectively. 

\subsection{Initial conditions}

Many core-collapse simulations with a fast-rotating progenitor have shown that the PNS rotates differentially for several hundred milliseconds \citep[e.g.,][]{2003AkiyamaMRI,2006OttRotPNS,2018Obergaulinger,2020Bugli}. In order to sustain differential rotation in our simulations for a similar duration, we forced the outer boundary to rotate according to the initial rotation profile, with a similar method as we used in our incompressible study (paper I). 
The rotation profile was then evolved dynamically inside the simulation domain.
The initial cylindrical rotation profile was inspired by the simulations of \citet{2020Bugli} and was composed of an inner part in solid-body rotation and an outer part in differential rotation with a cylindrical profile, 
\begin{equation}
\Omega(s) = \frac{\Omega_\mathrm{i}}{\left(1 + \left(\frac{s}{0.25r_\mathrm{o}}\right)^{20q_\mathrm{o}}\right)^{0.05}}
\,,
\label{e:omega}
\end{equation}
where $s$ is the cylindrical radius, $q_\mathrm{o}=1.5$ corresponds to the shear rate 
\begin{equation}
    q =  - \frac{s}{\Omega}\frac{d\Omega}{ds}
\end{equation}
at the outer boundary, and $\Omega_\mathrm{i}=\SI{3822}{rad.s^{-1}}$ is the rotation rate of the inner core.
This rotation rate $\Omega_\mathrm{i}$ was computed so that the ratio of total angular momentum over the moment of inertia was equal to the frame rotation rate $\Omega_0$ defined in Sect.~\ref{Eq:intro}.
This was computed with the following formula: 
\begin{equation}
    \Omega_{\rm i} = \frac{I \Omega_0}{\displaystyle\int_V \tilde{\rho} s^2 \left(1 + \left(\frac{s}{0.25r_\mathrm{o}}\right)^{20q_\mathrm{o}}\right)^{-0.05} dV },
\end{equation}
where $s$ is the cylindrical radius, $V$ is the volume of the domain, and $I$ is the moment of inertia of the simulation domain. This initial rotation profile was inspired by those found in the fast-rotating and magnetized core-collapse supernova simulations by \citet{2020Bugli}. The shear rate $q_o=1.5$ might seem high for protoneutron stars, but the shear rate $q$ is lower inside the domain than at the boundary, as already shown in paper I.

For the initial magnetic field, the toroidal component was set to zero and the poloidal component was initialized with a random superposition of modes with spherical harmonics indices $(l,m)$ with a radial dependence based on Fourier modes as in paper I, but with a radial modulation of their amplitude to keep a constant AlfvÃ©n speed, defined by 
\begin{equation}
    v_\mathrm{A} = \frac{B}{\sqrt{\mu_0 \tilde{\rho}}}.
\end{equation}
This initialization implies that the initial magnetic field is stronger in the inner region, which has a higher density (see the left panel of Fig.~\ref{Bphi_snap_vs_init}).  
For all the simulations, the Fourier and spherical harmonic modes were selected so that their wavelength was between $\left[0.3 D, 0.5 D\right],$ with $D$ the shell width, 
\begin{equation}
D= \ro -\ri = \SI{13.7}{km}.
\end{equation}
The initial root mean-square magnetic field strength at the outer boundary ranged from $B_{\rm o} = \SI{1.5e14}{G}$ to $B_{\rm o} =\SI{3.3e14}{G}$ depending on the model. 
This strong magnetic field allowed us to be sure that the MRI was well resolved: the wavelength of the fastest-growing mode in ideal MHD with $B_{\rm o} = \SI{1.5e14}{G}$ reads \citep{1991BalbusMRI} 
\begin{gather}
    \lambda_{\rm MRI} = \frac{8\pi}{q_\mathrm{o} (4-q_\mathrm{o})\Omega_0}\frac{B_{\rm o}}{\sqrt{\mu_0\tilde{\rho}_{\rm o}}} \simeq 0.69 D \simeq 9.5 \times 10^{5}\ \text{cm}.
\end{gather}
Our strong initial magnetic field may represent the magnetic field generated by the first MRI amplification on small scales found in local models of PNS \citep{2009ObergaulingerMRI,2015GuiletBuoy,2016RembiazMRIMaxB}, as explained in paper I.

\subsection{Boundary conditions}

We assumed nonpenetrating boundary conditions ($u_r=0$). At the inner boundary, we used a stress-free condition, where the viscous stress vanished. For the outer boundary, we forced $u_{\phi}$ to match the initial profile at all times, and the other components of the velocity were set to zero, exactly like in our previous setup (paper I).
For the magnetic field, we used insulating boundary conditions (matching a potential field outside the domain), while the entropy perturbations were set to zero at both boundaries.

\subsection{Physical parameters and dimensionless numbers}

\begin{figure}[h]
\centering
\resizebox{\hsize}{!}
            {
     \includegraphics{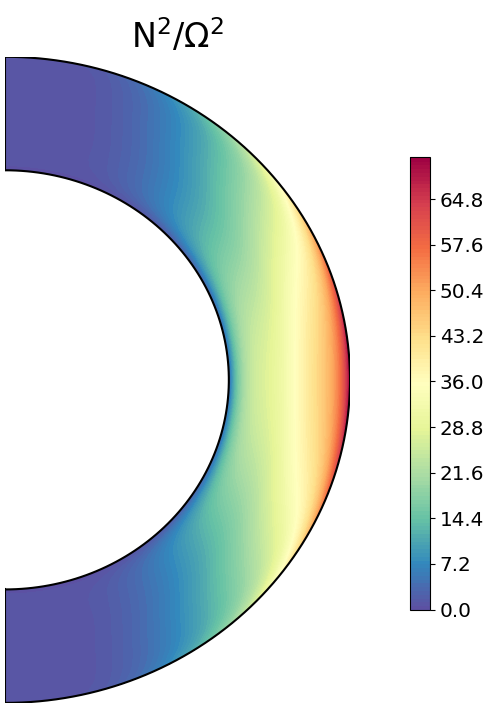}}
      \caption{Ratio of the squared Brunt-V\"ais\"al\"a frequency $N^2$ (Eq.~\ref{e:brunt}) to the squared initial rotation frequency $\Omega^2$ (Eq.~\ref{e:omega}). The buoyancy influence is strongest in the red regions.}
         \label{N2Omega}
\end{figure}

In fast-rotating fluids, the impact of stable stratification on the MRI is characterized by the ratio of the Brunt-V\"ais\"al\"a frequency squared $N^2$ over the rotation frequency squared $\Omega^2$, with 
\begin{equation}
    N^2 \equiv - \frac{\tilde{g}}{\tilde{\rho}} \left(\left. \frac{\partial \tilde{\rho}}{\partial \tilde{S}}\right|_{\tilde{P},Y_e} \frac{d \tilde{S}}{d r} + \left. \frac{\partial \tilde{\rho}}{\partial Y_e}\right|_{\tilde{P},\tilde{S}} \frac{d Y_e}{d r} \right)
    \,,
    \label{e:brunt}
\end{equation}
where $Y_e$ is the electron fraction and $\tilde{P}$ is the pressure of the reference state of the PNS model.  
In our model, this ratio is higher than for the model studied in \citet{2015GuiletBuoy}, especially at the equator (see Fig.~\ref{N2Omega}).
Our stratification remains small compared to typical values of ${N^2}/{\Omega^2} \sim 10^{3}-10^{4}$ in the radiative zone of intermediate and massive stars \citep{2019FullerTaylerSpruit}.   
The Brunt-V\"ais\"al\"a frequency at the outer boundary is $N_{\rm o} = \SI{4136}{s^{-1}}$.
Accordingly, we might expect the buoyancy to dampen MRI-driven turbulence \citep{2015GuiletBuoy}, but the thermal diffusivity $\kappa$ also plays an important role as it reduces the impact of
the stable stratification on radial scales smaller than a critical length $L_{\rm c}$. One way to estimate $L_{\rm c}$ is to compare the timescale for thermal diffusion to the timescale of gravity waves, from which we obtain
\begin{equation}
    L_{\rm c} = \sqrt{\frac{\kappa}{N_{\rm o}}}. 
    \label{e:Lc}
\end{equation}
For the uniform thermal and viscous diffusivities, we used the values from the 1D CCSN model taken in the middle of the stable zone $r\simeq \SI{3.3e6}{km}$, which are $\kappa= \SI{1.61e14}{cm.s^{-2}}$ and $\nu=\SI{8.03e11}{cm.s^{-2}}$, respectively.
While all our models share the same constant kinematic viscosity, we considered three different values for the thermal diffusivity, 
$\kappa = \{\num{1.61e14}, \num{4.02e13},\num{8.03e12}\}\,\si{cm.s^{-2}}$,
which means that the thermal Prandtl number is
\begin{equation}
    Pr = \frac{\nu}{\kappa} = \left\{0.005,\,0.02,\,0.1\right\}
    \,.
\end{equation}
Our standard value of the thermal Prandtl number $Pr = 0.005$ corresponds to the value resulting from the 1D core-collapse supernova model \citep{2014Hudepol}.
With these values, the critical length ranges from $L_{\rm c} = 0.14 L = \SI{1.97e5}{\cm}$ to $L_{\rm c}=0.031 L = \SI{4.4e4}{\cm}$. In our simulations, the ratio of the Brunt-V\"ais\"al\"a frequency $N_{\rm o}$ at the outer boundary to frame rotation $\Omega_0$ was fixed by the following formula:
\begin{equation}
    \frac{N_{\rm o}}{\Omega_0} = \sqrt{\frac{-Ra E^2}{Pr}},
\end{equation}
where the Ekman number $E$ (characterizing the importance of viscosity over the Coriolis force) reads
\begin{equation}
E = \frac{\nu}{\Omega_0 \ D^2} = 4.44 \times 10^{-4},
\end{equation}
and the Rayleigh number $Ra$ is defined by
\begin{equation}
    Ra = -\frac{\tilde{\alpha}_{T,\mathrm{o}} T_\mathrm{o} g_{\rm o} D^4}{c_p \nu \kappa} \left | \frac{d \tilde{S}}{dr} \right|_{\rm o} 
    \,.\label{eq:ra_N}
\end{equation}
The Ekman number was kept identical for all the models presented in this article. 
Therefore, to keep the Brunt-V\"ais\"al\"a frequency constant between our models with buoyancy effects while varying the thermal Prandtl number,  
the Rayleigh number was also varied from $Ra=-\SI{4.73e5}{}$ to $Ra=-\SI{9.46e6}{}$.

With these dimensionless numbers, we can estimate the regime of boundary forcing in our simulations. The viscous timescale $\tau_\nu = L^2/\nu$ can be compared to the Ekman spin-up timescale $\tau_E = \sqrt{{L^2}/({\Omega_0 \nu})}$ and the Eddington-Sweet circulation timescale 
$\tau_{ED} = L^2  N_o^2/({\kappa \Omega_0^2})$ \citep{2021Gouhier,2022GouhierMHD}. This gives 
\begin{equation}
    \frac{\tau_E}{\tau_\nu} = \sqrt{E} = 0.021 \ll \frac{\tau_{ED}}{\tau_\nu} = Pr\left(\frac{N_o}{\Omega_0}\right)^2 = 0.093  <  \frac{\tau_\nu}{\tau_\nu}=1
\end{equation}
for the simulation \texttt{Standard}.
This order (i.e., $\tau_E \ll \tau_{ED} \ll \tau_\nu$) corresponds to the Eddington-Sweet regime, but as Pr increases for some simulations, it approaches the viscous regime (i.e., $\tau_E \ll \tau_\nu \ll \tau_{ED}$).

We used a constant resistivity of $\eta=\SI{5.0e10}{cm.s^{-2}}$, which gives a magnetic Prandtl number 
\begin{equation}
Pm = \frac{\nu}{\eta} = 16 \,.     
\end{equation}
As in paper I, the choice of this
value for our simulations was forced by the high cost of simulating high $Pm$ models, and it lies well below a realistic parameter regime. 
The neutrino viscosity is very high compared to the typical resistivity of a degenerate electron gas inside a PNS. Realistic estimates of the magnetic Prandtl number predict  $Pm \approx 10^{13}$ \citep{1993ThompsonPNS,2007MasadaMRIneutrinolinear},
which cannot be reached in direct 
numerical simulations.
The magnetic Reynolds number that characterizes the relative importance of induction to magnetic diffusion is
\begin{equation}
Rm = \frac{D^2 \ \Omega_0}{\eta} = 3.6 \times 10^{4},
\end{equation}
and, for the same reasons, it is quite small compared to realistic estimates.
Overall, the only parameters that were varied in this study were the thermal Prandtl number $Pr$ and the Rayleigh number $Ra$, which in two simulations was also set to zero to remove the thermal stratification and allow a clearer assessment of the impact of buoyancy in our models.

\section{Typical anelastic simulation}

\label{results}

\subsection{Quasi-stationary state of an MRI-driven dynamo}

\begin{figure*}[h]
\centering
\resizebox{\hsize}{!}
            {
\includegraphics{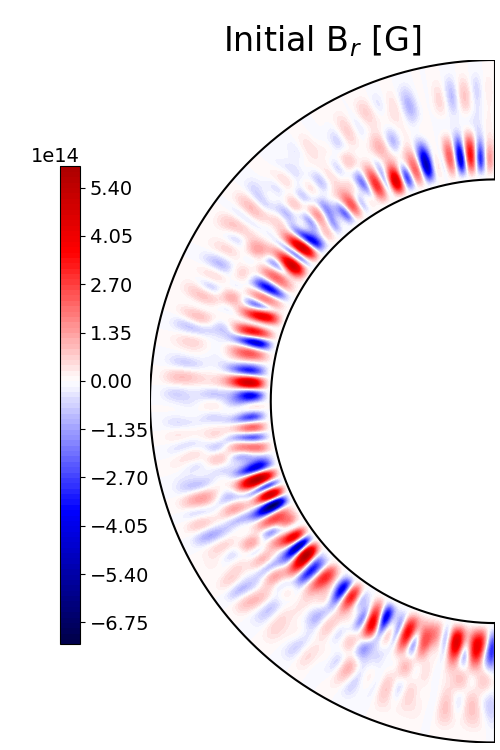}
\includegraphics{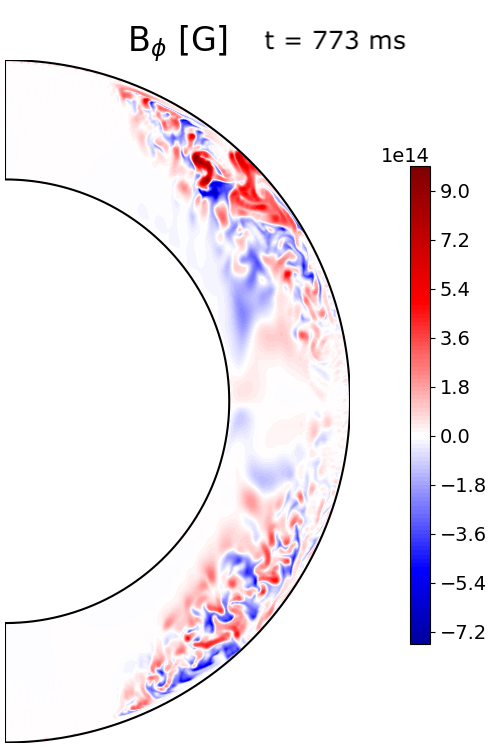}}
     \caption{Meridional slices at $\phi = 0$  of the initial poloidal magnetic field $B_\mathrm{r}$ (left) and of the toroidal magnetic field $B_\phi$ at $t=\SI{773}{ms}$ (right) for model \texttt{Standard}.}
    \label{Bphi_snap_vs_init}
\end{figure*}

\begin{figure}[h]
\centering
\resizebox{\hsize}{!}
            {
     \includegraphics{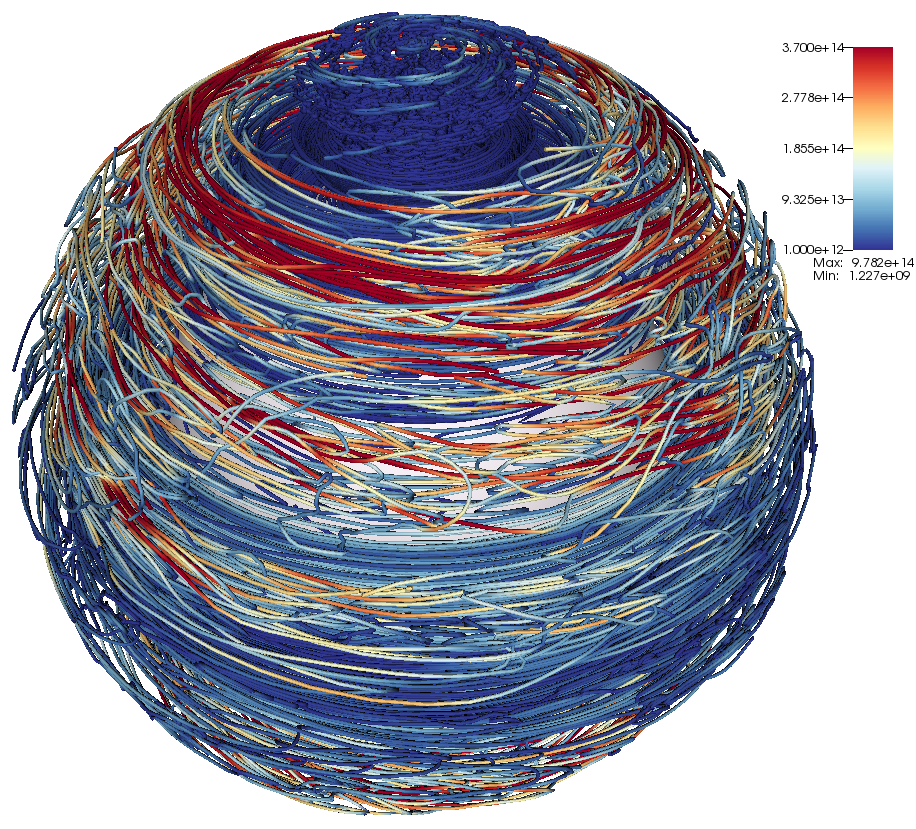}}
      \caption{3D rendering of the magnetic field lines in simulation \texttt{Standard} at $t=\SI{773}{ms}$. The color represents the magnetic field amplitude in Gauss.}
    \label{3D_snap}
\end{figure}

We start by describing the magnetic field produced by one fiducial simulation in which we obtain an MRI-driven dynamo, hereafter called model \texttt{Standard}. For this model, we used the standard diffusivities and an initial magnetic field intensity of $B_{\rm o} = \SI{1.5e14}{G}$ (see Table \ref{table:annex}). 
Figures \ref{Bphi_snap_vs_init} (right panel) and \ref{3D_snap}  show selected snapshots of the toroidal magnetic field and the magnetic field lines that highlight the complex geometry of the MRI-driven turbulence. 
On the meridional cuts in Fig.~\ref{Bphi_snap_vs_init},
the spatial distribution of the small-scale turbulence is particularly striking for several reasons. First, almost no turbulence is observed in the equatorial plane, while strong turbulence develops in mid-latitude regions. The same feature can be seen for the kinetic turbulence on the radial velocity~$u_r$.
This phenomenon can be explained by the influence of the buoyancy force (see Sect.~\ref{ss:buo} for more details). Second, the high-density regions at lower radii are devoid of MRI turbulence because only wide patches of magnetic field can be observed there without small-scale turbulence. This is more unexpected and is discussed in Sect.~\ref{ss:loc_turb}. Moreover, the region close to the rotation axis is in solid-body rotation and is therefore stable to the MRI, which leads to no turbulence, as expected. Finally, the comparison 
between the initial and saturated magnetic field shown in Fig.~\ref{Bphi_snap_vs_init} suggests that the magnetic field has lost the memory of its initial configuration.
The contrast between the equatorial plane and mid-latitude regions is present in the magnetic field lines as well, where strong and elongated field lines in the azimuthal direction can be seen in the mid-latitude regions of Fig.~\ref{3D_snap}. These field lines are generated by the expected winding of the magnetic field by the shear.
Some small-scale loops of field lines can also be seen, with a weaker magnetic field in the mid-latitude regions.

\begin{figure}[h]
\centering
\resizebox{\hsize}{!}
            {
     \includegraphics{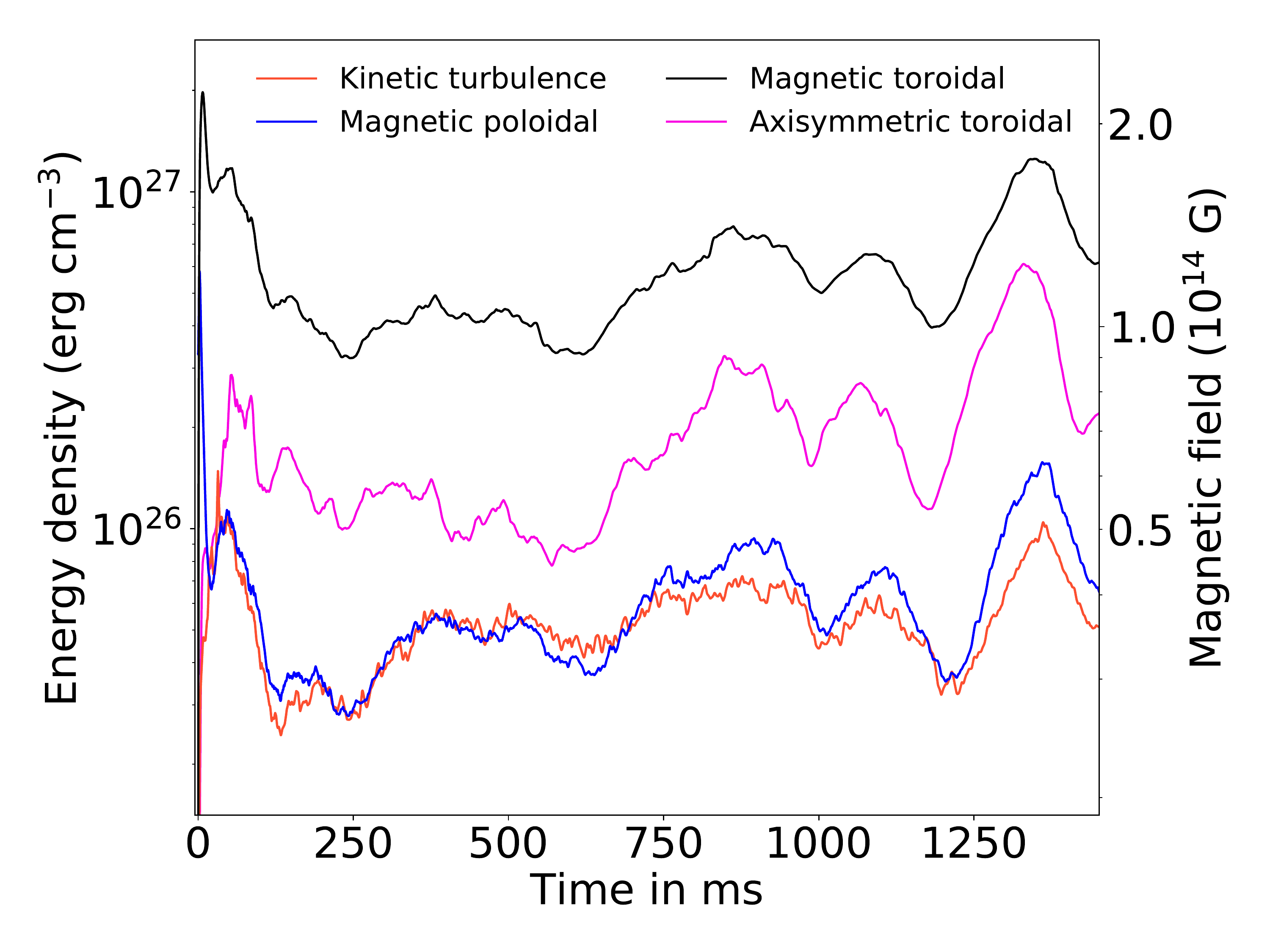}}
      \caption{Time evolution of the magnetic and turbulent kinetic energy densities for model \texttt{Standard} with $B_\mathrm{o} = \SI{1.5e14}{G}$. The black and blue lines are the toroidal and poloidal contributions of the magnetic energy density, and the red line is the turbulent kinetic energy density (i.e., the energy of the nonaxisymmetric component of the velocity). The magenta line is the axisymmetric contribution to the toroidal magnetic energy density.}
         \label{Ts_energies}
\end{figure}

For a more quantitative assessment, we now examine the energetics of this MRI-driven dynamo. 
Figure \ref{Ts_energies} shows the time evolution of the toroidal and poloidal magnetic energy densities and the turbulent kinetic energy density. 
The latter was computed by subtracting the contribution of the axisymmetric rotation from the averaged kinetic energy density in order to separate the differential rotation and the MRI-driven turbulent flow. After several hundred milliseconds, we obtain a statistically stationary state 
with a mean magnetic field of $B\simeq\SI{1.4e14}{G}$ in the full simulation volume. 
If we reduce the volume to take the localization of the turbulence into account, the resulting mean magnetic field is $B\simeq \SI{2.25e14}{G}$. 

In order to compare the intensity in the full volume to our previous incompressible study (paper I), where the density $\rho_o = \SI{4e13}{g \ cm^{-3}}$ was higher, we used the dimensionless magnetic field strength defined by the Lorentz number,
\begin{equation}
\mathcal{B} = \frac{B}{\sqrt{\tilde{\rho}_{\rm o} \mu_0} D \Omega}.
\end{equation}
We have $\mathcal{B} \simeq 0.096$ for model \texttt{Standard} and $\mathcal{B} \simeq 0.064$ for paper I, which is lower but on the same order of magnitude. This difference is discussed in further details in Sect.~\ref{ss:incompressible}. 
Compared to paper I, most of the ratios of the energies or magnetic fields are similar.
The kinetic energy is about ten times lower than the total magnetic energy, and the toroidal magnetic field is about $\sim 2.5$ times larger than the poloidal magnetic field. The main difference with paper I is the increased axisymmetric component of the toroidal magnetic field, which is higher than the total poloidal magnetic field and is almost equal to the nonaxisymmetric toroidal magnetic field, as we discuss in Sect.~\ref{ss:dens}. 
An interesting new feature compared to paper I are synchronous oscillations of the magnetic and kinetic energy densities. These oscillations suggest a dynamo cycle, as we show in Sect.~\ref{alphaomega}.

\begin{figure}[h]
\centering
\resizebox{\hsize}{!}
    {
     \includegraphics{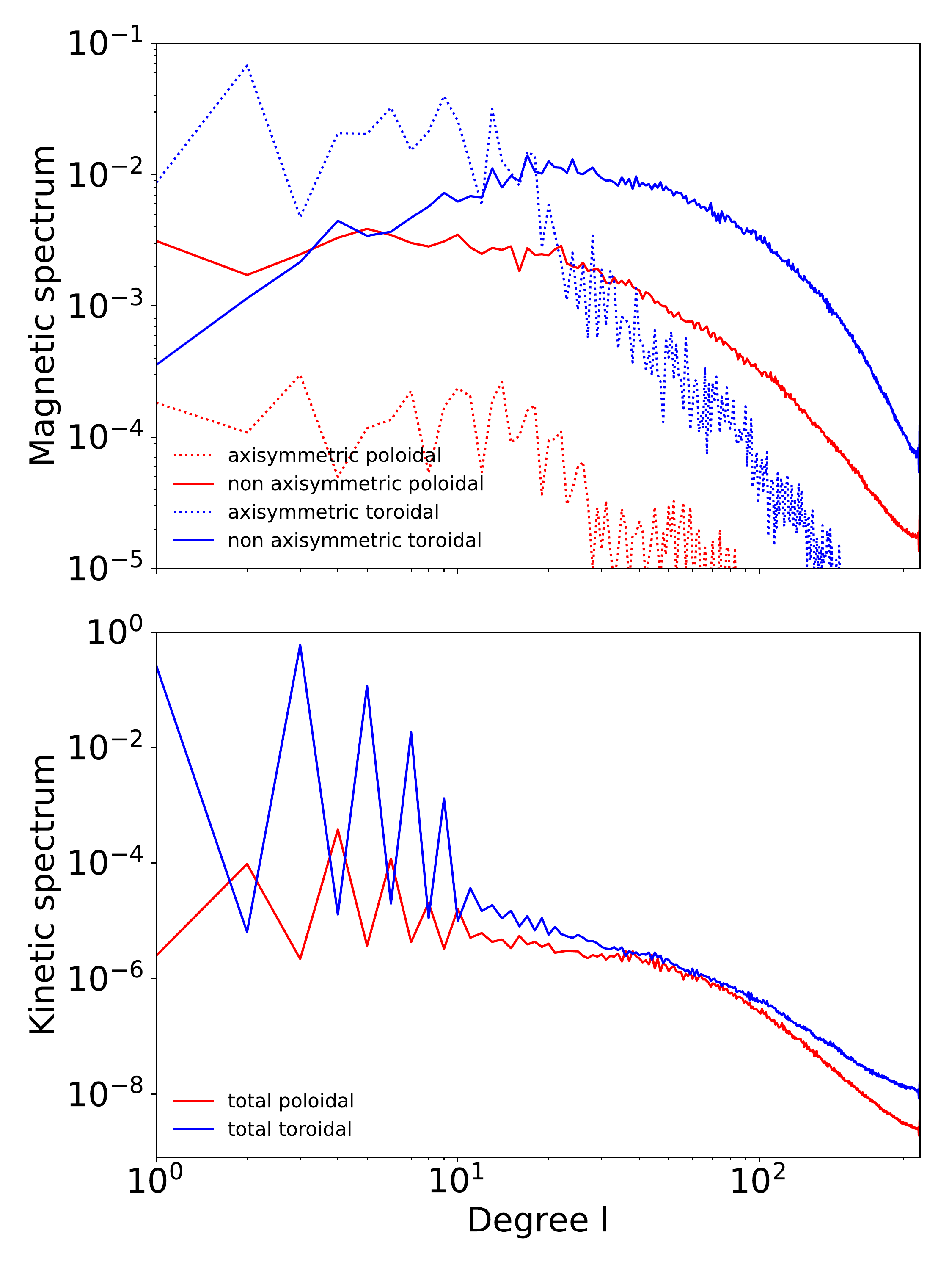}
     }
    \caption{Instantaneous volume-averaged spectra of magnetic and kinetic energies at $t=\SI{773}{ms}$. Top: Spectrum of the poloidal magnetic energy (red) and the toroidal magnetic energy (blue) normalized by the total magnetic energy as a function of the spherical harmonics degree~$l$. The dotted (solid) lines correspond to the axisymmetric (nonaxisymmetric) contributions to these energies. 
    Bottom: Spectrum of the poloidal kinetic energy (red) and the toroidal kinetic energy (blue) normalized by the total kinetic energy as a function of the spherical harmonics order $l$.}
    \label{Standard_spectra}
\end{figure}

To understand how the magnetic and kinetic energies are distributed over different scales, instantaneous toroidal and poloidal spectra are presented in Fig.~\ref{Standard_spectra}. 
On the small scales, the nonaxisymmetric poloidal and toroidal magnetic spectra are similar to those of paper I.
For the poloidal component, the nonaxisymmetric contribution dominates at all scales, as in paper I.
By contrast, at large scales, the toroidal magnetic spectrum is dominated by the axisymmetric component, with a particularly strong quadrupole. These strong axisymmetric modes are a new feature of the toroidal spectrum and are linked to the stronger axisymmetric toroidal energy and the mean-field dynamo described in the next section. 
Overall, these spectra show that the total magnetic energy comes essentially from the axisymmetric contribution at large scales and the nonaxisymmetric contribution at small scales.

Axisymmetric structures that are symmetric with respect to the equatorial plane, such as the rotation and meridional circulation, translate into oscillations between odd and even harmonic order~$l$ visible in the kinetic spectra \citep{1993GubbinsSymmetryProperties}.
The kinetic spectra are comparable to those of paper I because small and intermediate scales are dominated by turbulence, while differential rotation dominates at large scales.

\begin{figure}[h]
\centering
\resizebox{\hsize}{!}
    {
     \includegraphics{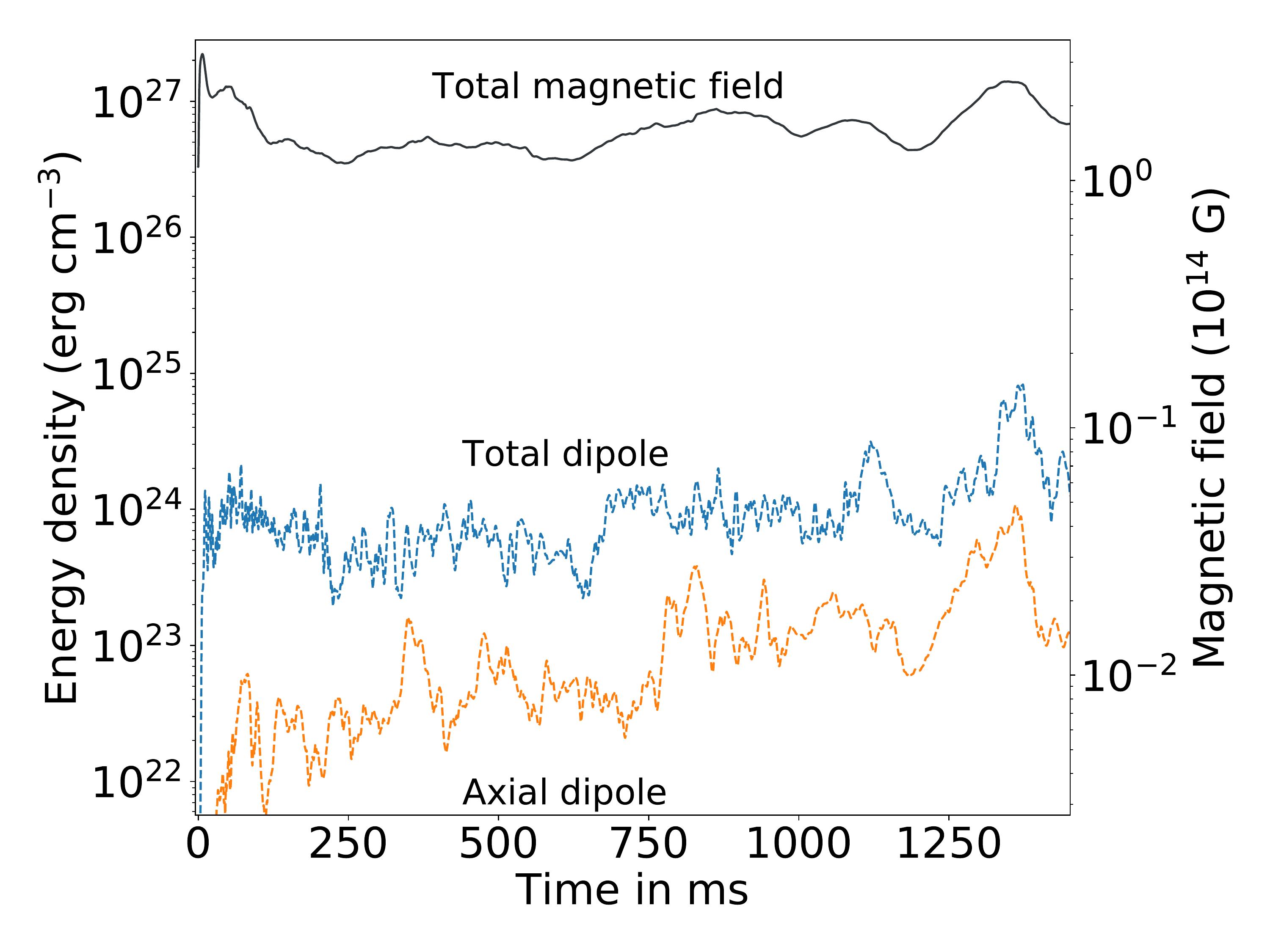}}
    \caption{Time evolution of the axial dipole (dashed orange), total dipole (dashed blue), and total magnetic (solid black) energy densities of the model \texttt{Standard}.}
    \label{Typical_dipole_ts}
\end{figure}

We now highlight the time evolution of the magnetic dipole because it is the component inferred by observations. 
The dipole energy translates into an intensity $B_\mathrm{dip} \simeq \SI{6.3e12}{G}$,
which represents $\simeq 4.5\%$ of the total magnetic field (Fig.~\ref{Typical_dipole_ts}). This intensity may seem weak, but the dimensionless magnetic field strength~$\mathcal{B}_{\rm dip}$ corresponding to this dipole intensity gives $\mathcal{B}_{\rm dip} = 0.0043 $, which is higher than the value $\mathcal{B}_{\rm dip}\simeq 0.0032$ obtained for paper I. 
Furthermore, the evolution of the dipole energy suggests that the dipole is tilted toward the equator because the axial dipole is twice lower than the average dipole.
With the same method as paper I, an averaged tilt angle of $\theta_\mathrm{dip} \approx 100 \degree$ can be computed from the magnetic dipole moment.
Overall, this model produces results that are qualitatively consistent with those in paper I, with a stronger dimensionless dipole. 

\begin{figure}[h]
\centering
\resizebox{\hsize}{!}
            {
     \includegraphics[width=0.9\textwidth]{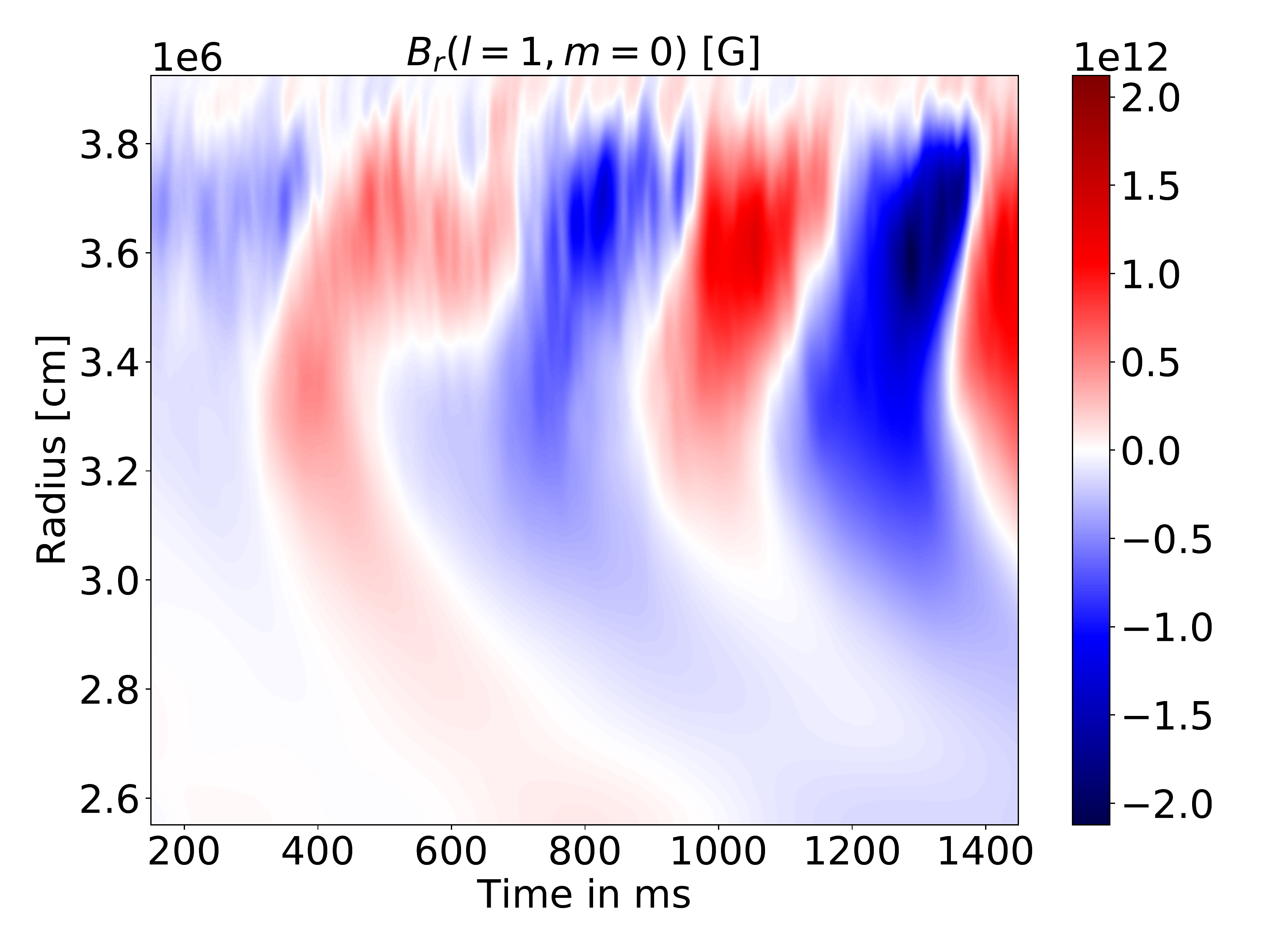}
     }
\resizebox{\hsize}{!}
            {
     \includegraphics[width=0.9\textwidth]{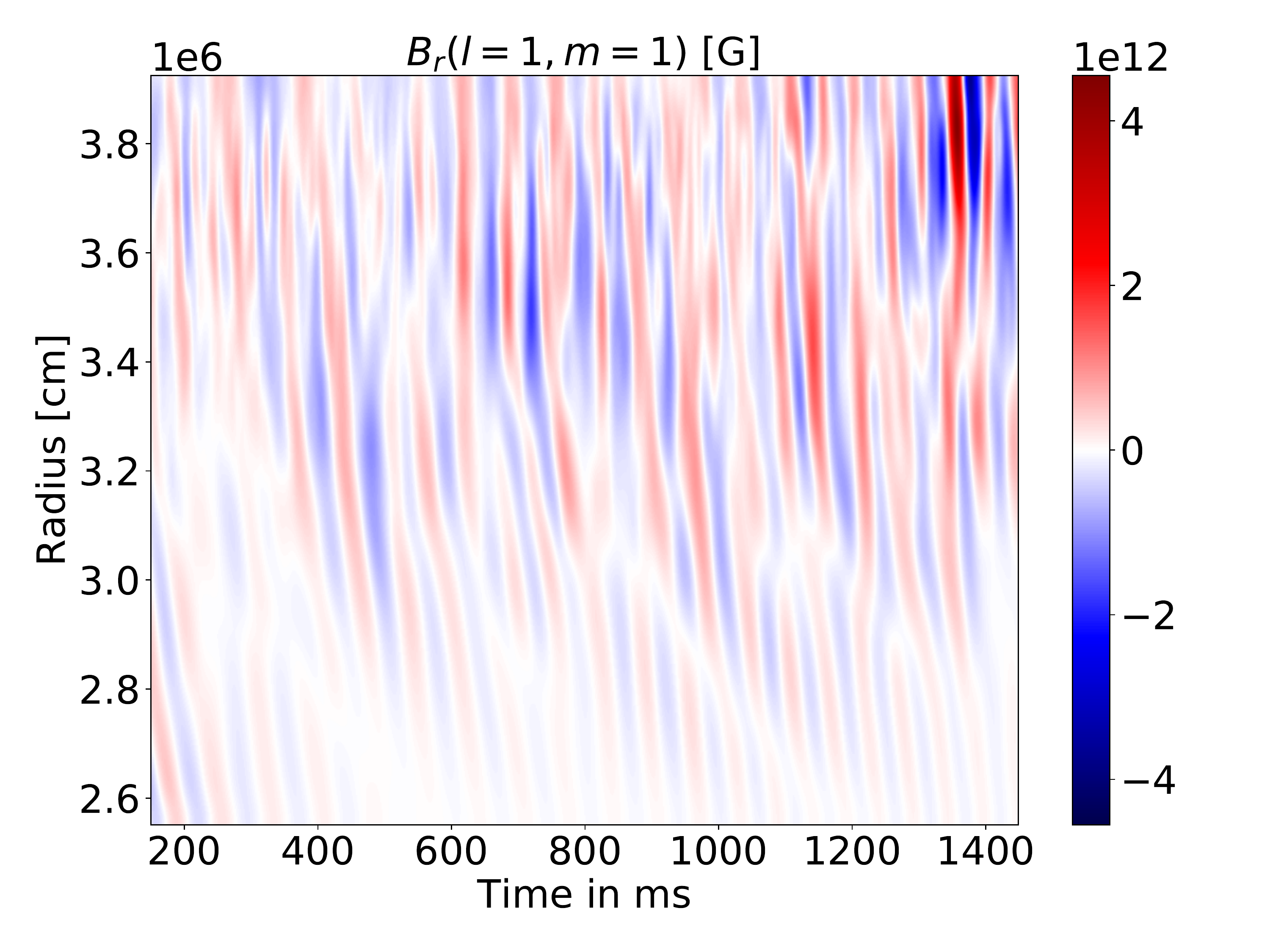}
     } 
    \caption{Radius-time diagram of the dipolar modes.
    Top: Radius-time diagram of the axial dipolar mode $B_r(l=1,m=0)$.
        The period of the dipole reversal is $P_\mathrm{dip} \simeq \SI{410}{ms}$. Bottom: Radius-time diagram of the equatorial dipolar mode amplitude $B_r(l=1,m=1)$ at a given azimuth $\phi=0$ in the rotating frame of the simulation.
     The fast oscillations are due to the rotation of this dipole mode.}
     \label{Dipole_diagram}
\end{figure}

To study the dipole geometry in more detail, we show in Fig.~\ref{Dipole_diagram} the time evolution and radial dependence of the modes $(l=1,m=0)$ and $(l=1, m=1)$.
The dipole is mainly present in the low-density region (outer radii), where the turbulence is the strongest. Both modes are radially coherent and form large-scale structures. The equatorial dipole rotates at a different rotation speed than the simulation frame, which leads to oscillations of the equatorial dipole. The timescale of these oscillations (about $\SI{50}{ms}$) is much longer than the rotation period of the frame, which means that the magnetic dipole rotates with a frequency close to that of the simulation frame. 
The time evolution of the axisymmetric component of magnetic dipole is characterized by slower periodic reversals, which again suggests that a mean-field oscillatory behavior is present in the simulations.

\subsection{ $\alpha\Omega$ dynamo}
\label{alphaomega}

The previous subsection indicated a mean-field dynamo by pointing out an oscillatory behavior.
The general mean-field theory has been developed by \citet{1978moffattfield} and \citet{1980KrauseMeanField} and has been widely used to study dynamos. We used the mean-field concept to understand which processes dominate the generation of the mean magnetic field.
The basic idea of a mean-field dynamo is that a large-scale magnetic field is generated by small-scale turbulence. The velocity and magnetic fields are therefore decomposed into a mean and a small-scale component, which we represent using the following notation: $\vec{X}=\overline{\vec{X}}^{\phi}+\vec{X}'$. The definition of \emph{mean} here is the axisymmetric average operator noted $\overline{\cdot}^{\phi}$, which verifies the Reynolds averaging rules. 
The approach of the mean-field theory is to expand the electromotive force (EMF) $\vec{\mathcal{E}}$ only in terms of the mean quantities ($\overline{\vec{u}}^\phi$ and $\overline{\vec{B}}^\phi$) and the statistical properties of the fluctuating quantities ($\vec{u}'$ and $\vec{B}'$). 
We now focus in more detail on the characterization of this mechanism.  
The most common realization of a mean-field dynamo with differential rotation is the so-called $\alpha\Omega$ dynamo. The $\Omega$ effect corresponds to the shearing of the magnetic field by differential rotation that generates a toroidal magnetic field from a poloidal field. With our cylindrical differential rotation, the $\Omega$ effect reads
\begin{equation}
      \frac{\partial\overline{B_\phi}^{\phi}}{\partial t} = s \overline{B_s}^{\phi}  \frac{d \Omega}{d s}
,\end{equation}
and it should induce an anticorrelation between the radial field $\overline{B_s}^{\phi}$ in cylindrical coordinates and the azimuthal magnetic field~$\overline{B_\phi}^{\phi}$.

The $\alpha$ effect comes instead from the closure relation of the mean EMF,
\begin{equation}
    \vec{\mathcal{E}}=\overline{\vec{u}' \times \vec{B}'}^\phi
    \,,    
\end{equation}
where $\vec{u}'=\vec{u} - \overline{\vec{u}}^\phi$ and $\vec{B}'=\vec{B} - \overline{\vec{B}}^\phi$,
expressed as a function of the mean magnetic field $\overline{\vec{B}}^\phi,$
\begin{equation}
    \mathcal{E}_i = \alpha_{ij} \overline{{B}}^\phi_j + \beta_{ij} \left(\overline{\vec{\nabla} \times \vec{B}}^{\phi}\right)_j 
    \,,
    \label{eq:emf}
\end{equation}
where $\alpha_{ij}$ and $\beta_{ij}$ are tensors that do not depend on $\overline{\vec{B}}^\phi$ , and $i,j$ refer to spherical $r,\ \theta,\ \phi$ or cylindrical coordinates $s, \phi,\ z$. 
The diagonal components of the $\alpha$ tensor correspond to the component of the EMF in the direction of the mean magnetic field, and their effect is physically described as 
the twisting of the mean magnetic field lines by the cyclonic turbulence, which forms magnetic field loops that can generate poloidal magnetic field from the toroidal magnetic field and vice versa.

In our case, the generation of the poloidal magnetic field by this effect can be seen as a correlation between the toroidal component of the EMF and the toroidal component of the magnetic field $B_\phi$ in the form of ${\mathcal{E_\phi}} = \alpha_{\phi \phi} \overline{{B_\phi}}^\phi$. 
The diagonal components of the $\beta$ tensor are in the direction of the mean current $\overline{\vec{J}}^{\phi} = \mu_0^{-1} \nabla \times \overline{\vec{B}}^{\phi}$ , and their effect is physically described as a turbulent diffusivity, which adds to the magnetic diffusivity $\eta$. 
Another effect that has been proposed to complete the dynamo loop in the case of the MRI is the nondiagonal resistivity $\beta_{\phi s}$, which could generate a poloidal field from the toroidal field \citep{2008LesurMRINondiagonalresistivity}.  
In this case, the azimuthal component of the electromotive force $\mathcal{E}_\phi$ would be correlated with the radial current $\overline{J_s}^{\phi}$ in cylindrical coordinates. 
One (or both) of these two effects, in combination with the $\Omega$ effect, may therefore be expected to complete the dynamo cycle, leading to a self-sustained magnetic field. 

\begin{figure}[h]
\centering
\resizebox{\hsize}{!}
            {
     \includegraphics{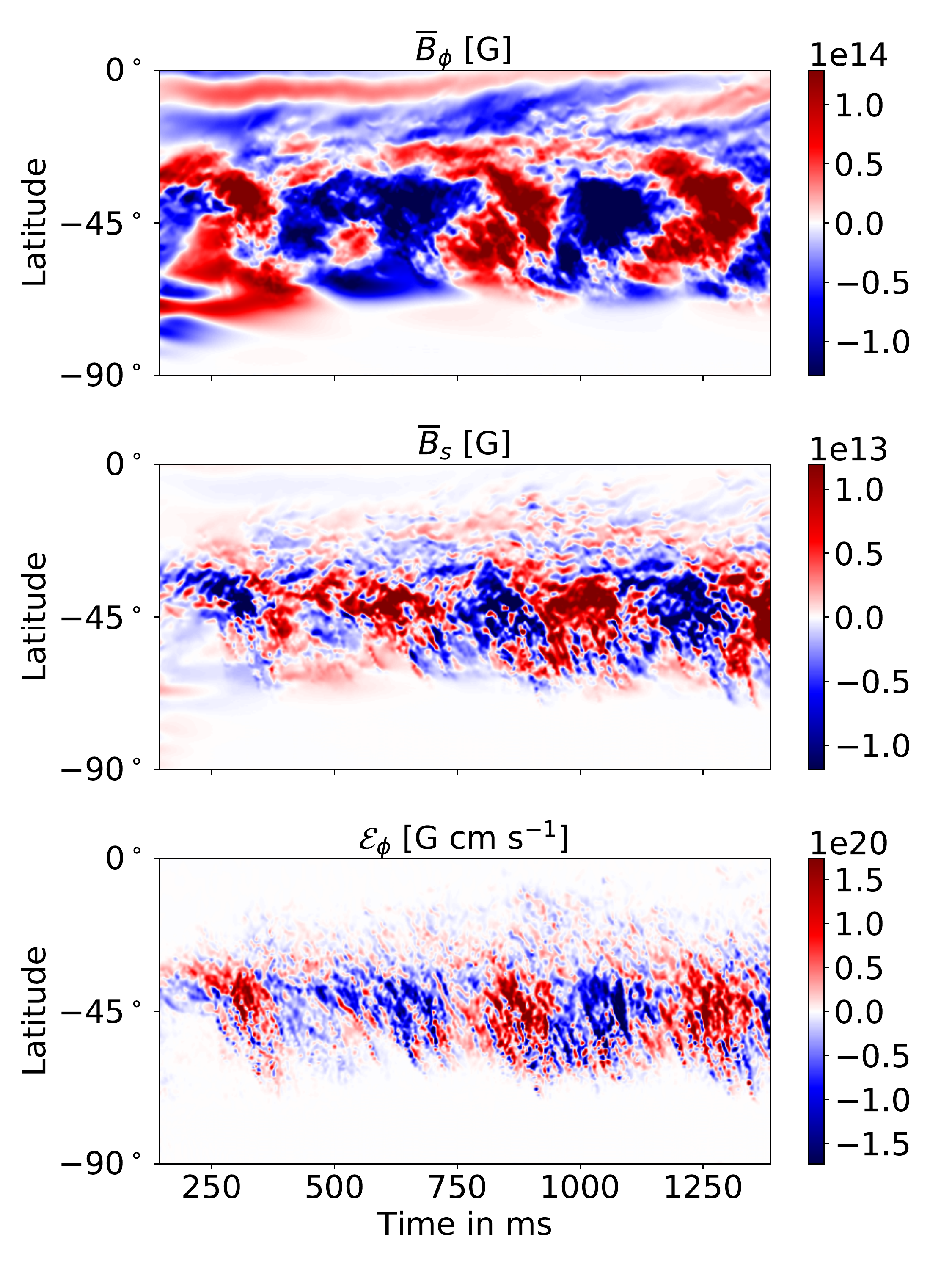}}
     \caption{Butterfly diagrams of $\overline{B_\phi}^\phi$ (top), $\overline{B_s}^\phi$ (middle), and $\mathcal{E}_\phi$ (bottom) in the southern hemisphere of model \texttt{Standard}. Each butterfly diagram is computed at 
     the same spherical radius $r=0.86\, \ro$ in the turbulent region.}
    \label{Butterfly_diagram}
\end{figure}

The usual way to characterize an $\alpha\Omega$ dynamo is to compute space-time diagrams of $B_\phi$, $B_s$ , and $\mathcal{E}_\phi$, which are often referred to as \emph{\textup{butterfly}} diagrams. 
Figure \ref{Butterfly_diagram} shows these azimuthally averaged quantities in the southern hemisphere at $r=0.86\, \ro$ (which lies in the middle of the turbulent region). First, the butterfly diagrams show large coherent structures in the mid-latitudes, which
is consistent with a mean-field oscillatory dynamo with a period of 
$P \simeq \SI{410}{ms}$. This period is estimated by taking the peaks of the mean toroidal magnetic field at different colatitudes and averaging the frequencies for each hemisphere.
The visual inspection of the butterfly diagrams suggests that the dynamo can be interpreted as an $\alpha\Omega$ dynamo.
The components $\overline{B_\phi}^{\phi}$ and $\overline{B_s}^{\phi}$ are anticorrelated, which shows that $\overline{B_\phi}^{\phi}$ is mainly generated by the $\Omega$ effect.
On the other hand, the electromotive force $\mathcal{E}_\phi$ is correlated with $\overline{B_\phi}^{\phi}$, which suggests that the $\alpha$ effect plays an important role.

\begin{figure}[h]
\centering
\resizebox{\hsize}{!}
    {
    \includegraphics{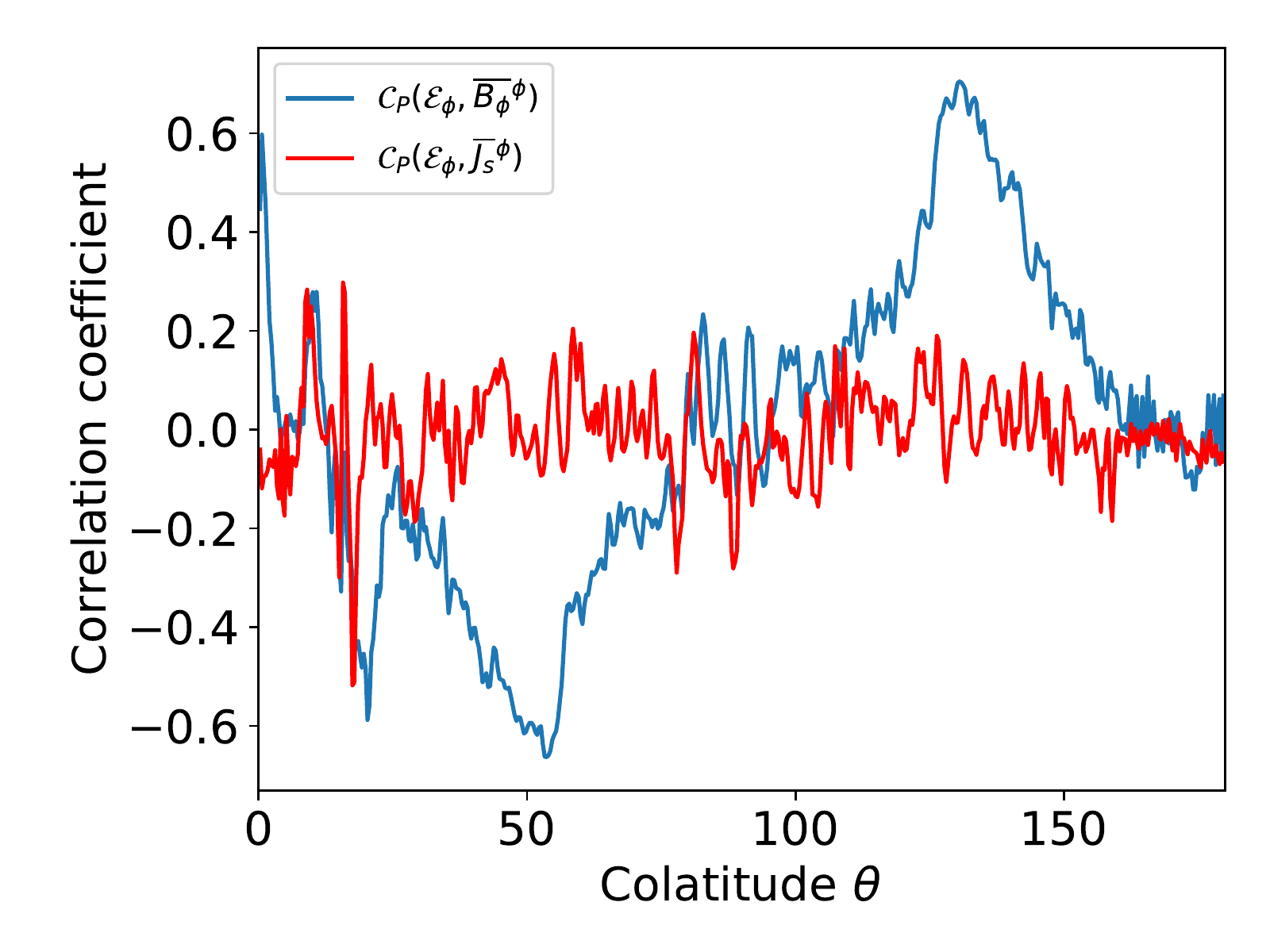}}
    \caption{Time-averaged correlation coefficients (Eq.~\ref{eq:correlation}) between $\mathcal{E}_\phi$ and $\overline{B_\phi}^\phi$ (blue) and between $\mathcal{E}_\phi$ and $\overline{J_s}^{\phi}$ (red) taken at $r=0.86\,\ro$ in model \texttt{Standard}.}
    \label{alpha_correlations}
\end{figure}

To corroborate the visual correlations of the butterfly diagrams, we computed the Pearson correlation coefficient between two quantities $X$ and $Y$ with the following formula:
\begin{equation}
    \mathcal{C}_P(X,Y) = \frac{\int_t (X - \langle X \rangle_t) dt \int_t (Y - \langle Y \rangle_t) dt }{\sqrt{(\int_t (X - \langle X \rangle_t)^2 dt)}\sqrt{(\int_t (Y - \langle Y \rangle_t)^2 dt)}}
    \label{eq:correlation}
,\end{equation}
where $\langle \cdot \rangle_t$ represents a time average.
Fig.~\ref{alpha_correlations} shows the correlation coefficients between $\mathcal{E}_\phi$ and $\overline{B_\phi}^\phi$ and the radial current $\overline{J_s}^\phi$ in order to test the nondiagonal resistivity hypothesis. 
We find that $\overline{B_\phi}^\phi$ and $\mathcal{E}_\phi$ are well correlated, while no correlation is found between $\overline{J_s}$ and $\mathcal{E}_\phi$. This means that the $\alpha$ effect is prominent in our simulations. Moreover, the antisymmetry of the correlation matches the expected symmetry of the components of the tensor $\alpha$. 

\begin{figure}[h]
\centering
\resizebox{\hsize}{!}
            {
     \includegraphics{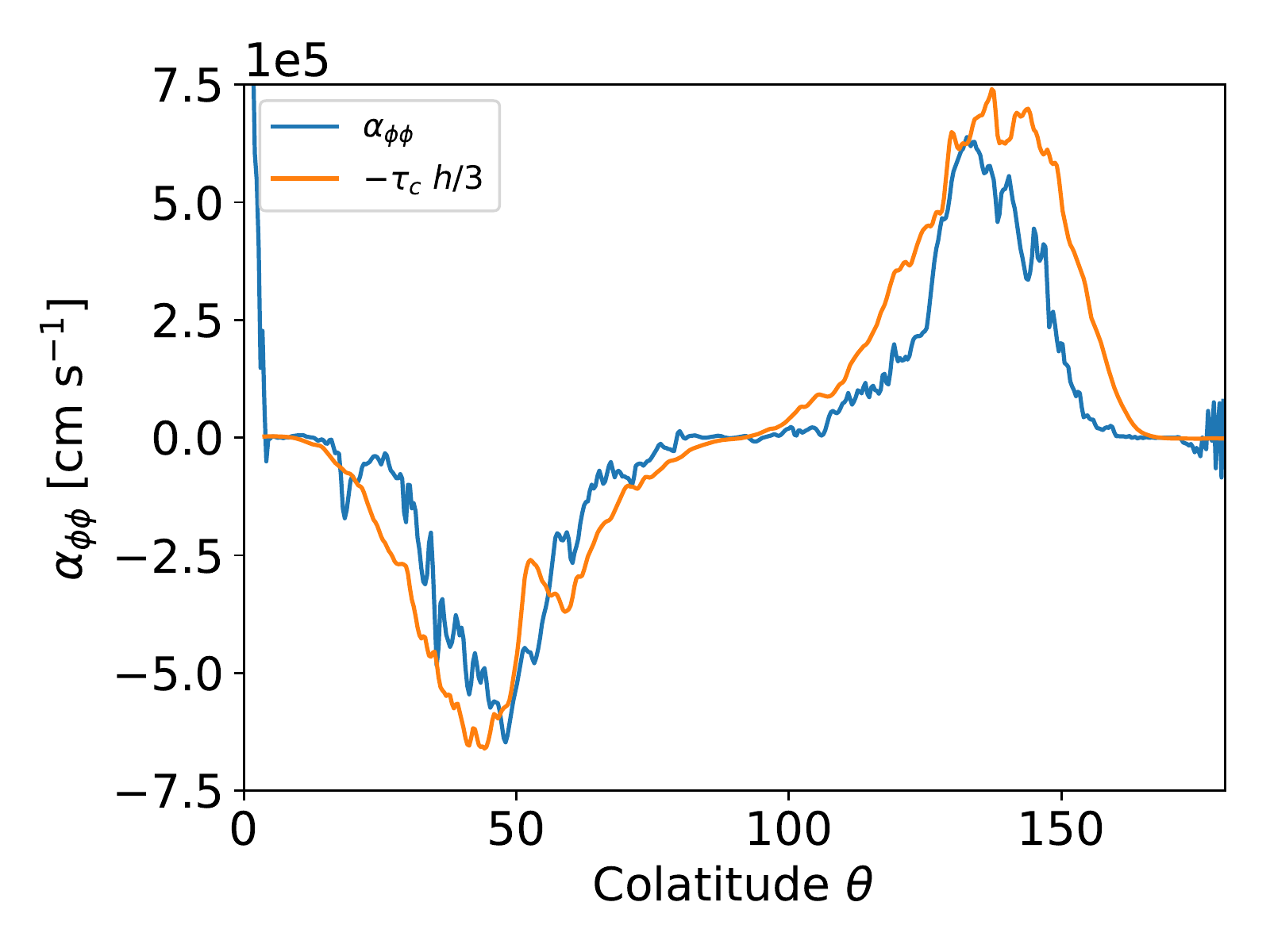}}
    \caption{ Time-averaged values of the $\alpha_{\phi \phi}$ component estimated 
    with Eq.~\ref{e:alpha} (blue) and with the turbulent kinetic helicity (orange) (Eq.~\ref{e:helicity}) taken at $r=0.86\, \ro$ and with $\tau_c = \SI{2.5}{ms}$.}
    \label{alpha_values}
\end{figure}

We can estimate the value of the diagonal components of the $\alpha$ tensor with the formula
\begin{equation}
    \alpha_{ii} = \frac{\langle \overline{B_i}^{\phi}  \ \overline{\mathcal{E}_i}^{\phi}\rangle_t}{\langle  (\overline{ B_{i}}^\phi)^2 \rangle_t}\,.
    \label{e:alpha}
\end{equation}
This estimation assumes that the EMF is only due to the $\alpha$ effect, which is a good approximation in the case of high correlation values.
A theoretical estimation is possible under the second-order correlation approximation, which considers only second-order fluctuating quantities, in addition to several hypotheses \citep{1978moffattfield,1980Krause}. The turbulence is, in fact, assumed to be statistically homogeneous and isotropic, and the mean flow $\overline{u}^\phi$ is usually also neglected. The second-order correlation approximation is valid when one of two dimensionless numbers are small: the magnetic Reynolds number $Rm,$ or the Strouhal number $St=\mathcal{V} \tau/\mathcal{L}$, where $\mathcal{V}$, $\tau$ and $\mathcal{L}$ are typical values of the velocity, time variation, and length scale of the turbulence.
Under these hypotheses, the diagonal components of the $\alpha$ tensor are proportional to the kinetic helicity $h,$ 
\begin{equation}
    \alpha_{ii} = - \frac{\tau_c}{3} \overline{\vec{u}' \cdot \nabla \times \vec{u}'}^\phi
    = - \frac{\tau_c}{3} h\,,
    \label{e:helicity}
\end{equation}
where $\tau_c$ is the correlation timescale of the turbulence.
Although the conditions that are theoretically necessary for the validity of this formula are not satisfied in our simulation \footnote{The Strouhal number is on the order of unity and the magnetic Reynolds number is on the order of a few hundred.}, we find a quantitative agreement between Eqs.~\eqref{e:alpha} and \eqref{e:helicity} for $\tau_c \simeq \SI{2.5}{ms} =0.38\times{2\pi}/{\Omega_0}$ with a peak value of $\alpha_{\phi\phi} \simeq \SI{6e5}{cm.s^{-1}}$ (see Fig.~\ref{alpha_values}). This empirical turbulent correlation time is consistent with the timescale of MRI turbulence, which is typically a fraction of the rotation period. 

Finally, the value of $\alpha_{\phi \phi}$ can be used to estimate the theoretical frequency of an $\alpha\Omega$ dynamo using the relation
\citep[e.g.,][]{2006BusseFrequencyDynamo,2012GastineDipolarMultipolar,2015GresselMRImeanfield} 
\begin{gather}
    \omega_\mathrm{\alpha\Omega}
    = \left|\frac{1}{2} \alpha_{\phi \phi} \frac{d \Omega }{d \ln s} k_z\right|^{1/2}
    = 
    \left|\frac{1}{2} \alpha_{\phi \phi} q \Omega k_z\right|^{1/2}, 
    \label{dynamo_frequency}
\end{gather}
where $k_z$ is the vertical wavenumber. We computed the shear rate~$q=-0.65$ and the rotation rate~$\Omega=\SI{688}{rad.s^{-1}}$ in the middle of the turbulent region from a vertically and azimuthally averaged rotation profile, which gives lower values than~$q_{\rm o}$ and~$\Omega_0$. 
We estimated $k_z \simeq \SI{2.79e-06}{cm^{-1}}$ by taking the longest vertical length in the turbulent region at mid-latitudes and obtained a period of $P_{\alpha\Omega}= {2 \pi}/{\omega_{\alpha\Omega}} = \SI{324}{ms}$, which is roughly agrees with the period inferred from the butterfly diagrams. 
The small difference between the two values is compatible with the uncertainties in our estimate of the different parameters and may, for example, be due to an overestimation of the $\alpha$ effect, as we take its peak value in the turbulent region. If we consider an average on the angles $\theta \in[30\degree,60\degree]$, we obtain a period $P_{\alpha\Omega} = \SI{393}{ms}$, 
which is closer to the measured period $P\simeq \SI{410}{ms}$.
Moreover, according to the theoretical Parker-Yoshimura rule \citep{1955AParkerDynamo,1975YoshimuraDynamo}, the dynamo wave should propagate in the direction of $-\alpha_{\phi\phi}\vec{e_\phi} \times \vec{\nabla} \Omega$, that is, from the equator toward the pole in our case. Unfortunately, a difficulty arises here because the full pattern does not have a clear propagation direction, as shown in Fig.~\ref{Butterfly_diagram}.
Overall, all these results show that the MRI-driven dynamo we obtained in our simulations can be described as an $\alpha\Omega$ dynamo. 

\section{Comparison with other models}
\label{comp}

After demonstrating with a realistic PNS setup that the MRI can produce a subdominant dipole and a turbulent dynamo with a mean-field behavior, we aim at understanding the influence of the different physical ingredients we added in the anelastic model (shell aspect ratio, buoyancy, thermodynamic background, etc.). We discuss in particular the impact of the density profile (Sect.~\ref{ss:dens}), the entropy profile (Sect.~\ref{ss:buo}), and the thermal diffusivity (Sect.~\ref{ss:pr}). We compare these results with our previous incompressible study (Sect.~\ref{ss:incompressible}).

\subsection{Impact of the density gradient}\label{ss:dens}

\begin{figure*}[h]
\centering
\begin{subfigure}{0.245\textwidth}
     \includegraphics[width=1.0\textwidth]{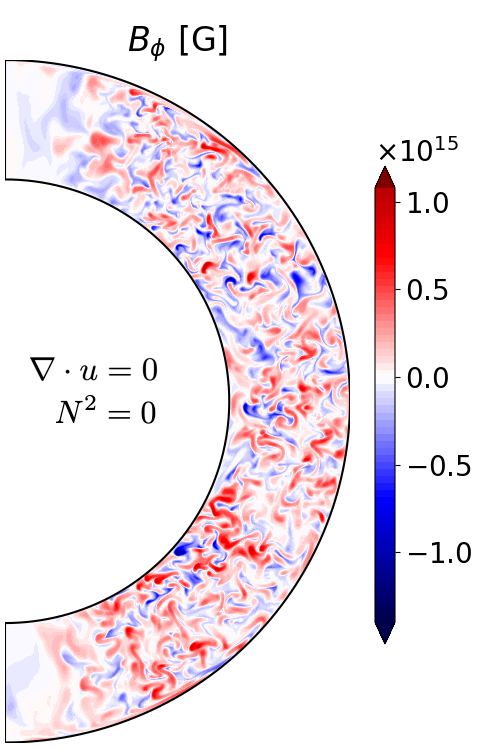}
\caption{}            
\end{subfigure}
\begin{subfigure}{0.245\textwidth}
     \includegraphics[width=1.0\textwidth]{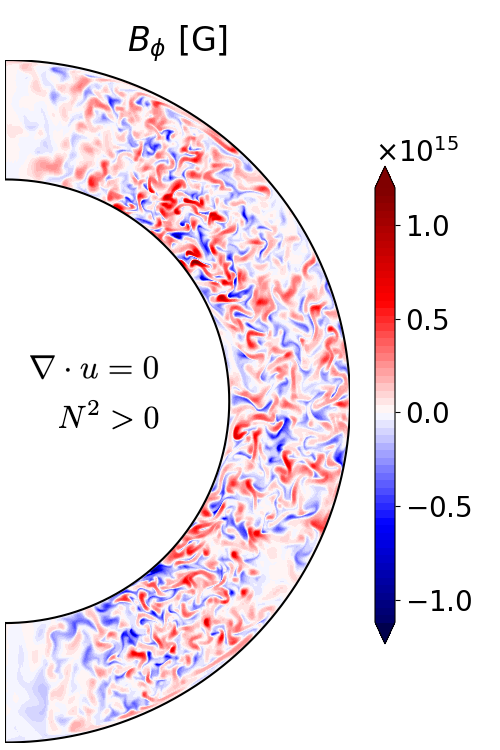}
\caption{}            
\end{subfigure}
\begin{subfigure}{0.245\textwidth}
      \includegraphics[width=1.0\textwidth]{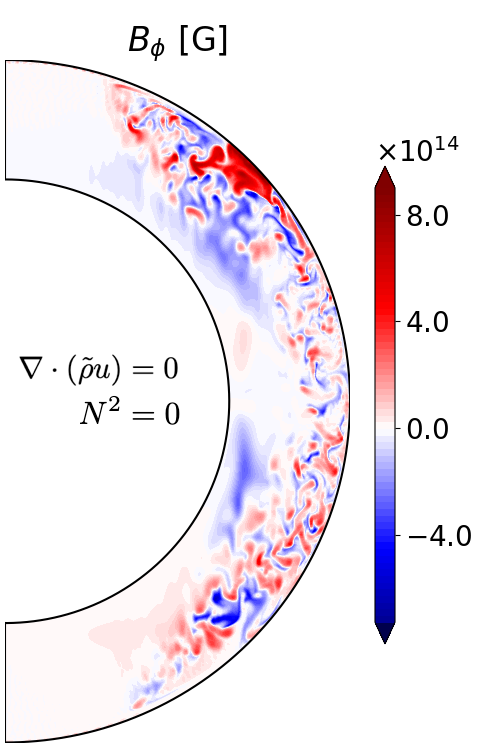}
\caption{}            
\end{subfigure}
\begin{subfigure}{0.245\textwidth}
      \includegraphics[width=1.0\textwidth]{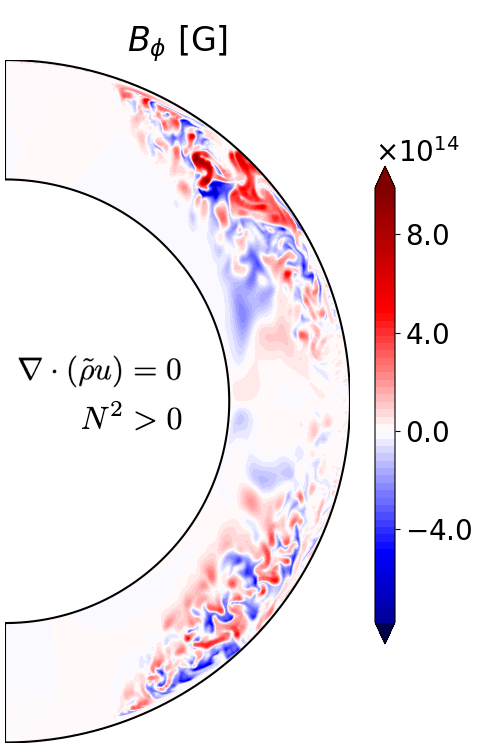}
\caption{}            
\end{subfigure}
\begin{subfigure}{0.245\textwidth}
     \includegraphics[width=1.0\textwidth]{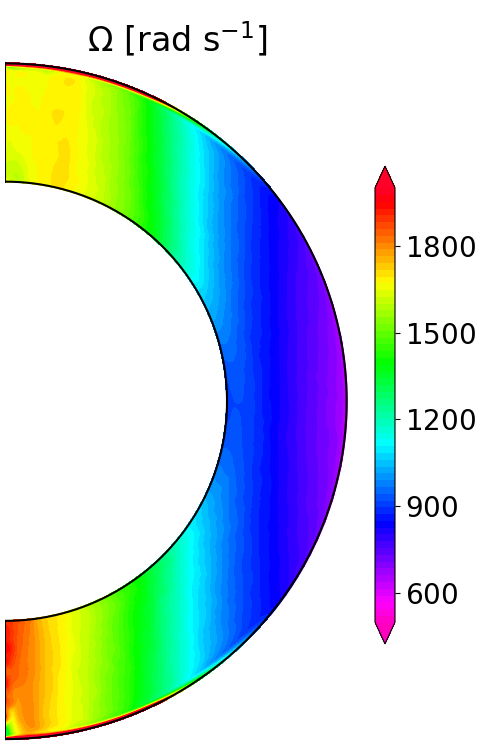}
\caption{}            
\end{subfigure}
\begin{subfigure}{0.245\textwidth}
     \includegraphics[width=1.0\textwidth]{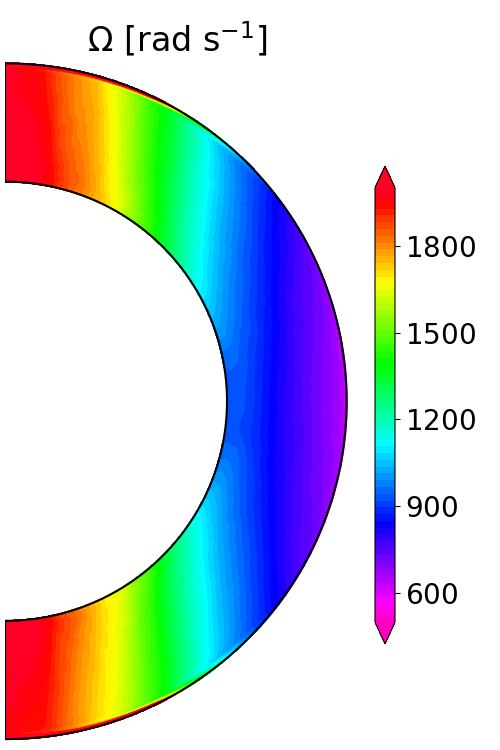}
\caption{}            
\end{subfigure}
\begin{subfigure}{0.245\textwidth}
      \includegraphics[width=1.0\textwidth]{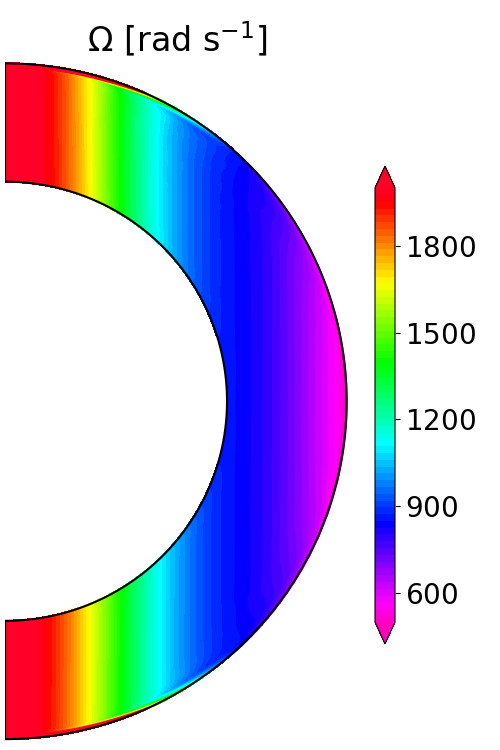}
\caption{}            
\end{subfigure}
\begin{subfigure}{0.245\textwidth}
      \includegraphics[width=1.0\textwidth]{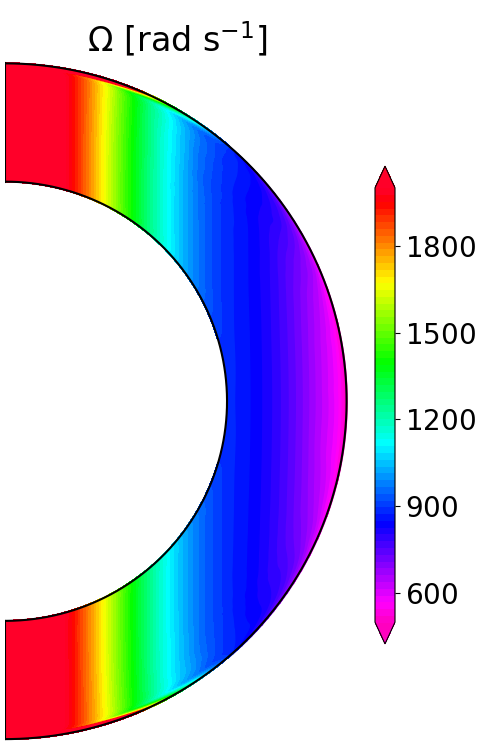}
\caption{}            
\end{subfigure}
      \caption{Meridional cuts of $B_\phi$ (top) and azimuthal average of $\Omega$ (bottom) at a statistically stationary state for incompressible (panels a and e), Boussinesq (panels b and f), anelastic without buoyancy (panels c and g), and anelastic including buoyancy (panels d and h) models. }
     \label{Snapshots_approx}
\end{figure*}
We compared the results of four simulations under different approximations: the Boussinesq approximation (i.e., no density gradient) or the anelastic approximation (i.e., with a density gradient), both with or without buoyancy ($N^2>0$ or $N^2=0$). For Boussinesq models, constant density was taken to be equal to the density $\rho_o=\SI{e11}{g cm^{-3}}$ at the outer boundary of the anelastic model. All the other simulation parameters were kept identical. In order to simplify the interpretation of the results, we start by describing the effect of the density gradient.

The comparison of the snapshots of $B_\phi$ in Fig.~\ref{Snapshots_approx} gives qualitative insights into the effect of the density gradient.
While the maximum intensity of the magnetic field is slightly higher for Boussinesq simulations, the striking difference between Boussinesq and anelastic simulations is their structure: for the
former models, MRI-driven turbulence develops throught the domain, except at the poles. It is restricted to the outer low-density layers for the latter. 
It also seems that at mid-latitudes, the magnetic field has structures at slightly larger scales in the anelastic cases. These differences are due to the background density gradient because both anelastic simulations (with or without buoyancy) have a similar magnetic field structure. 

The presence of the background density gradient also impacts the angular velocity. The bottom panels of Fig.~\ref{Snapshots_approx} show that the flow inside the domain rotates more slowly with a density gradient. This can be understood as a consequence of the lower efficiency of the inward transport of the angular moment by the outer boundary forcing in the presence of a density gradient due to the low density at the outer boundary.

\begin{figure}[h]
\centering
\resizebox{\hsize}{!}{
    \includegraphics{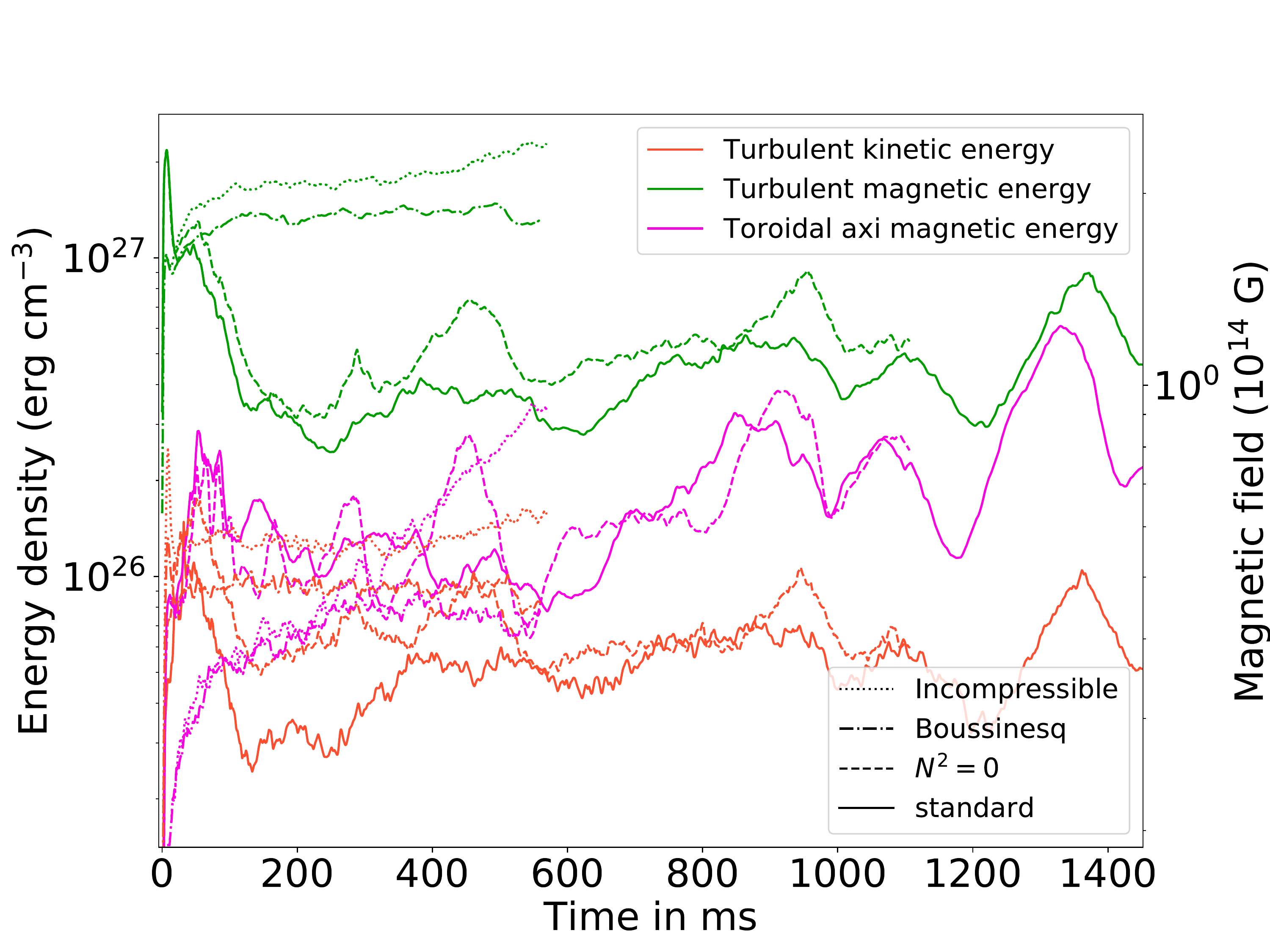}}
      \caption{Time evolution of the turbulent kinetic (orange), nonaxisymmetric (green), and axisymmetric toroidal (purple) magnetic energy densities for incompressible (dotted line), Boussinesq (dash-dotted line), anelastic without buoyancy (dashed line), and anelastic (solid line) models.
      }\label{Ts_energies_approx}
\end{figure}
Figure~\ref{Ts_energies_approx} compares the time evolution of the turbulent magnetic (green) and kinetic (orange) energy densities and the axisymmetric
magnetic energy density (purple).
Both Boussinesq simulations have a higher turbulent magnetic energy and kinetic energy. This is at least partly due to the presence of turbulence in only half of the domain for anelastic simulations compared to the full domain for Boussinesq simulations because the maximum field strength difference in the snapshots is too low to explain the difference in magnetic energy. On the other hand, the Boussinesq model with buoyancy has a similar axisymmetric magnetic energy density such that the ratio of the axisymmetric magnetic energy and turbulent magnetic energy is higher for anelastic simulations. The axisymmetric component for the Boussinesq model without buoyancy, called model \texttt{Incompressible}, is quite peculiar because it increases faster than the turbulent magnetic energy and reaches a higher energy than all other models. The ratio of axisymmetric magnetic energy to turbulent magnetic energy also saturates at higher levels. 
Another important point, however, is that clear oscillations occur only for anelastic simulations, which suggests that there is no mean-field dynamo with a constant density. 

\begin{figure}[h]
\centering
\resizebox{\hsize}{!}
            {
    \includegraphics{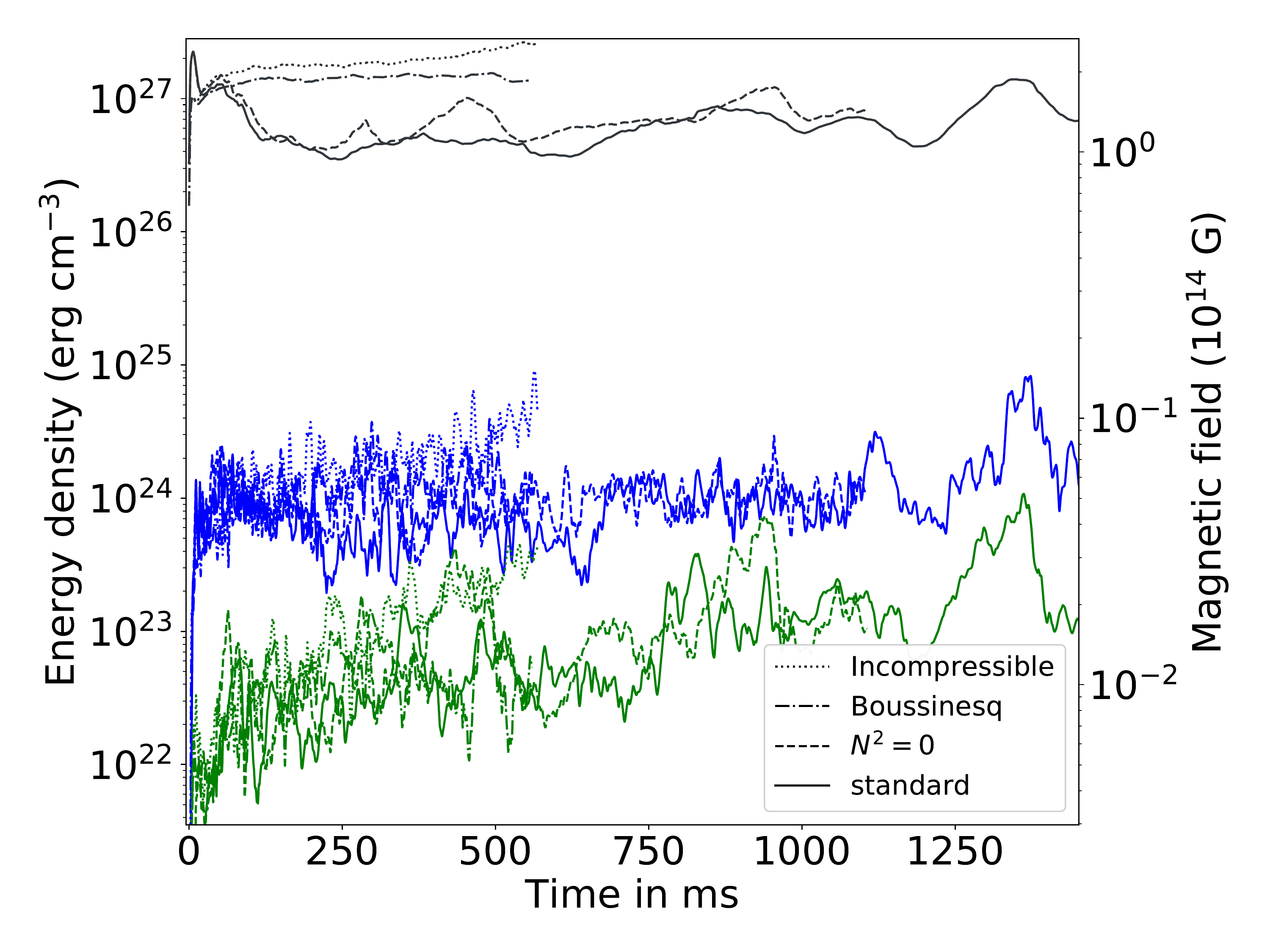}}
      \caption{ 
     Time evolution of the axial dipole (green), total dipole (blue), and total magnetic (black) energy densities of the same models as displayed in Fig.~\ref{Ts_energies_approx}.}
    \label{Dipole_comp_approx}
\end{figure}
In a similar fashion to axisymmetric magnetic energy, the dipole energy density seems to be on the same order for all simulations, except for model \texttt{Incompressible,} which is higher at late times (Fig.~\ref{Dipole_comp_approx}). Because their total magnetic energy also increases, both Boussinesq models have a similar ratio of dipole field to total magnetic field of $\simeq 3.4 \%$, which is lower than the ratios of the anelastic models $\simeq 4.3 \%,$ as we discuss in section \ref{ss:incompressible}.
The dipole energy density is also dominated by its equatorial component in all simulations, which indicates that the inclination of the dipole toward the equator is a robust feature of the MRI.

\begin{figure*}[h]
\centering
\begin{subfigure}[b]{0.43 \textwidth}
   \includegraphics[width=1.0\textwidth]{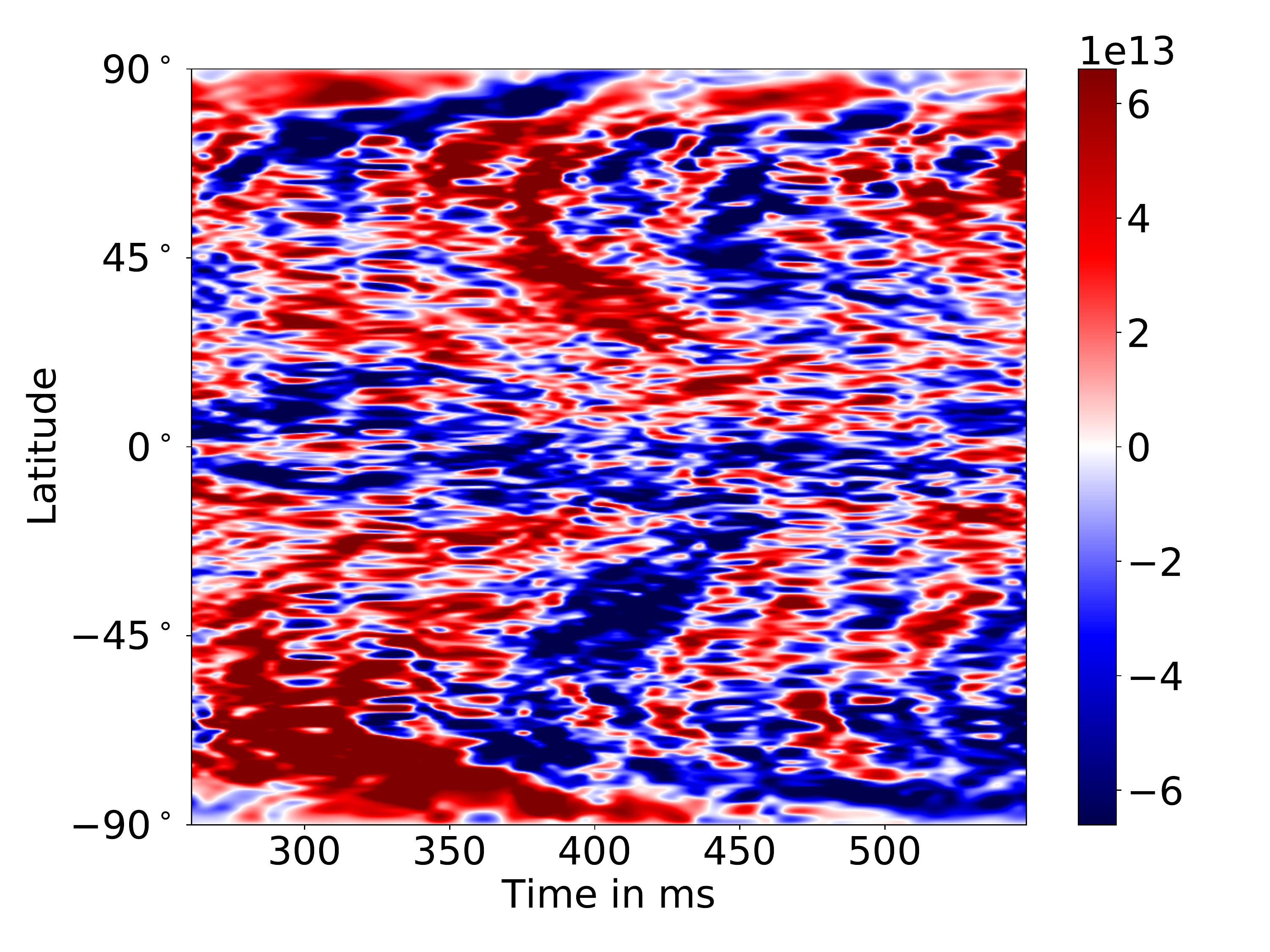}
     \caption{$\nabla \cdot u = 0,\, N^2 >0$, $Pr=0.005$}
\end{subfigure}
\begin{subfigure}[b]{0.43\textwidth}
      \includegraphics[width =1.0 \textwidth]{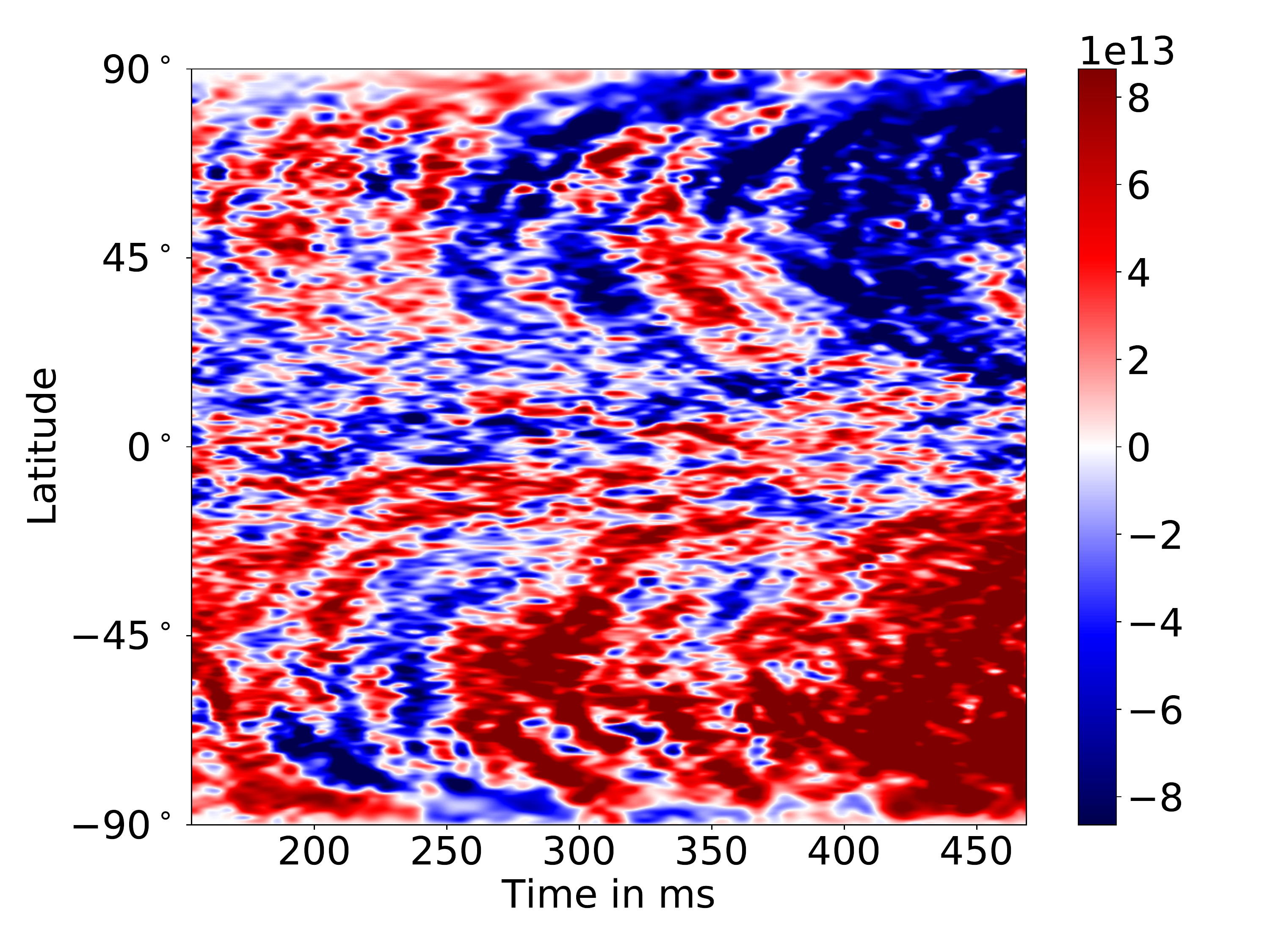}
      \caption{$\nabla \cdot u = 0,\, N^2 =0$, $Pr=0.005$}
\end{subfigure}
\begin{subfigure}[b]{0.43 \textwidth}
    \includegraphics[width=1.0\textwidth]{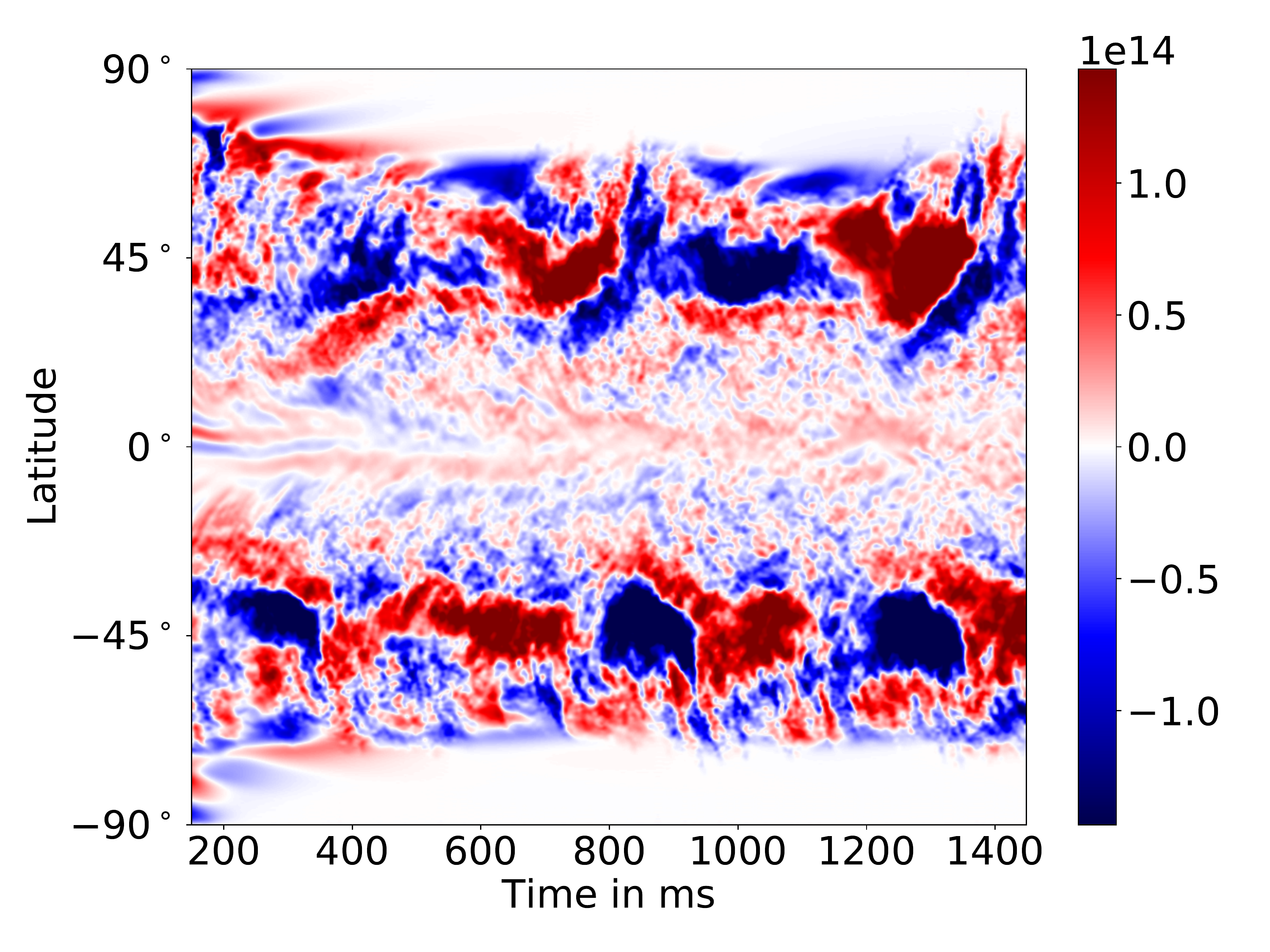}
     \caption{$\nabla \cdot (\tilde{\rho} u) = 0,\, N^2 >0$, $Pr=0.005$}
\end{subfigure}
\begin{subfigure}[b]{0.43 \textwidth}
\includegraphics[width=1.0\textwidth]{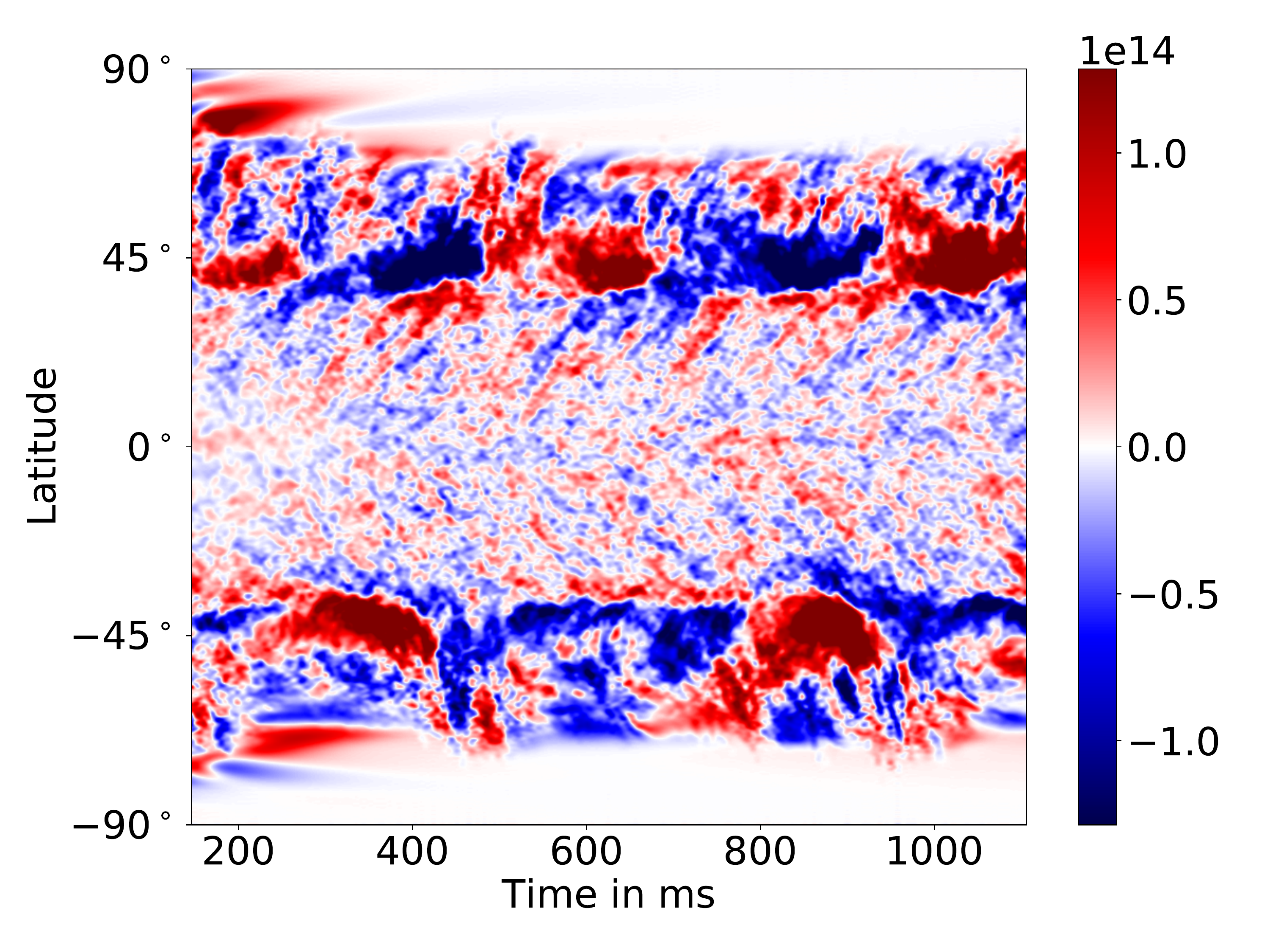}
\caption{$\nabla \cdot (\tilde{\rho} u) = 0,\, N^2 =0$, $Pr=0.005$}
\end{subfigure}
\begin{subfigure}[b]{0.43 \textwidth}
    \includegraphics[width=1.0\textwidth]{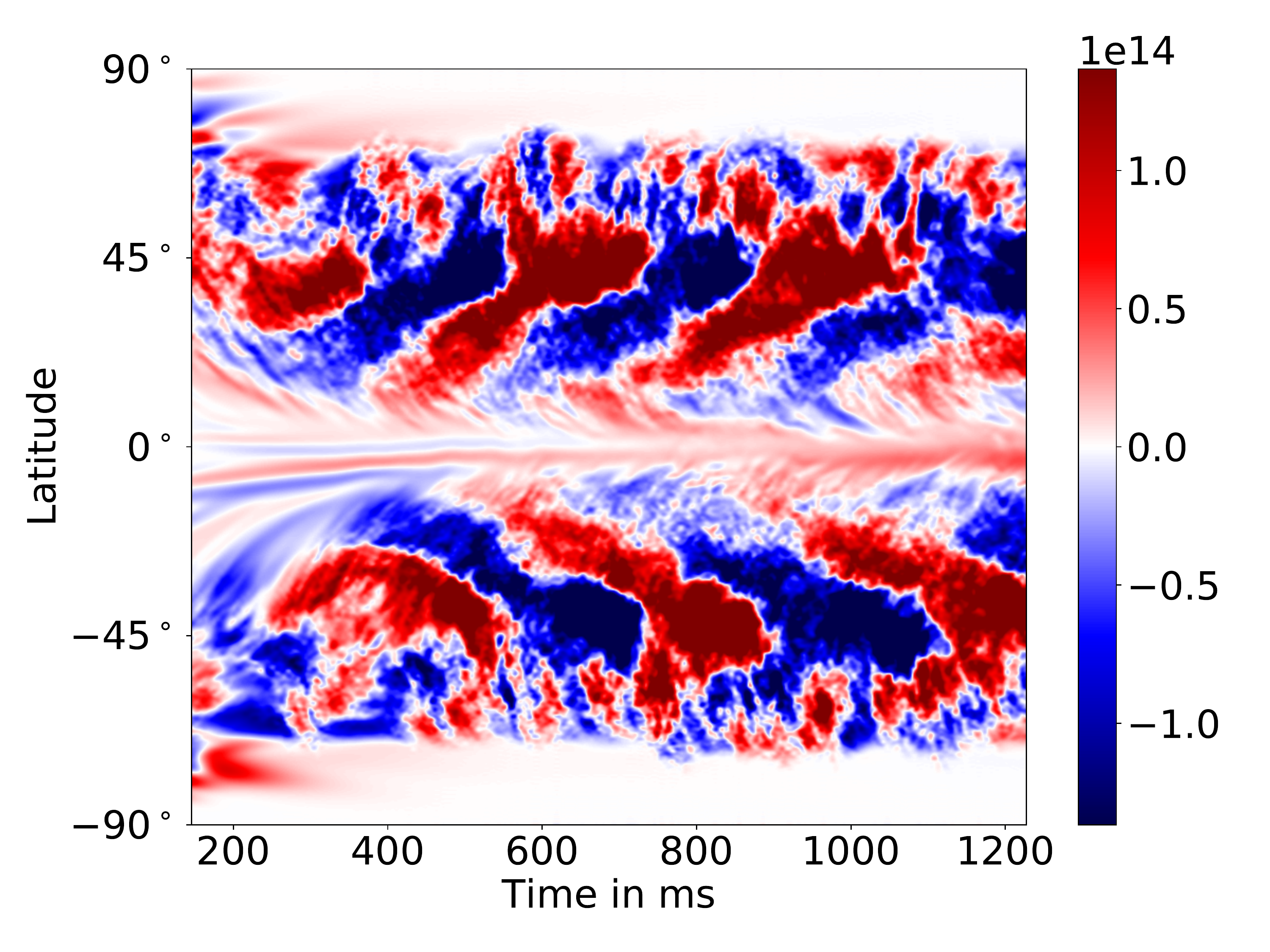}
     \caption{$\nabla \cdot (\tilde{\rho} u) = 0,\, N^2 >0$, $Pr=0.02$}
\end{subfigure}
\begin{subfigure}[b]{0.43 \textwidth}
      \includegraphics[width=1.0\textwidth]{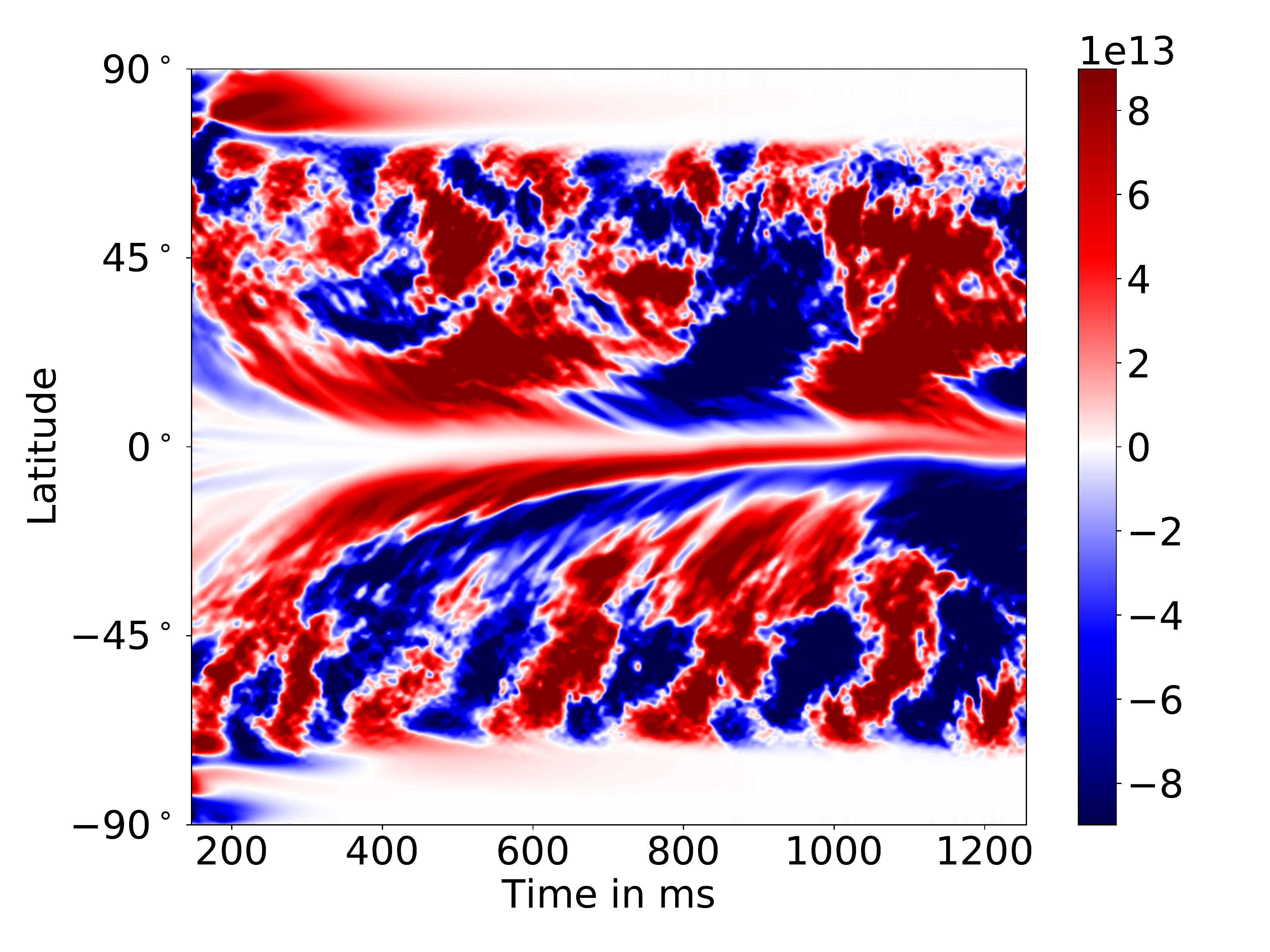}  
     \caption{$\nabla \cdot (\tilde{\rho} u) = 0,\, N^2 >0$, $Pr=0.1$}
\end{subfigure}
      \caption{Butterfly diagrams of $\overline{B_\phi}^\phi$ at $r=0.92\, \ro$ for the models shown in Fig.~\ref{Snapshots_approx} for Boussinesq with buoyancy (panel a), Boussinesq without buoyancy (panel b), anelastic with buoyancy (panel c), anelastic without buoyancy (panel d), and anelastic with buoyancy with $Pr=0.02$ (panel e) and $Pr=0.1$ (panel f).
      Each butterfly diagram is computed at the same spherical radius $r=0.92\,\ro$ (in the turbulent region).}
    \label{Butterfly_comp_all}
\end{figure*}

To confirm the mean-field behavior, we show in Fig.~\ref{Butterfly_comp_all} the butterfly diagrams of $\overline{B_\phi}^\phi$. 
For Boussinesq simulations (panels a and b), the MRI-unstable region is noisier. It is hard to observe clear signs of coherent mean-field patterns with buoyancy (panel a), while without buoyancy (panel b), an increasing quadrupole develops after $\SI{400}{ms}$. This quadrupole is rather puzzling because its strength is higher than the incompressible models of paper I, and it does not match the $\alpha\Omega$ behavior. There is no clear signal in the toroidal component of the EMF $\mathcal{E}_\phi$ , and we find that the correlation coefficients between the EMF $\mathcal{E}_\phi$ and $\overline{B_\phi}^{\phi}$ are low, with an average $C_P(\mathcal{E}_\phi,\overline{B_\phi}^{\phi}) \simeq 0.04$ in both hemispheres. We find no correlations either between the EMF $\mathcal{E}_\phi$ and $\overline{J_s}^{\phi}$. 
The physical origin of this toroidal quadrupole is so far uncertain. 
We would like to point out that this region is not turbulent, and the patterns found there (which can also be seen in the Boussinesq model with buoyancy in panel a) are probably not caused by a true mean-field mechanism.
These results suggest that it is more difficult to have a mean-field behavior without density stratification. The stronger ratio of dipole to total magnetic field in the case of anelastic models might be due to the mean-field dynamo boosting the generation of the dipole.

\subsection{Impact of buoyancy}\label{ss:buo}

To study the influence of buoyancy, we focused first on the comparison of model \texttt{Incompressible} to the Boussinesq model with buoyancy. 
The comparison between the snapshots of $B_\phi$ shows that the structure of the turbulence is rather similar, while the turbulent magnetic field is stronger without buoyancy (Fig.~\ref{Snapshots_approx}). This result is also found in terms of turbulent energy density for the magnetic and kinetic energies (Fig.~\ref{Ts_energies_approx}). 
It can be expected that the higher magnetic field increases the angular momentum transport, thus decreasing the angular frequency near the axis.
As mentioned in the previous section, the physical origin of the high axisymmetric energy density of model \texttt{Incompressible} is uncertain, and this feature is not discussed. 

For anelastic models, we see a difference in the equatorial plane where MRI-driven turbulence is damped with buoyancy (panels (c) and (d) of Fig.~\ref{Snapshots_approx}). 
Moreover, the butterfly diagrams in Fig.~\ref{Butterfly_comp_all} suggest that the model with buoyancy (panel c) persistently suppresses small-scale fluctuations in the equatorial region, while without buoyancy (panel d), turbulence is fully developed.
This effect may be expected because in the equatorial plane, the buoyancy quenches the motions in the direction of the differential rotation gradient, which reduces the MRI turbulence \citep{2004MenouMRI,2015GuiletBuoy}. 
The weaker turbulence in the equatorial plane with buoyancy may partly explain the lower turbulent magnetic and kinetic energy densities of model \texttt{Standard} compared to the model without buoyancy (Fig.~\ref{Ts_energies_approx}).

By contrast, the mid-latitude regions are less impacted by the buoyancy. The magnetic field $B_\phi$ is similar in strength and
structure outside of the equatorial plane (Fig.~\ref{Snapshots_approx}).  
In the anelastic model without buoyancy, a mean-field dynamo also produces a field with a similar amplitude to model \texttt{Standard} at the mid-latitudes (see panel (c) and (d) in Fig.~\ref{Butterfly_comp_all}). This leads to a similar axisymmetric toroidal energy density (Fig.~\ref{Ts_energies_approx}) and dipole energy density (Fig.~\ref{Dipole_comp_approx}) for both anelastic models.
In terms of dynamo periods, both theoretical dynamo periods are similar, even though the patterns have a longer cycle without buoyancy. The dynamo frequencies agree within $15\%$ with buoyancy and $24\%$ without buoyancy (see Table \ref{table:annex}). For these two models, the dynamo periods are more difficult to measure due to the combination of short and long patches of magnetic fields, and this can lead to differences between the hemispheres (see Table \ref{tab:dyn_period} for model \texttt{standard}).
Considering the different uncertainties in our estimate, the two models give consistent results with theoretical values, which shows that buoyancy has a weak influence on the dynamo mechanism at $Pr=0.005$.  

\subsection{Influence of the thermal Prandtl number}\label{ss:pr}

The previous section shows that except in the equatorial plane,  buoyancy has a rather small impact on the anelastic simulations overall, which might be due to the high thermal diffusivity.
To further study the influence of the thermal diffusion, we ran two anelastic simulations with a lower thermal diffusivity corresponding to larger Prandtl numbers, $Pr=0.02$ and $Pr= 0.1$, respectively. 
By comparing the timescale for thermal diffusion to compensate for the entropy fluctuations to the timescale of gravity waves, we may expect the thermal diffusion to reduce the effects of buoyancy on scales smaller than the critical length $L_{\rm c}$ defined by Eq.~\eqref{e:Lc}.
For our simulation \texttt{Standard} at $Pr = 0.005$, the critical length at mid-latitudes $L_\mathrm{c} \simeq 0.14 D$ is on the same order of magnitude as the turbulent scale of the radial velocity, such that buoyancy effects are expected to be marginal. 
By contrast, the critical length decreases to $L_\mathrm{c} \simeq 0.03 D$ at $Pr=0.1$, hence decreasing the range of scales where turbulent motions are suppressed by buoyancy (i.e., for scales larger than $L_\mathrm{c}$). 
As a consequence, the typical scale of the radial velocity is expected to be smaller at higher $Pr$.

\begin{figure*}[h]
\centering
\resizebox{\hsize}{!}
            {
     \includegraphics{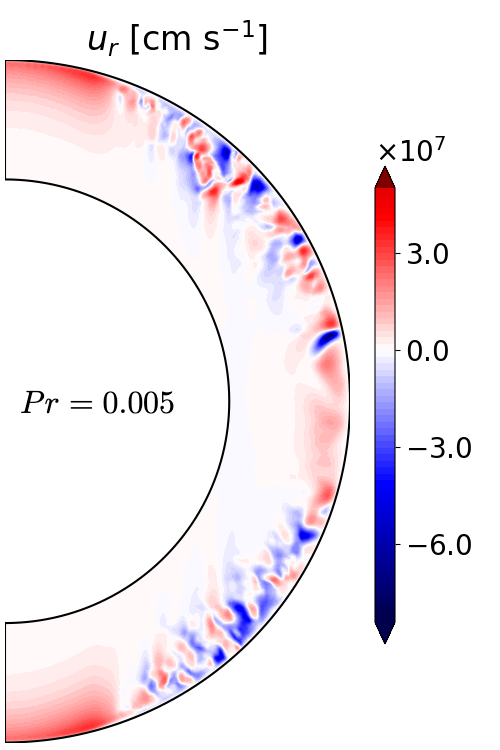}
      \includegraphics{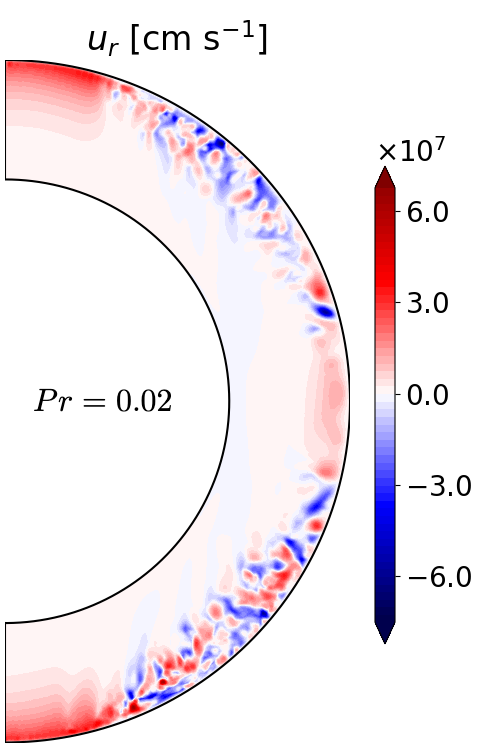}
      \includegraphics{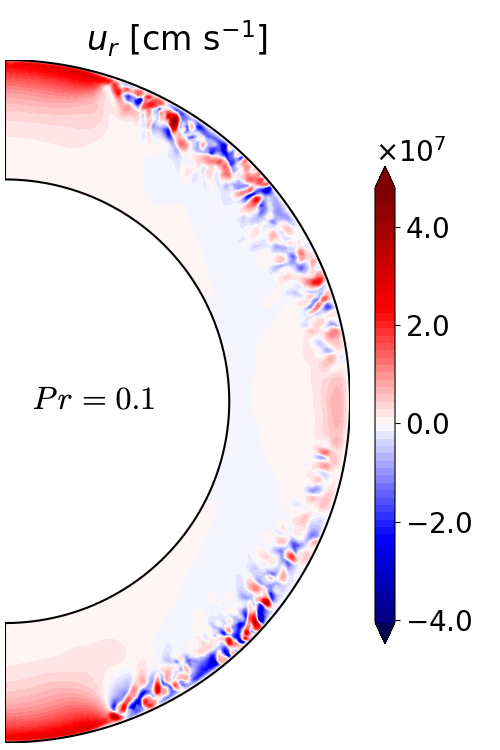}}
\resizebox{\hsize}{!}
            {
     \includegraphics{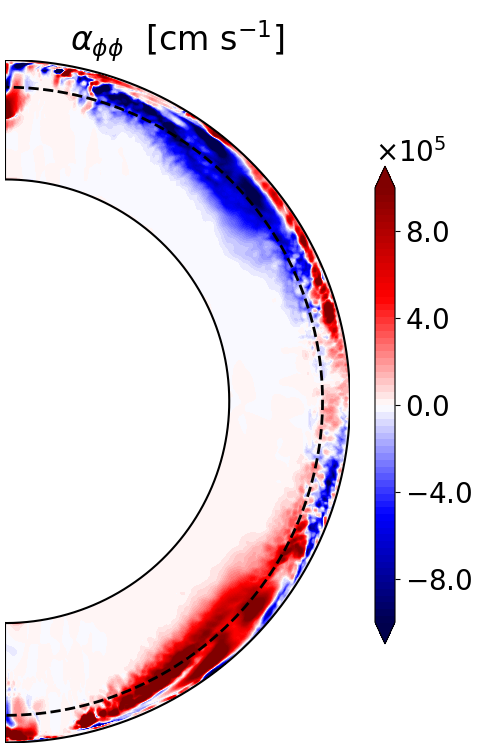}
      \includegraphics{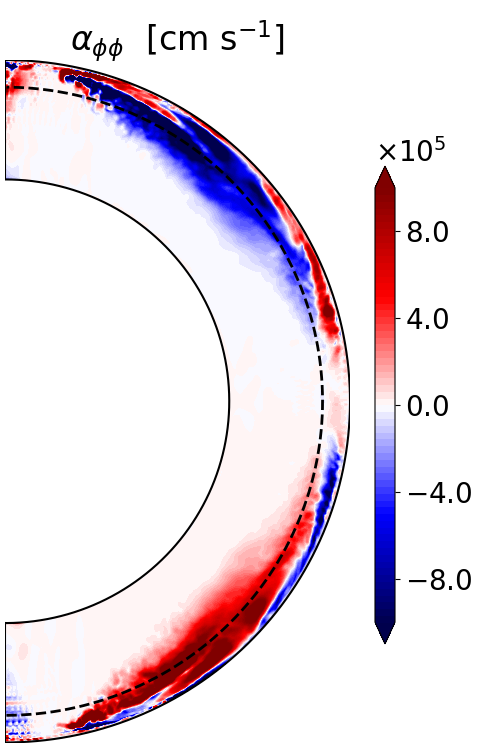}
      \includegraphics{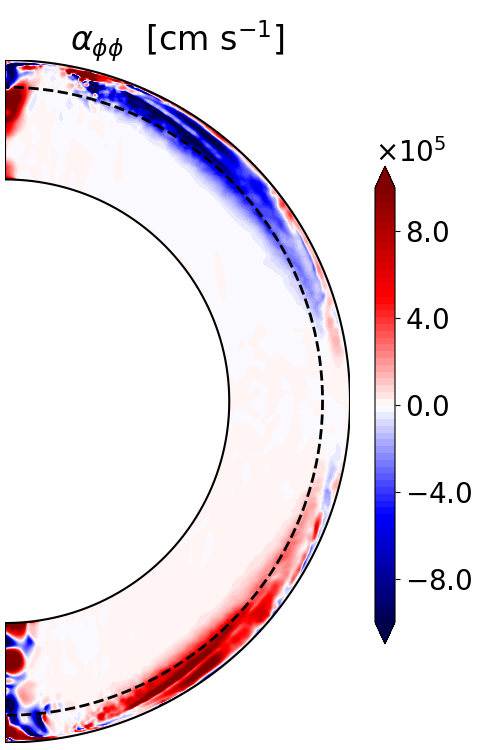}}
      \caption{Radial velocity $u_r$ and $\alpha_{\phi\phi}$ for anelastic models with buoyancy and for increasing Prandtl numbers.
      Top: Snapshots of the radial velocity $u_r$. From left to right, the slices correspond to $Pr=0.005$ at $t=\SI{773}{ms}$, $Pr=0.02$ at $t=\SI{1128}{ms}$, and $Pr=0.1$ at $t=\SI{1128}{ms}$. Bottom: 2D distribution of $\alpha_{\phi\phi}$ for $Pr=0.005$ (left), $Pr=0.02$ (middle), and $Pr=0.1$ (right). The dashed line represents the spherical radius $r=0.92\,r_o$.}
         \label{Snapshots_Pr}
\end{figure*}

\begin{figure}[h]
\centering
     \includegraphics[width =0.45\textwidth]{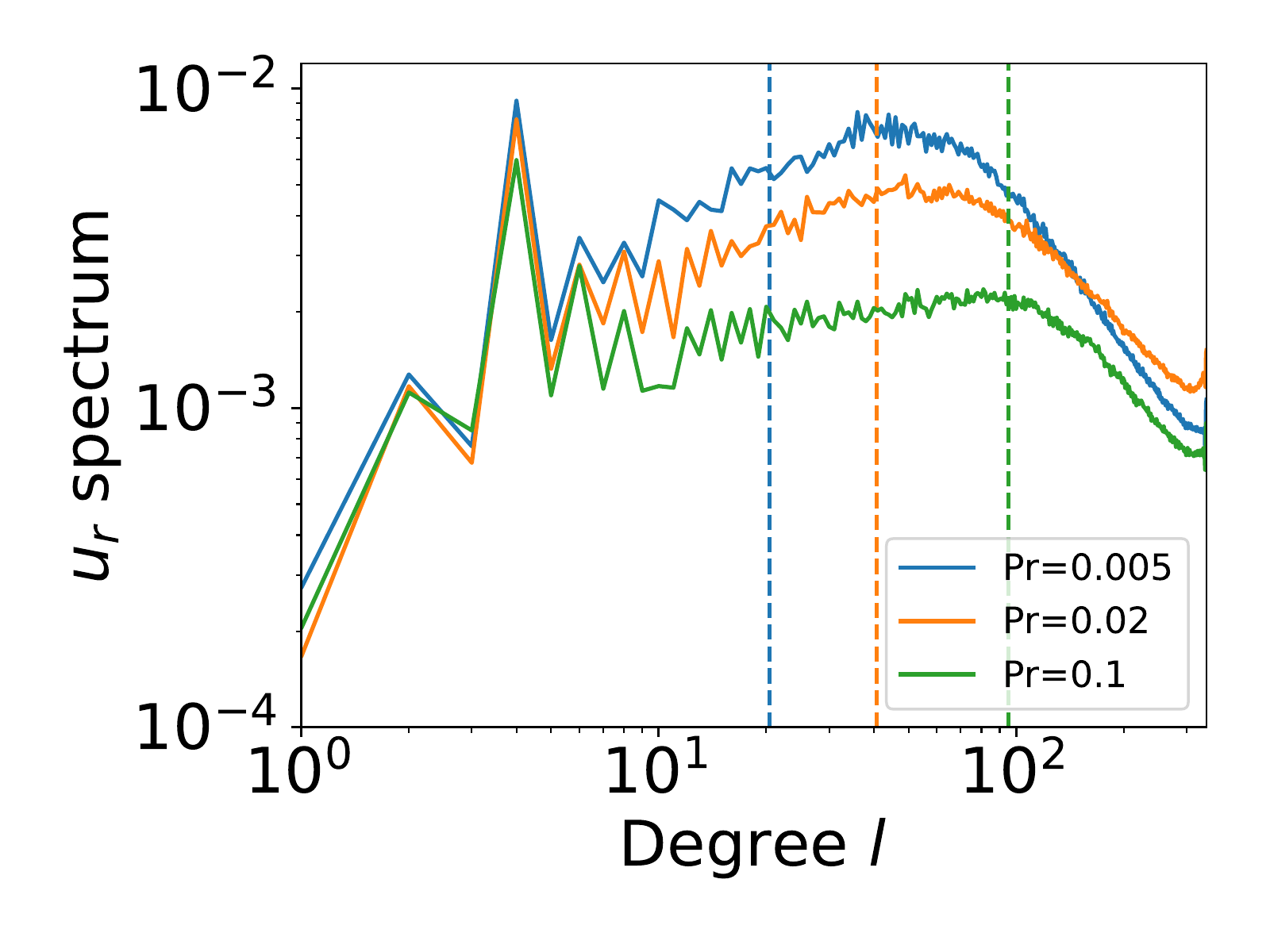}
      \caption{Normalized volume-averaged spectrum of the radial velocity squared $u_r$ for increasing Prandtl numbers $Pr=0.005$ (blue), $Pr=0.02$ (orange), and $Pr=0.1$ (green). All the spectra are normalized by the integral over l of the \texttt{Standard} model spectrum. The dashed lines correspond to the critical degree $l_\mathrm{c} \approx{R}/{L_\mathrm{c}}$ above which the scales are impacted by buoyancy.}
         \label{Spec_Pr}
\end{figure}

\begin{figure}[h]
\centering
\resizebox{\hsize}{!}
            {
    \includegraphics{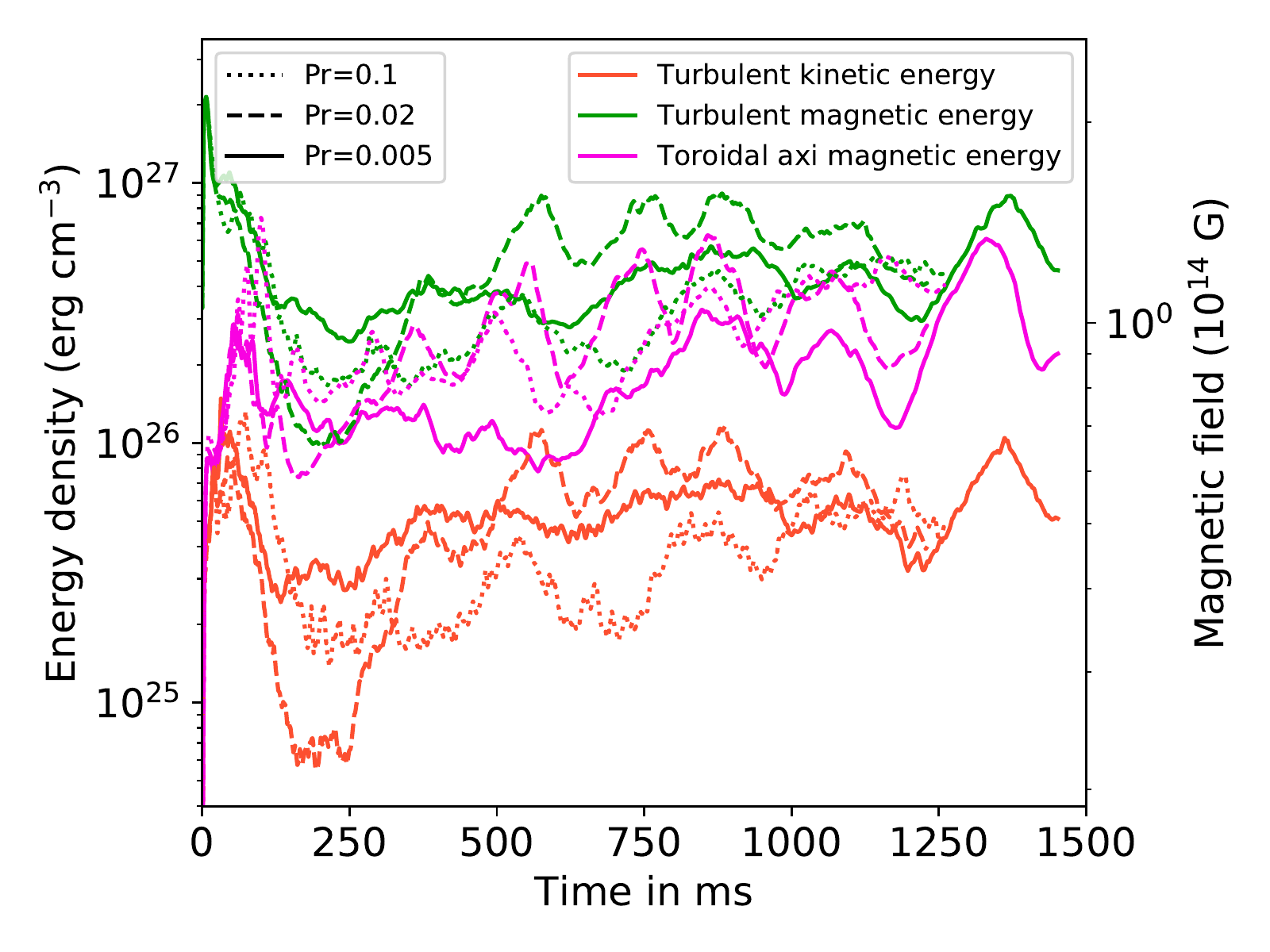}}
      \caption{Time evolution of the magnetic and turbulent kinetic energy densities for models with $Pr=0.1$ (dotted), $Pr=0.02$ (dashed), and $Pr=0.005$ (solid). The colors represent the same quantities as in Fig.~\ref{Ts_energies_approx}. }
    \label{Ts_energies_Pr}
\end{figure}

To verify this expectation, we compare snapshots of the radial velocity in Fig.~\ref{Snapshots_Pr}.
For increasing $Pr,$ we observe a decrease in the size of the velocity structures and in the maximum amplitude of the radial velocity. 
This trend is also confirmed in the radial velocity spectrum of the different simulations (Fig.~\ref{Spec_Pr}). The peak values of the spectrum in the small scales for $l>20$ are lower and shifted toward smaller scales with increasing $Pr$. For models with $Pr=0.02$ and $Pr=0.1$, they also seem to match the theoretical critical degree $l_\mathrm{c} \approx \frac{R}{L_\mathrm{c}}$ well, corresponding to the critical length.
The suppression of small scale-turbulence can also be seen to some extent in the corresponding kinetic energy time series in Fig.~\ref{Ts_energies_Pr}. 
The kinetic energy of the simulation at $Pr=0.1$ (dotted lines) is indeed the lowest during most of the simulation time. 
The turbulent magnetic energy follows the same trend with Pr as the turbulent kinetic energy (green curves).

Despite the differences in the turbulence properties, the axisymmetric toroidal magnetic energy is not strongly dependent on $Pr$ and the characteristic oscillations of a mean-field dynamo are present in all three simulations. This suggests that buoyancy does not strongly impact the mean-field behavior at mid-latitudes. The main difference is the weaker turbulence in the equatorial plane for model \texttt{Standard}, while no turbulence can be seen for the other models.
The butterfly diagrams in the turbulent region at mid-latitudes are similar for $Pr=0.02$ (panel (e) of Fig.~\ref{Butterfly_comp_all}) and the south hemisphere of the \texttt{Standard} simulation at $Pr=0.005$ (panel (c)), with a similar dynamo period (see Table~\ref{tab:dyn_period}). One noticeable difference between panels (c) and (e) in Fig.~\ref{Butterfly_comp_all}
is the poleward propagation of the dynamo wave, which is consistent with the Parker-Yoshimura rule for the simulation at $Pr=0.02$, probably due to the lack of turbulence in the equatorial plane.

\begin{figure}[h]
\centering
\resizebox{\hsize}{!}
            {
    \includegraphics{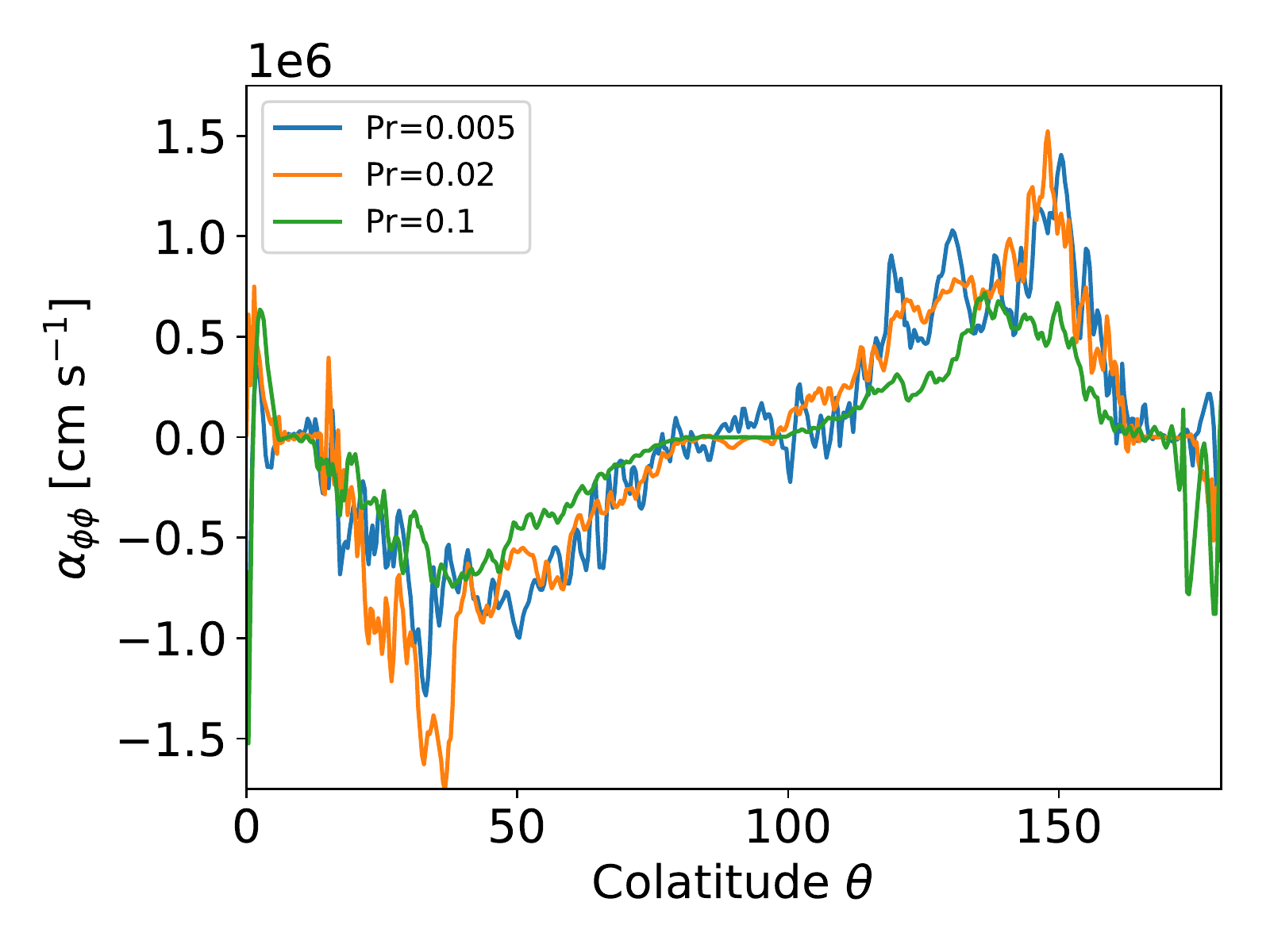}}
      \caption{Time-averaged values of the $\alpha_{\phi\phi}$ component estimated using Eq. (\ref{e:alpha}) at $r=0.92\, \ro$ in the turbulent region for three different Prandtl numbers: $Pr=0.005$ (blue), $Pr=0.02$ (orange), and $Pr=0.1$ (green).}
    \label{alpha_values_Pr}
\end{figure}

To determine whether the specific behavior at different $Pr$ is still an $\alpha\Omega$ dynamo, we compare in Fig.~\ref{alpha_values_Pr} the $\alpha_{\phi\phi}$ component for the different Prandtl numbers.
At $Pr=0.02$, we find a mean $\alpha_{\phi\phi} \simeq \SI{8.0e5}{cm .s^{-1}}$ in the turbulent region, which translates into a period~$P_{\alpha\Omega}\simeq \SI{300}{ms}$ using the same $k_z$ and Eq.~\eqref{dynamo_frequency}. This agrees with the measured dynamo frequency within $17\%$. 
Moreover, both values are relatively close to those measured and estimated for the southern hemisphere of model \texttt{Standard}.

The 2D meridional distribution of $\alpha_{\phi\phi}$ (bottom left and bottom center panels of Fig.~\ref{Snapshots_Pr}) also shows that there is little difference between these two models for the mean-field dynamo. This can be interpreted as follows: most of the kinetic turbulence is still at smaller scales than the critical wavelength for the buoyancy, as seen in the $u_r$ spectra (Figure \ref{Spec_Pr}), and the MRI is not as much impacted by buoyancy at mid-latitudes. Therefore, the change in the velocity spectrum is not important enough to modify the $\alpha$ effect strongly, and both models have a similar mean-field dynamo.

At the higher Prandtl number $Pr=0.1$, oscillating dynamo cycles are still present, but have different characteristics. In the northern hemisphere, the patterns look like those in the other anelastic simulations, albeit with a lower frequency $P_\mathrm{North} \simeq \SI{493}{ms}$. However, in the southern hemisphere, we observe mean-field patterns of similar amplitude, but higher frequency (with a period of $P_\mathrm{South} \simeq \SI{213}{ms}$). 
Moreover, the pattern propagates toward the equator, which is in the opposite direction to the Parker-Yoshimura rule. 
The meridional distribution of $\alpha_{\phi\phi}$ (bottom left panel of Fig.~\ref{Snapshots_Pr}) also highlights the smaller turbulent region and the lower amplitude of $\alpha_{\phi\phi}$ than in the other two models.
\begin{table}[]
    \renewcommand{\arraystretch}{1.2}
    \caption{Periods of the different mean-field patterns for $Pr=0.005, \ Pr=0.02,$ and $Pr=0.1$.}
    \label{tab:dyn_period}
    \centering
    \begin{tabular}{lcccc}
    \hline\hline
    Model & $Pr$ & $P_{North}$ &$P_\mathrm{South}$ & $P_{\alpha\Omega}$ \\
     & - & [\si{ms}] & [\si{ms}] & [\si{ms}]\\
    \hline
    \texttt{Standard}& 0.005 & 530 & 401 & 342 \\
    \texttt{Anel Pr0 02}& 0.02  & 357 & 366 & 300\\
    \texttt{Anel Pr0 1} & 0.1 & 493 & 213 & 366\\
    \hline \hline
    \end{tabular}
\end{table}
At $Pr=0.1$, we measure $\alpha_{\phi \phi} \simeq \SI{4.57e5}{cm.s^{-1}}$ in the northern and $\alpha_{\phi \phi} \simeq \SI{4.47e5}{cm.s^{-1}}$ in the southern hemisphere (Fig.~\ref{alpha_values_Pr}). 
By assuming that the turbulent region is $20 \%$ smaller than the region from simulation \texttt{Standard}, we obtain a theoretical prediction of the period~$P_{\alpha\Omega} = \SI{366}{ms}$ in the northern hemisphere, which agrees within $27\%$ with the numerical value.   
The similar $\alpha_{\phi\phi}$ value in the southern hemisphere would lead to a similar period and therefore disagrees more strongly with the measured numerical value. This result, combined with the opposite direction of propagation, may suggest a significant deviation from the $\alpha\Omega$ dynamo formalism.
This new behavior of the southern hemisphere with $Pr=0.1$ remains obscure and might result from being close to the dynamo threshold. 
This idea is also supported by the fact that a more typical $\alpha\Omega$ dynamo occurs in the northern hemisphere, but with a weaker $\alpha$ effect.
All in all, the results presented in this section show that a low Prandtl number diminishes the impact of buoyancy on the MRI dynamo, whereas the buoyancy force can limit the MRI-driven dynamo for Prandtl numbers closer to unity.

\subsection{Comparison to our previous incompressible models}\label{ss:incompressible}

One of the important results of paper I is the robust linear relation between the dipole intensity and the averaged magnetic field strength (see their Fig.~16).
In order to directly compare the ratios measured in simulations that have different densities and parameters, we reproduce the same figure by using the dimensionless dipole strength defined by $\mathcal{B}_\mathrm{dip} = B_\mathrm{dip}/(\sqrt{\mu_0 \tilde{\rho}_o}D \Omega)$ and the dimensionless total magnetic field strength $\mathcal{B}_\mathrm{tot} = B_\mathrm{tot}/(\sqrt{\mu_0 \tilde{\rho}_o} D \Omega)$ instead of the magnetic intensities. 
For anelastic simulations, we find a greater magnetic strength $\mathcal{B}_\mathrm{tot}$ and dipole strength $\mathcal{B}_\mathrm{dip}$ than in paper I. However, the ratio of $\mathcal{B}_\mathrm{dip}$ to $\mathcal{B}_\mathrm{tot}$ of anelastic simulations is quite similar, if slightly lower ($\simeq 4.3 \%$) than the linear relation of paper I.

For our Boussinesq simulations, the differences with paper I are stronger. The magnetic strength $\mathcal{B}_\mathrm{tot}$ is higher by a factor of approximately$~\text{three}$ due to the stronger turbulence, and the dipole strength $\mathcal{B}_\mathrm{dip}$ is twice stronger. This gives a different linear relation of $\simeq 3.4\%$ between the magnetic dipole and the magnetic field strength.
This change in linear relation between the Boussinesq models here and the incompressible models of Paper I is probably due to the difference in the aspect ratio between the models.
With a smaller shell gap $D$, the forcing of the differential rotation by the outer boundary is expected to be more efficient, which might lead to stronger turbulence for a similar dipole intensity.  
This result highlights the importance of developing global models that take the full PNS structure into account. 

\begin{figure}[h]
\centering
\resizebox{\hsize}{!}
            {
    \includegraphics{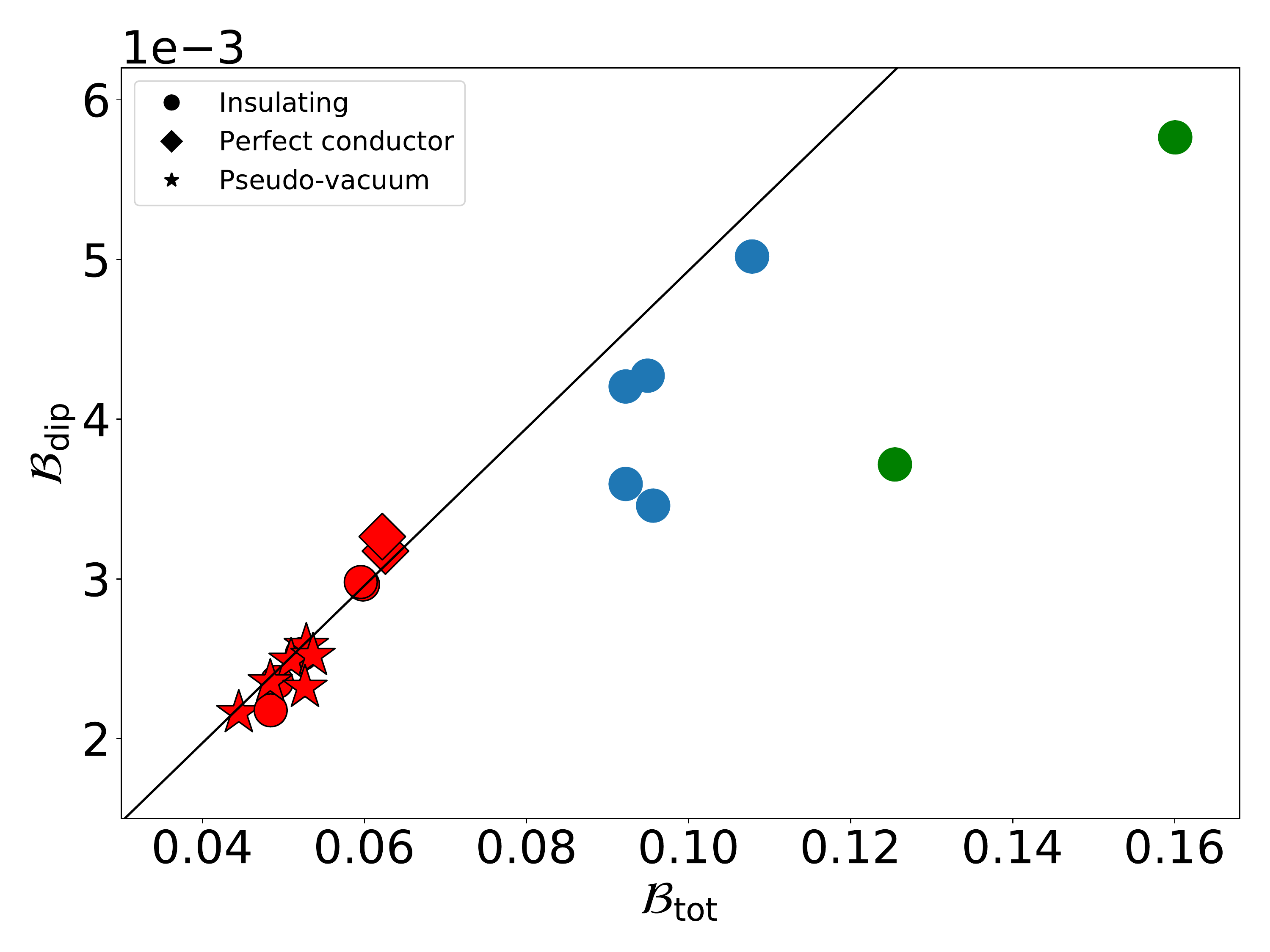}}
      \caption{Dimensionless dipole strength $\mathcal{B}_{\rm dipole}$ as a function of the dimensionless magnetic field strength $\mathcal{B}_{\rm tot}$. The simulations of this work are plotted in blue (anelastic models) and green (no density stratification), and the incompressible simulations from paper I are shown in red. Dimensionless magnetic field strengths $\mathcal{B}$ were used in order to compare results with different densities $\tilde{\rho}_{\rm o}$ and parameters.}
         \label{Comp_dip_alfven}
\end{figure}

\section{Discussion}
\label{discuss}

\subsection{Spatial distribution of the turbulence}
\label{ss:loc_turb}

One puzzling feature of the turbulence in the anelastic models is its concentration in the low-density region. The MRI modes can be damped by the viscosity or the resistivity if the initial magnetic field is lower than a critical value.
In order to investigate whether the lack of turbulence in the high-density region is due to the initial magnetic field, we ran a simulation with a higher resolution $(n_r,n_{\theta},n_{\phi})=(385,768,1536)$ and a magnetic field that was twice as high $B_\mathrm{o} = \SI{3.3e14}{G}$. 
We find similar results for the magnetic field as model \texttt{Standard} as shown in Table~\ref{table:annex}, which confirms that our results converge well and are not an artifact of the initial magnetic field. 
Our interpretation of the absence of turbulence in the high-density region is that the forcing of the differential rotation is not efficient enough to sustain the turbulence in this region. As shown in the bottom panels of Fig.~\ref{Snapshots_approx}, the angular velocity in the domain is slower than at the outer boundary. This result might change with a different forcing method or when the magnetic Prandtl number is increased with a lower physical resistivity, which would make the turbulence easier to sustain.

\subsection{$\alpha\Omega$ versus $\alpha^2\Omega$ dynamo}

As shown in the previous sections, the dynamos observed in our anelastic models are consistent with an $\alpha\Omega$ mechanism. Although the anticorrelation of $B_s$ and $B_\phi$ in the butterfly diagrams (Fig.~\ref{Butterfly_diagram}) suggests that the $\Omega$ effect generates the toroidal field, we would like to assess this more quantitatively. For this purpose, we compared the $\Omega$ effect to the generation of toroidal field by the diagonal components $\alpha_{rr}$ and $\alpha_{\theta \theta}$ (see Appendix \ref{an:alpha}), which would be relevant in the case of an $\alpha^2\Omega$ dynamo.  
To distinguish between an $\alpha\Omega$ and $\alpha^2\Omega$ dynamo, we computed the ratio of the two dynamo numbers $C_\alpha = {\rm max(\alpha_{rr},\alpha_{\theta\theta}) R}/{\eta}$ and $C_\Omega={\Omega R^2}/{\eta}$ in the turbulent region, which gives
\begin{gather}
    \frac{C_\Omega}{C_\alpha} = \frac{\Omega R}{\alpha} \approx 100, 
\end{gather}
where $R$ is taken as the middle radius $(\ri + \ro)/2$. 
This estimate therefore confirms that the $\Omega$ effect largely dominates the generation of the toroidal field and that the dynamo in our simulations is an $\alpha\Omega$ dynamo. 

\subsection{Properties of the $ \alpha$ tensor components}

The MRI-driven $\alpha\Omega$ dynamo in our simulations was characterized by estimating the tensor components $\alpha_{ij}$.
Our method assumes that the EMF component $\mathcal{E}_i$ is only due to the contribution of the mean magnetic field component $B_j$ and neglects the contribution of the other components of the $\alpha$ tensor and the turbulent resistivity tensor $\beta$ components. 
In the azimuthal EMF $\mathcal{E}_\phi$, the contribution from $\alpha_{\phi r}$ and $\alpha_{\phi\theta}$ can be neglected because the Pearson correlation coefficients between $\mathcal{E}_\phi$ and $B_\theta$ or $B_r$ are much lower in the turbulent region than they are for $B_\phi$ (panels (c) and (d) of Appendix \ref{an:corre}). 
 The $\alpha_{\phi\phi}$ component is well estimated and is therefore the main contribution to the generation of the poloidal magnetic field. Due to their weaker correlation in the turbulent region, the other $\alpha_{\phi i}$ components (Appendix \ref{an:alpha}) may be incorrectly estimated, and they do not drive the generation of the poloidal field. The other $\alpha$ components do not dominate the generation of toroidal field because the dynamo cannot be described as an $\alpha^2 \Omega$ dynamo.

In the case of a statistically symmetric velocity field with respect to the equator, mean-field theory predicts that the components of the $\alpha_{ij}$ tensor (Eq. \ref{eq:emf}) are either equatorial-symmetric ($\alpha_{\mathrm{r}\theta},\ \alpha_{\theta \mathrm{r}},\ \alpha_{\phi\theta},\ \alpha_{\theta \phi}$) or equatorial-antisymmetric ($\alpha_{\mathrm{rr}},\ \alpha_{\phi \mathrm{r}},\ \alpha_{\phi\phi},\ \alpha_{\mathrm{r} \phi},\ \alpha_{\theta \theta}$) \citep{1993GubbinsSymmetryProperties}. 
The estimation of the components $\alpha_{ij}$ (Fig.~\ref{alpha_values} and Appendix \ref{an:alpha}) shows that these properties are verified in our simulations.

We also compared the signs of the $\alpha$ tensor components to the study of the MRI by \citet{2015GresselMRImeanfield}, who used the test field method to measure them, taking the contribution of all components into account.
For the diagonal components, we also find that the sign of $\alpha_{\phi\phi}$ is opposite to that of $\alpha_{rr}$ (see Appendix~\ref{an:alpha}), which corresponds to the radial component $\alpha_{xx}$ in \citet{2015GresselMRImeanfield}.
Interestingly, in our models the off-diagonal elements of the $\alpha$ tensor have opposite signs across the domain with respect to their conjugate, but they do not have the same amplitude, contrary to the expectations of mean-field theory in the case of isotropic turbulence \citep{1980Krause}. 
By contrast, in local simulations of the MRI with stratification, they are found to have the same sign, which has been interpreted as a significant anisotropy of the MRI turbulence \citep{2008BrandenburgMRImeanfield,2015GresselMRImeanfield}. 
This different behavior might be due to the difference in geometry between a local model with a vertical stratification and a spherical global model with a radial stratification.

\subsection{Diffusive processes}

Ideal MHD without any explicit diffusion has sometimes been used to study the MRI, but convergence studies have shown that the strength of MRI turbulence then depends on resolution \citep{2007FromangMRIA,2007PessahMRI}. To avoid this issue, we considered explicit diffusivities and used pseudo-spectral methods (which have low numerical dissipation), making sure that the diffusive scales are well resolved. 
This constraint forced us to use diffusivities that were sometimes higher than the expected values in a PNS. In particular, the value of the magnetic Prandtl number $Pm$ we adopted in our simulations is much lower than the value expected in a PNS. 
This is a significant limit to our models because the quantitative results of the MRI turbulence are heavily dependent on $Pm$, as shown by studies of accretion disks \citep{2007LesurMRIPm,2007FromangMRIB,2010LongarettiMRIPm,2015MeheutMRIPmlimit} and in paper I.
The magnetic field strength that would be reached at higher $Pm$ is probably higher than the field strengths presented in this work, which should therefore be considered as lower bounds. 
The turbulent region may also be larger and occupy the entire domain in this regime, if its size is indeed affected by the magnetic diffusivity in the denser region.  

Last, we assumed that neutrinos are in the diffusive regime, which is valid when the considered scales are longer than the neutrino mean free path. As the simulation domain is close to the PNS surface, the neutrino mean free path is large ($\simeq \SI{5e5}{cm}$ at $r=\SI{33}{km}$) and most of the turbulence would therefore be impacted by a neutrino drag. In this regime, \citet{2015GuiletVisc} showed that the neutrino drag, weaker in the outer layer of the PNS (for a radius $\ge \SI{30}{km}$), does not impact the linear modes of the MRI. Therefore, the MRI is able to grow freely on small scales in this region.
However, the impact of the neutrino drag has been studied only in the linear phase, but never in nonlinear simulations. It would be especially important to assess it if we consider that the turbulence in our simulations is close to the PNS surface, where the diffusive approximation is less valid.
The evolution of the MRI in this regime is postponed to future studies. 

\subsection{Validity of the anelastic approximation}

The anelastic approximation is used in our models to include density stratification with a reduced computational cost with respect to standard MHD, as it allows filtering out sound waves. 
The filtering of sound waves is justified when the different velocities of our system, such as fluid velocity $u$, AlfvÃ©n velocity $v_\mathrm{A}$ , and gravity wave $v_\mathrm{g}$ velocity, are much lower than the sound speed $c_\mathrm{s}$ of the fluid. From the 1D CCSN simulation used to compute our anelastic reference state, we can compute the sound speed, which ranges from $c_\mathrm{s} \simeq \SI{4e9}{cm.s^{-1}}$ at the inner boundary to $c_\mathrm{s} \simeq \SI{2.9e9}{cm.s^{-1}}$ at the outer boundary. Using the values of AlfvÃ©n speed shown in Fig.~\ref{Comp_dip_alfven}, we can verify a posteriori that ${v_\mathrm{A}^2}/{c_\mathrm{s}^2} \le 10^{-2}$. Because the turbulent kinetic energy is lower than the magnetic energy, we also have ${u^2}/{c_\mathrm{s}^2}\le {v_\mathrm{A}^2}/{c_\mathrm{s}^2} \le 10^{-2}$.
For the gravity waves, we ran some hydrodynamic tests in the anelastic model with $Pr=0.1$ and a rotation that was a thousand times slower. The results agree well with the frequency and damping rate expected from large-scale gravity modes ($l\leq 6$). To justify the filtering of sound waves, we also verified a posteriori in our anelastic simulations that the oscillation frequency of these modes is lower than the Lamb frequency $w_\mathrm{Lamb} = c_\mathrm{s} k$, where $k = \sqrt{l(l+1)/r^2}$ is the horizontal wavenumber of the spherical harmonics. Furthermore, the LBR anelastic approximation \citep{1992LantzPHDanelastic,1995BraginskyAnelastic} that we used describes the propagation of gravity modes well when compared to compressible equations \citep{2012BrownAnelasticEnergyConservation}.
Last, we also verified that relative density perturbations $\delta \rho$ due to entropy perturbations are small: $\delta \rho/\tilde{\rho} \sim \tilde{\alpha}_T \tilde{T} s'/(c_p\tilde{\rho}) \ll 1$. 

The main limitation in our implementation of the anelastic approximation is that the thermodynamical background was assumed not to evolve over the timescale of the simulation. Our simulations lasted approximately one second, but the 1D CCSN simulation shows that during this time, the width of the outer stably stratified region shrinks from approximately $\SI{15}{km}$ to $\SI{5}{km}$,  
while the PNS radius contracts from $\SI{40}{km}$ to $\SI{20}{km}$. It is difficult to take this structural evolution in our models into account. We therefore postpone the study of MRI-driven dynamos at later times to future works. 

\section{Conclusions}
\label{conclusion}

We have investigated the effect of the density gradient and stable stratification on the generation of large-scale magnetic fields by the MRI in 3D spherical PNS models. 
We developed anelastic models using a thermodynamic background describing the stably stratified region of a PNS based on 1D CCSN simulations and compared our results to those obtained with incompressible and Boussinesq models.
The main findings of our study are summarized below. \\

  We obtain a quasi-stationary state with a MRI-driven dynamo in presence of density stratification and stable thermal stratification. 
  The averaged magnetic field strength is $\sim \SI{1.4e14}{G}$ and the (mostly) equatorial magnetic dipole represents $\sim 4.3\%$ of the averaged magnetic field. These results are consistent with our previous incompressible study when they are rescaled for the different densities.

  The MRI-driven dynamo shows a qualitatively different behavior in presence of density stratification, with a prominent axisymmetric component of the magnetic field displaying oscillations on a timescale significantly longer than the rotation period. This mean-field dynamo and its oscillation period can be described by an $\alpha\Omega$ mechanism. 
  The dynamo frequencies in our simulations agree relatively well with the theoretical dynamo frequency explained by an $\alpha$ effect, which is consistent with the theoretical calculations based on the local kinetic helicity. 
 
  A low thermal Prandtl number prevents the buoyancy from damping the MRI turbulence, except in the equatorial plane. When $Pr$ is increased, the radial velocity is distributed at smaller scales, which can impact the MRI-driven dynamo and its $\alpha\Omega$ mechanism. \\

These findings support the hypothesis that the MRI is able to generate a strong large-scale magnetic field in presence of realistic physical ingredients. This means that it is an important mechanism that can help explain magnetar formation.

Although the magnetic dipole of $\SI{6e12}{G}$ obtained in our simulations may seem weak to form a magnetar, we note that these results are obtained with a large PNS with a radius of $r_o =\SI{39.25}{km}$, that is, before the contraction to the final size of a cold neutron star with a radius close to $\SI{12}{km}$. Under the plausible hypothesis that magnetic flux is conserved during this contraction, the magnetic field could be amplified by a factor of $\sim10$ and the dipole would then be close to the lower end of the magnetar range $10^{14}-\SI{e15}{G}$. 
In addition to this dipolar magnetic field, the MRI in our anelastic models generates a large-scale toroidal field of $B \simeq \SI{8e13}{G}$ that can be amplified to $\sim \SI{e15}{G}$ by flux conservation.
Finally, we note that there are reasons to expect that our simulations may underestimate the intensity of the magnetic field generated by the MRI. Importantly, as highlighted in the discussion, a higher magnetic Prandtl number will quantitatively change the results, leading most likely to an increase in the magnetic field strength. Higher magnetic Prandtl numbers may also enable a fully turbulent state extending to the higher-density region at smaller radii, where we may expect a stronger magnetic field.

An important limitation of our approach concerns the evolution of the PNS structure.
The 1D CCSN simulation we used shows that the PNS contracts in about \SI{5}{s} and becomes almost fully convective with a thin stably stratified outer layer of a few kilometers. 
We did not take this evolution into account so far because the required developments are beyond the scope of the present study.
Forthcoming improvements will consist of modeling the entire PNS, which includes the stably stratified and convective zones. 
In the latter, the onset of a convective dynamo with fast rotation can generate magnetar-like magnetic fields \citep{2020ScienceRaynaud}. 
This leads to a strong magnetic field that is buried below the stably stratified zone, which could impact the MRI-driven dynamo. 
The stably stratified region may also influence the convective dynamo, as shown in some studies of planetary dynamos \citep{2020GastineStableStrat}. 
The interplay between the convective dynamo and the MRI-driven dynamo is therefore a key question of magnetorotational explosions because the magnetic field at the PNS surface has the strongest impact on the launch of the explosion \citep{2017Obergaulingerprotomagnetar,2020Obergaulinger,2020Bugli,2021Bugli3D}.  

The core-collapse dynamics may be impacted by the large-scale magnetic field that we obtain in our simulations. The dipole field in our models may be slightly low for directly launching jets, especially because the equatorial component is less efficient than the aligned component, as shown by \citet{2021Bugli3D}.
Nonetheless, other large-scale mean-field structures, such as the toroidal mean magnetic field, may still impact the supernova dynamics. 
The mean-field dynamo in our simulations opens exciting perspectives for modeling the generation of large-scale magnetic fields in core-collapse simulations. Our results can be used to calibrate a subgrid model of the MRI-driven turbulence with an $\alpha\Omega$ dynamo mechanism. This would allow models to describe an in-situ\textit{} amplification of the PNS magnetic field instead of relying on an unrealistically strong initial magnetic field. 

Magnetic fields are also important in the context of binary neutron star mergers because magnetars are invoked to power short gamma-ray bursts and kilonovae, such as the GW170817 event \citep{2018MetzgerMagnetarOrigin}.  
A stable magnetar could indeed explain the high luminosity of the kilonova associated with the recent short gamma-ray burst GRB200522A \citep{2021FongGRB200522A}. This interpretation was also invoked to explain an X-ray transient as the aftermath of a binary neutron star merger \citep{2019XueXBNS}.
To support this scenario, the MRI has been invoked to amplify the magnetic field, as similar conditions in terms of neutrino radiation and differential rotation can be found in neutron star mergers \citep{2017GuiletMergers}.
Realistic simulations have shown that the MRI amplifies the magnetic field from an initial strong axial dipole \citep{2013SiegelMRIBNS,2018KiuchiGRMHD,2019Ciolfi,2020MostaMagnetarKilo}. It is difficult to study large-scale field generation in realistic models of neutron star mergers because it requires taking many different physical ingredients into account: general relativity, treatment of neutrino physics, equation of state of hot and dense matter, and MHD. The development of idealized models with methods similar to those employed in this study will help to understand the effects of different physical processes and provide a useful reference for comparison and calibration of $\alpha\Omega$ dynamos in merger simulations \citep{2021ShibataGRMHDalphaOmega}.

Investigating the different scenarios of magnetar formation is a promising avenue of research as more statistics will be available for transients events in the multi-messenger era.  
For instance, new statistics on FRBs and their host galaxies may give new insights on magnetar formation in further galaxies because magnetars are at least one of the FRB progenitors.

\begin{acknowledgements}
We thank Hans-Thomas Janka who provided the data of 1D CCSN supernova simulations. This research was supported by the European Research Council through the ERC starting grant MagBURST No. 715368. The computations were performed on the supercomputer Occigen of the CINES (Applications A0050410317, A0070410317 and A0090410317). The authors are grateful to Thomas Gastine, Thierry Foglizzo and SÃ©bastien Fromang for their valuable insight and expertise.
\end{acknowledgements}

\bibliographystyle{aa}
\bibliography{biblio,supplement}

\begin{thebibliography}{108}
\expandafter\ifx\csname natexlab\endcsname\relax\def\natexlab#1{#1}\fi

\bibitem[{{Acheson} \& {Gibbons}(1978)}]{1978Acheson}
{Acheson}, D.~J. \& {Gibbons}, M.~P. 1978, Philosophical Transactions of the
  Royal Society of London Series A, 289, 459

\bibitem[{{Akiyama} {et~al.}(2003){Akiyama}, {Wheeler}, {Meier}, \&
  {Lichtenstadt}}]{2003AkiyamaMRI}
{Akiyama}, S., {Wheeler}, J.~C., {Meier}, D.~L., \& {Lichtenstadt}, I. 2003,
  \apj, 584, 954

\bibitem[{{Balbus}(1995)}]{1995BalbusMRI}
{Balbus}, S.~A. 1995, \apj, 453, 380

\bibitem[{{Balbus} \& {Hawley}(1991)}]{1991BalbusMRI}
{Balbus}, S.~A. \& {Hawley}, J.~F. 1991, \apj, 376, 214

\bibitem[{{Bochenek} {et~al.}(2020){Bochenek}, {Ravi}, {Belov}, {Hallinan},
  {Kocz}, {Kulkarni}, \& {McKenna}}]{2020BochenekGalaticFRB}
{Bochenek}, C.~D., {Ravi}, V., {Belov}, K.~V., {et~al.} 2020, \nat, 587, 59

\bibitem[{{Braginsky} \& {Roberts}(1995)}]{1995BraginskyAnelastic}
{Braginsky}, S.~I. \& {Roberts}, P.~H. 1995, Geophysical and Astrophysical
  Fluid Dynamics, 79, 1

\bibitem[{{Brandenburg}(2008)}]{2008BrandenburgMRImeanfield}
{Brandenburg}, A. 2008, Astronomische Nachrichten, 329, 725

\bibitem[{{Brown} {et~al.}(2012){Brown}, {Vasil}, \&
  {Zweibel}}]{2012BrownAnelasticEnergyConservation}
{Brown}, B.~P., {Vasil}, G.~M., \& {Zweibel}, E.~G. 2012, \apj, 756, 109

\bibitem[{{Bucciantini} {et~al.}(2012){Bucciantini}, {Metzger}, {Thompson}, \&
  {Quataert}}]{2012BucciantiniSGRB}
{Bucciantini}, N., {Metzger}, B.~D., {Thompson}, T.~A., \& {Quataert}, E. 2012,
  \mnras, 419, 1537

\bibitem[{{Bugli} {et~al.}(2021){Bugli}, {Guilet}, \&
  {Obergaulinger}}]{2021Bugli3D}
{Bugli}, M., {Guilet}, J., \& {Obergaulinger}, M. 2021, \mnras, 507, 443

\bibitem[{{Bugli} {et~al.}(2020){Bugli}, {Guilet}, {Obergaulinger},
  {Cerd{\'a}-Dur{\'a}n}, \& {Aloy}}]{2020Bugli}
{Bugli}, M., {Guilet}, J., {Obergaulinger}, M., {Cerd{\'a}-Dur{\'a}n}, P., \&
  {Aloy}, M.~A. 2020, \mnras, 492, 58

\bibitem[{{Busse} \& {Simitev}(2006)}]{2006BusseFrequencyDynamo}
{Busse}, F.~H. \& {Simitev}, R.~D. 2006, Geophysical and Astrophysical Fluid
  Dynamics, 100, 341

\bibitem[{{CHIME/FRB Collaboration} {et~al.}(2020){CHIME/FRB Collaboration},
  {Andersen}, {Bandura}, {Bhardwaj}, {Bij}, {Boyce}, {Boyle}, {Brar},
  {Cassanelli}, {Chawla}, {Chen}, {Cliche}, {Cook}, {Cubranic}, {Curtin},
  {Denman}, {Dobbs}, {Dong}, {Fandino}, {Fonseca}, {Gaensler}, {Giri}, {Good},
  {Halpern}, {Hill}, {Hinshaw}, {H{\"o}fer}, {Josephy}, {Kania}, {Kaspi},
  {Landecker}, {Leung}, {Li}, {Lin}, {Masui}, {McKinven}, {Mena-Parra},
  {Merryfield}, {Meyers}, {Michilli}, {Milutinovic}, {Mirhosseini},
  {M{\"u}nchmeyer}, {Naidu}, {Newburgh}, {Ng}, {Patel}, {Pen},
  {Pinsonneault-Marotte}, {Pleunis}, {Quine}, {Rafiei-Ravandi}, {Rahman},
  {Ransom}, {Renard}, {Sanghavi}, {Scholz}, {Shaw}, {Shin}, {Siegel}, {Singh},
  {Smegal}, {Smith}, {Stairs}, {Tan}, {Tendulkar}, {Tretyakov}, {Vanderlinde},
  {Wang}, {Wulf}, \& {Zwaniga}}]{2020ChimeGalacticMagnetar}
{CHIME/FRB Collaboration}, {Andersen}, B.~C., {Bandura}, K.~M., {et~al.} 2020,
  \nat, 587, 54

\bibitem[{Christensen \& Wicht(2015)}]{2015CHRISTENSEN245}
Christensen, U. \& Wicht, J. 2015, in Treatise on Geophysics (Second Edition),
  second edition edn., ed. G.~Schubert (Oxford: Elsevier), 245 -- 277

\bibitem[{{Ciolfi} {et~al.}(2019){Ciolfi}, {Kastaun}, {Kalinani}, \&
  {Giacomazzo}}]{2019Ciolfi}
{Ciolfi}, R., {Kastaun}, W., {Kalinani}, J.~V., \& {Giacomazzo}, B. 2019, \prd,
  100, 023005

\bibitem[{{Coti Zelati} {et~al.}(2021){Coti Zelati}, {Borghese}, {Israel},
  {Rea}, {Esposito}, {Pilia}, {Burgay}, {Possenti}, {Corongiu}, {Ridolfi},
  {Dehman}, {Vigan{\`o}}, {Turolla}, {Zane}, {Tiengo}, \&
  {Keane}}]{2021CotiZelatiOutbursts}
{Coti Zelati}, F., {Borghese}, A., {Israel}, G.~L., {et~al.} 2021, \apjl, 907,
  L34

\bibitem[{{Coti Zelati} {et~al.}(2018){Coti Zelati}, {Rea}, {Pons}, {Campana},
  \& {Esposito}}]{2018CotiZelati}
{Coti Zelati}, F., {Rea}, N., {Pons}, J.~A., {Campana}, S., \& {Esposito}, P.
  2018, \mnras, 474, 961

\bibitem[{{Davis} {et~al.}(2010){Davis}, {Stone}, \&
  {Pessah}}]{2010DavisButterfly}
{Davis}, S.~W., {Stone}, J.~M., \& {Pessah}, M.~E. 2010, \apj, 713, 52

\bibitem[{Deng {et~al.}(2019)Deng, Mayer, Latter, Hopkins, \&
  Bai}]{2019DengMRI}
Deng, H., Mayer, L., Latter, H., Hopkins, P.~F., \& Bai, X.-N. 2019, The
  Astrophysical Journal Supplement Series, 241, 26

\bibitem[{{Dessart} {et~al.}(2008){Dessart}, {Burrows}, {Livne}, \&
  {Ott}}]{2008DessartHyper}
{Dessart}, L., {Burrows}, A., {Livne}, E., \& {Ott}, C.~D. 2008, \apj, 673, L43

\bibitem[{{Drout} {et~al.}(2011){Drout}, {Soderberg}, {Gal-Yam}, {Cenko},
  {Fox}, {Leonard}, {Sand}, {Moon}, {Arcavi}, \& {Green}}]{2011Drout}
{Drout}, M.~R., {Soderberg}, A.~M., {Gal-Yam}, A., {et~al.} 2011, \apj, 741, 97

\bibitem[{{Duncan} \& {Thompson}(1992)}]{1992DuncanLGRB}
{Duncan}, R.~C. \& {Thompson}, C. 1992, \apj, 392, L9

\bibitem[{{Esposito} {et~al.}(2021){Esposito}, {Rea}, \&
  {Israel}}]{2021EspositoMagnetarReview}
{Esposito}, P., {Rea}, N., \& {Israel}, G.~L. 2021, {Magnetars: A Short Review
  and Some Sparse Considerations}, ed. T.~M. {Belloni}, M.~{M{\'e}ndez}, \&
  C.~{Zhang}, Vol. 461, 97--142

\bibitem[{{Fong} {et~al.}(2021){Fong}, {Laskar}, {Rastinejad}, {Escorial},
  {Schroeder}, {Barnes}, {Kilpatrick}, {Paterson}, {Berger}, {Metzger}, {Dong},
  {Nugent}, {Strausbaugh}, {Blanchard}, {Goyal}, {Cucchiara}, {Terreran},
  {Alexander}, {Eftekhari}, {Fryer}, {Margalit}, {Margutti}, \&
  {Nicholl}}]{2021FongGRB200522A}
{Fong}, W., {Laskar}, T., {Rastinejad}, J., {et~al.} 2021, \apj, 906, 127

\bibitem[{{Fromang} \& {Papaloizou}(2007)}]{2007FromangMRIA}
{Fromang}, S. \& {Papaloizou}, J. 2007, \aap, 476, 1113

\bibitem[{{Fromang} {et~al.}(2007){Fromang}, {Papaloizou}, {Lesur}, \&
  {Heinemann}}]{2007FromangMRIB}
{Fromang}, S., {Papaloizou}, J., {Lesur}, G., \& {Heinemann}, T. 2007, \aap,
  476, 1123

\bibitem[{{Fuller} {et~al.}(2019){Fuller}, {Piro}, \&
  {Jermyn}}]{2019FullerTaylerSpruit}
{Fuller}, J., {Piro}, A.~L., \& {Jermyn}, A.~S. 2019, \mnras, 485, 3661

\bibitem[{{Gastine} {et~al.}(2012){Gastine}, {Duarte}, \&
  {Wicht}}]{2012GastineDipolarMultipolar}
{Gastine}, T., {Duarte}, L., \& {Wicht}, J. 2012, \aap, 546, A19

\bibitem[{Gastine \& Wicht(2012)}]{2012GastineCode}
Gastine, T. \& Wicht, J. 2012, Icarus, 219, 428

\bibitem[{Gastine \& Wicht(2021)}]{2020GastineStableStrat}
Gastine, T. \& Wicht, J. 2021, Icarus, 368, 114514

\bibitem[{{Gilman} \& {Glatzmaier}(1981)}]{1981GilmanAnel}
{Gilman}, P.~A. \& {Glatzmaier}, G.~A. 1981, \apjs, 45, 335

\bibitem[{{Gompertz} {et~al.}(2014){Gompertz}, {O'Brien}, \&
  {Wynn}}]{2014GompertzSGRB}
{Gompertz}, B.~P., {O'Brien}, P.~T., \& {Wynn}, G.~A. 2014, \mnras, 438, 240

\bibitem[{{G{\"o}tz} {et~al.}(2006){G{\"o}tz}, {Mereghetti}, {Molkov},
  {Hurley}, {Mirabel}, {Sunyaev}, {Weidenspointner}, {Brand t}, {del Santo},
  {Feroci}, {G{\"o}{\u{g}}{\"u}{\textcommabelow s}}, {von Kienlin}, {van der
  Klis}, {Kouveliotou}, {Lund}, {Pizzichini}, {Ubertini}, {Winkler}, \&
  {Woods}}]{2006Gotz}
{G{\"o}tz}, D., {Mereghetti}, S., {Molkov}, S., {et~al.} 2006, \aap, 445, 313

\bibitem[{{Gouhier} {et~al.}(2022){Gouhier}, {Jouve}, \&
  {Ligni{\`e}res}}]{2022GouhierMHD}
{Gouhier}, B., {Jouve}, L., \& {Ligni{\`e}res}, F. 2022, \aap, 661, A119

\bibitem[{{Gouhier} {et~al.}(2021){Gouhier}, {Ligni{\`e}res}, \&
  {Jouve}}]{2021Gouhier}
{Gouhier}, B., {Ligni{\`e}res}, F., \& {Jouve}, L. 2021, \aap, 648, A109

\bibitem[{{Gressel} \& {Pessah}(2015)}]{2015GresselMRImeanfield}
{Gressel}, O. \& {Pessah}, M.~E. 2015, \apj, 810, 59

\bibitem[{{Gubbins} \& {Zhang}(1993)}]{1993GubbinsSymmetryProperties}
{Gubbins}, D. \& {Zhang}, K. 1993, Physics of the Earth and Planetary
  Interiors, 75, 225

\bibitem[{{Guilet} {et~al.}(2017){Guilet}, {Bauswein}, {Just}, \&
  {Janka}}]{2017GuiletMergers}
{Guilet}, J., {Bauswein}, A., {Just}, O., \& {Janka}, H.-T. 2017, \mnras, 471,
  1879

\bibitem[{{Guilet} \& {M{\"u}ller}(2015)}]{2015GuiletBuoy}
{Guilet}, J. \& {M{\"u}ller}, E. 2015, \mnras, 450, 2153

\bibitem[{{Guilet} {et~al.}(2015){Guilet}, {M{\"u}ller}, \&
  {Janka}}]{2015GuiletVisc}
{Guilet}, J., {M{\"u}ller}, E., \& {Janka}, H.-T. 2015, \mnras, 447, 3992

\bibitem[{{Hawley} \& {Balbus}(1992)}]{1992HawleyMRI}
{Hawley}, J.~F. \& {Balbus}, S.~A. 1992, \apj, 400, 595

\bibitem[{H\"udepohl(2014)}]{2014Hudepol}
H\"udepohl, L. 2014, PhD thesis, Technische Universit\"at, M\"unchen

\bibitem[{{Hurley} {et~al.}(2005){Hurley}, {Boggs}, {Smith}, {Duncan}, {Lin},
  {Zoglauer}, {Krucker}, {Hurford}, {Hudson}, {Wigger}, {Hajdas}, {Thompson},
  {Mitrofanov}, {Sanin}, {Boynton}, {Fellows}, {von Kienlin}, {Lichti}, {Rau},
  \& {Cline}}]{2005HurleyFlare}
{Hurley}, K., {Boggs}, S.~E., {Smith}, D.~M., {et~al.} 2005, \nat, 434, 1098

\bibitem[{{Inserra} {et~al.}(2013){Inserra}, {Smartt}, {Jerkstrand}, {Valenti},
  {Fraser}, {Wright}, {Smith}, {Chen}, {Kotak}, {Pastorello}, {Nicholl},
  {Bresolin}, {Kudritzki}, {Benetti}, {Botticella}, {Burgett}, {Chambers},
  {Ergon}, {Flewelling}, {Fynbo}, {Geier}, {Hodapp}, {Howell}, {Huber},
  {Kaiser}, {Leloudas}, {Magill}, {Magnier}, {McCrum}, {Metcalfe}, {Price},
  {Rest}, {Sollerman}, {Sweeney}, {Taddia}, {Taubenberger}, {Tonry},
  {Wainscoat}, {Waters}, \& {Young}}]{2013InserraSLSN}
{Inserra}, C., {Smartt}, S.~J., {Jerkstrand}, A., {et~al.} 2013, \apj, 770, 128

\bibitem[{{Israel} {et~al.}(2005){Israel}, {Belloni}, {Stella}, {Rephaeli},
  {Gruber}, {Casella}, {Dall'Osso}, {Rea}, {Persic}, \&
  {Rothschild}}]{2005Israel}
{Israel}, G.~L., {Belloni}, T., {Stella}, L., {et~al.} 2005, \apjl, 628, L53

\bibitem[{{Jones} {et~al.}(2011){Jones}, {Boronski}, {Brun}, {Glatzmaier},
  {Gastine}, {Miesch}, \& {Wicht}}]{2011JonesAnelasticBench}
{Jones}, C.~A., {Boronski}, P., {Brun}, A.~S., {et~al.} 2011, \icarus, 216, 120

\bibitem[{{Kaspi} \& {Beloborodov}(2017)}]{2017Kaspi}
{Kaspi}, V.~M. \& {Beloborodov}, A.~M. 2017, \araa, 55, 261

\bibitem[{{Kiuchi} {et~al.}(2018){Kiuchi}, {Kyutoku}, {Sekiguchi}, \&
  {Shibata}}]{2018KiuchiGRMHD}
{Kiuchi}, K., {Kyutoku}, K., {Sekiguchi}, Y., \& {Shibata}, M. 2018, \prd, 97,
  124039

\bibitem[{{Kouveliotou} {et~al.}(1998){Kouveliotou}, {Dieters}, {Strohmayer},
  {van Paradijs}, {Fishman}, {Meegan}, {Hurley}, {Kommers}, {Smith}, {Frail},
  \& {Murakami}}]{1998Kouveliotou}
{Kouveliotou}, C., {Dieters}, S., {Strohmayer}, T., {et~al.} 1998, \nat, 393,
  235

\bibitem[{{Krause} \& {Raedler}(1980{\natexlab{a}})}]{1980KrauseMeanField}
{Krause}, F. \& {Raedler}, K.~H. 1980{\natexlab{a}}, {Mean-field
  magnetohydrodynamics and dynamo theory}

\bibitem[{{Krause} \& {Raedler}(1980{\natexlab{b}})}]{1980Krause}
{Krause}, F. \& {Raedler}, K.~H. 1980{\natexlab{b}}, {Mean-field
  magnetohydrodynamics and dynamo theory} ({Pergamon Press})

\bibitem[{{Kuroda} {et~al.}(2020){Kuroda}, {Arcones}, {Takiwaki}, \&
  {Kotake}}]{2020KurodaMRexplosion3D}
{Kuroda}, T., {Arcones}, A., {Takiwaki}, T., \& {Kotake}, K. 2020, \apj, 896,
  102

\bibitem[{{Lantz}(1992)}]{1992LantzPHDanelastic}
{Lantz}, S.~R. 1992, PhD thesis, CORNELL UNIVERSITY.

\bibitem[{{Lattimer} \& {Swesty}(1991)}]{1991LattimerEOSdensematter}
{Lattimer}, J.~M. \& {Swesty}, D.~F. 1991, \nphysa, 535, 331

\bibitem[{{Lesur} \& {Longaretti}(2007)}]{2007LesurMRIPm}
{Lesur}, G. \& {Longaretti}, P.~Y. 2007, \mnras, 378, 1471

\bibitem[{{Lesur} \& {Ogilvie}(2008)}]{2008LesurMRINondiagonalresistivity}
{Lesur}, G. \& {Ogilvie}, G.~I. 2008, \aap, 488, 451

\bibitem[{{Longaretti} \& {Lesur}(2010)}]{2010LongarettiMRIPm}
{Longaretti}, P.~Y. \& {Lesur}, G. 2010, \aap, 516, A51

\bibitem[{{Lorimer} {et~al.}(2007){Lorimer}, {Bailes}, {McLaughlin},
  {Narkevic}, \& {Crawford}}]{2007LorimerfirstFRB}
{Lorimer}, D.~R., {Bailes}, M., {McLaughlin}, M.~A., {Narkevic}, D.~J., \&
  {Crawford}, F. 2007, Science, 318, 777

\bibitem[{{Margalit} {et~al.}(2018){Margalit}, {Metzger}, {Thompson},
  {Nicholl}, \& {Sukhbold}}]{2018MargalitSLSN}
{Margalit}, B., {Metzger}, B.~D., {Thompson}, T.~A., {Nicholl}, M., \&
  {Sukhbold}, T. 2018, \mnras, 475, 2659

\bibitem[{{Masada} {et~al.}(2007){Masada}, {Sano}, \&
  {Shibata}}]{2007MasadaMRIneutrinolinear}
{Masada}, Y., {Sano}, T., \& {Shibata}, K. 2007, \apj, 655, 447

\bibitem[{{Masada} {et~al.}(2015){Masada}, {Takiwaki}, \&
  {Kotake}}]{2015MasadaMRIdisk}
{Masada}, Y., {Takiwaki}, T., \& {Kotake}, K. 2015, \apj, 798, L22

\bibitem[{{Masada} {et~al.}(2020){Masada}, {Takiwaki}, \&
  {Kotake}}]{2020MasadaNSConvection}
{Masada}, Y., {Takiwaki}, T., \& {Kotake}, K. 2020, arXiv e-prints,
  arXiv:2001.08452

\bibitem[{{Masada} {et~al.}(2012){Masada}, {Takiwaki}, {Kotake}, \&
  {Sano}}]{2012MasadaRot}
{Masada}, Y., {Takiwaki}, T., {Kotake}, K., \& {Sano}, T. 2012, \apj, 759, 110

\bibitem[{{Meheut} {et~al.}(2015){Meheut}, {Fromang}, {Lesur}, {Joos}, \&
  {Longaretti}}]{2015MeheutMRIPmlimit}
{Meheut}, H., {Fromang}, S., {Lesur}, G., {Joos}, M., \& {Longaretti}, P.-Y.
  2015, \aap, 579, A117

\bibitem[{{Menou} {et~al.}(2004){Menou}, {Balbus}, \& {Spruit}}]{2004MenouMRI}
{Menou}, K., {Balbus}, S.~A., \& {Spruit}, H.~C. 2004, \apj, 607, 564

\bibitem[{{Metzger} {et~al.}(2018{\natexlab{a}}){Metzger}, {Beniamini}, \&
  {Giannios}}]{2018MetzgerLGRBs}
{Metzger}, B.~D., {Beniamini}, P., \& {Giannios}, D. 2018{\natexlab{a}}, \apj,
  857, 95

\bibitem[{{Metzger} {et~al.}(2011){Metzger}, {Giannios}, {Thompson},
  {Bucciantini}, \& {Quataert}}]{2011MetzgerLGRB}
{Metzger}, B.~D., {Giannios}, D., {Thompson}, T.~A., {Bucciantini}, N., \&
  {Quataert}, E. 2011, \mnras, 413, 2031

\bibitem[{{Metzger} {et~al.}(2008){Metzger}, {Quataert}, \&
  {Thompson}}]{2008MetzgerSGRB}
{Metzger}, B.~D., {Quataert}, E., \& {Thompson}, T.~A. 2008, \mnras, 385, 1455

\bibitem[{{Metzger} {et~al.}(2018{\natexlab{b}}){Metzger}, {Thompson}, \&
  {Quataert}}]{2018MetzgerMagnetarOrigin}
{Metzger}, B.~D., {Thompson}, T.~A., \& {Quataert}, E. 2018{\natexlab{b}},
  \apj, 856, 101

\bibitem[{Moffatt(1978)}]{1978moffattfield}
Moffatt, H.~K. 1978, Cambridge University Press, Cambridge, London, New York,
  Melbourne, 2, 5

\bibitem[{{Moiseenko} {et~al.}(2006){Moiseenko}, {Bisnovatyi-Kogan}, \&
  {Ardeljan}}]{2006MoiseenkoMHDjet}
{Moiseenko}, S.~G., {Bisnovatyi-Kogan}, G.~S., \& {Ardeljan}, N.~V. 2006,
  \mnras, 370, 501

\bibitem[{{M{\"o}sta} {et~al.}(2015){M{\"o}sta}, {Ott}, {Radice}, {Roberts},
  {Schnetter}, \& {Haas}}]{2015MostaMRI}
{M{\"o}sta}, P., {Ott}, C.~D., {Radice}, D., {et~al.} 2015, \nat, 528, 376

\bibitem[{{M{\"o}sta} {et~al.}(2020){M{\"o}sta}, {Radice}, {Haas}, {Schnetter},
  \& {Bernuzzi}}]{2020MostaMagnetarKilo}
{M{\"o}sta}, P., {Radice}, D., {Haas}, R., {Schnetter}, E., \& {Bernuzzi}, S.
  2020, \apjl, 901, L37

\bibitem[{{M{\"o}sta} {et~al.}(2014){M{\"o}sta}, {Richers}, {Ott}, {Haas},
  {Piro}, {Boydstun}, {Abdikamalov}, {Reisswig}, \&
  {Schnetter}}]{2014MostaHyper}
{M{\"o}sta}, P., {Richers}, S., {Ott}, C.~D., {et~al.} 2014, \apj, 785, L29

\bibitem[{{Nicholl} {et~al.}(2013){Nicholl}, {Smartt}, {Jerkstrand}, {Inserra},
  {McCrum}, {Kotak}, {Fraser}, {Wright}, {Chen}, {Smith}, {Young}, {Sim},
  {Valenti}, {Howell}, {Bresolin}, {Kudritzki}, {Tonry}, {Huber}, {Rest},
  {Pastorello}, {Tomasella}, {Cappellaro}, {Benetti}, {Mattila}, {Kankare},
  {Kangas}, {Leloudas}, {Sollerman}, {Taddia}, {Berger}, {Chornock}, {Narayan},
  {Stubbs}, {Foley}, {Lunnan}, {Soderberg}, {Sanders}, {Milisavljevic},
  {Margutti}, {Kirshner}, {Elias-Rosa}, {Morales-Garoffolo}, {Taubenberger},
  {Botticella}, {Gezari}, {Urata}, {Rodney}, {Riess}, {Scolnic}, {Wood-Vasey},
  {Burgett}, {Chambers}, {Flewelling}, {Magnier}, {Kaiser}, {Metcalfe},
  {Morgan}, {Price}, {Sweeney}, \& {Waters}}]{2013NichollSLSN}
{Nicholl}, M., {Smartt}, S.~J., {Jerkstrand}, A., {et~al.} 2013, \nat, 502, 346

\bibitem[{{Obergaulinger} \& {Aloy}(2017)}]{2017Obergaulingerprotomagnetar}
{Obergaulinger}, M. \& {Aloy}, M.~{\'A}. 2017, \mnras, 469, L43

\bibitem[{{Obergaulinger} \& {Aloy}(2020)}]{2020Obergaulinger}
{Obergaulinger}, M. \& {Aloy}, M.~{\'A}. 2020, \mnras, 492, 4613

\bibitem[{{Obergaulinger} {et~al.}(2009){Obergaulinger}, {Cerd{\'a}-Dur{\'a}n},
  {M{\"u}ller}, \& {Aloy}}]{2009ObergaulingerMRI}
{Obergaulinger}, M., {Cerd{\'a}-Dur{\'a}n}, P., {M{\"u}ller}, E., \& {Aloy},
  M.~A. 2009, \aap, 498, 241

\bibitem[{{Obergaulinger} {et~al.}(2018){Obergaulinger}, {Just}, \&
  {Aloy}}]{2018Obergaulinger}
{Obergaulinger}, M., {Just}, O., \& {Aloy}, M.~A. 2018, J. Phys. G Nucl. Part.
  Phys., 45, 084001

\bibitem[{{Ott} {et~al.}(2006){Ott}, {Burrows}, {Thompson}, {Livne}, \&
  {Walder}}]{2006OttRotPNS}
{Ott}, C.~D., {Burrows}, A., {Thompson}, T.~A., {Livne}, E., \& {Walder}, R.
  2006, \apjs, 164, 130

\bibitem[{{Parker}(1955)}]{1955AParkerDynamo}
{Parker}, E.~N. 1955, \apj, 122, 293

\bibitem[{{Pessah} \& {Chan}(2008)}]{2008PessahMRI}
{Pessah}, M.~E. \& {Chan}, C.-k. 2008, \apj, 684, 498

\bibitem[{{Pessah} {et~al.}(2007){Pessah}, {Chan}, \&
  {Psaltis}}]{2007PessahMRI}
{Pessah}, M.~E., {Chan}, C.-k., \& {Psaltis}, D. 2007, \apj, 668, L51

\bibitem[{{Rampp} \& {Janka}(2002)}]{2002RamppJankaVertex}
{Rampp}, M. \& {Janka}, H.~T. 2002, \aap, 396, 361

\bibitem[{{Raynaud} {et~al.}(2021){Raynaud}, {Cerd{\'a}-Dur{\'a}n}, \&
  {Guilet}}]{2021RaynaudGWPNS}
{Raynaud}, R., {Cerd{\'a}-Dur{\'a}n}, P., \& {Guilet}, J. 2021, arXiv e-prints,
  arXiv:2103.12445

\bibitem[{Raynaud {et~al.}(2020)Raynaud, Guilet, Janka, \&
  Gastine}]{2020ScienceRaynaud}
Raynaud, R., Guilet, J., Janka, H.-T., \& Gastine, T. 2020, Science Advances, 6

\bibitem[{{Reboul-Salze} {et~al.}(2021){Reboul-Salze}, {Guilet}, {Raynaud}, \&
  {Bugli}}]{2021ReboulSalze}
{Reboul-Salze}, A., {Guilet}, J., {Raynaud}, R., \& {Bugli}, M. 2021, \aap,
  645, A109

\bibitem[{{Rembiasz} {et~al.}(2016){Rembiasz}, {Guilet}, {Obergaulinger},
  {Cerd{\'a}-Dur{\'a}n}, {Aloy}, \& {M{\"u}ller}}]{2016RembiazMRIMaxB}
{Rembiasz}, T., {Guilet}, J., {Obergaulinger}, M., {et~al.} 2016, \mnras, 460,
  3316

\bibitem[{{Roberts} {et~al.}(2021){Roberts}, {Veres}, {Baring}, {Briggs},
  {Kouveliotou}, {Bissaldi}, {Younes}, {Chastain}, {DeLaunay}, {Huppenkothen},
  {Tohuvavohu}, {Bhat}, {G{\"o}{\v{g}}{\"u}{\c{s}}}, {van der Horst}, {Kennea},
  {Kocevski}, {Linford}, {Guiriec}, {Hamburg}, {Wilson-Hodge}, \&
  {Burns}}]{2021RobertsGiantFlare}
{Roberts}, O.~J., {Veres}, P., {Baring}, M.~G., {et~al.} 2021, \nat, 589, 207

\bibitem[{{Rodr{\'\i}guez Castillo} {et~al.}(2016){Rodr{\'\i}guez Castillo},
  {Israel}, {Tiengo}, {Salvetti}, {Turolla}, {Zane}, {Rea}, {Esposito},
  {Mereghetti}, {Perna}, {Stella}, {Pons}, {Campana}, {G{\"o}tz}, \&
  {Motta}}]{2016RodriguezCastillo}
{Rodr{\'\i}guez Castillo}, G.~A., {Israel}, G.~L., {Tiengo}, A., {et~al.} 2016,
  \mnras, 456, 4145

\bibitem[{{Rowlinson} {et~al.}(2013){Rowlinson}, {O'Brien}, {Metzger},
  {Tanvir}, \& {Levan}}]{2013RowlinsonSGRBR}
{Rowlinson}, A., {O'Brien}, P.~T., {Metzger}, B.~D., {Tanvir}, N.~R., \&
  {Levan}, A.~J. 2013, \mnras, 430, 1061

\bibitem[{{Schaeffer}(2013)}]{2013SHTNS}
{Schaeffer}, N. 2013, Geochem. Geophys. Geosyst., 14, 751

\bibitem[{{Schneider} {et~al.}(2019){Schneider}, {Ohlmann}, {Podsiadlowski},
  {R{\"o}pke}, {Balbus}, {Pakmor}, \& {Springel}}]{2019SchneiderStar}
{Schneider}, F. R.~N., {Ohlmann}, S.~T., {Podsiadlowski}, P., {et~al.} 2019,
  \nat, 574, 211

\bibitem[{{Shi} {et~al.}(2016){Shi}, {Stone}, \& {Huang}}]{2016Shi}
{Shi}, J.-M., {Stone}, J.~M., \& {Huang}, C.~X. 2016, \mnras, 456, 2273

\bibitem[{Shibata {et~al.}(2021)Shibata, Fujibayashi, \&
  Sekiguchi}]{2021ShibataGRMHDalphaOmega}
Shibata, M., Fujibayashi, S., \& Sekiguchi, Y. 2021, Phys. Rev. D, 104, 063026

\bibitem[{{Shibata} {et~al.}(2006){Shibata}, {Liu}, {Shapiro}, \&
  {Stephens}}]{2006ShibataHyper}
{Shibata}, M., {Liu}, Y.~T., {Shapiro}, S.~L., \& {Stephens}, B.~C. 2006, \prd,
  74, 104026

\bibitem[{{Siegel} {et~al.}(2013){Siegel}, {Ciolfi}, {Harte}, \&
  {Rezzolla}}]{2013SiegelMRIBNS}
{Siegel}, D.~M., {Ciolfi}, R., {Harte}, A.~I., \& {Rezzolla}, L. 2013, \prd,
  87, 121302

\bibitem[{{Simon} \& {Hawley}(2009)}]{2009SimonMRIPmeffectsonNetTor}
{Simon}, J.~B. \& {Hawley}, J.~F. 2009, \apj, 707, 833

\bibitem[{{Svinkin} {et~al.}(2021){Svinkin}, {Frederiks}, {Hurley}, {Aptekar},
  {Golenetskii}, {Lysenko}, {Ridnaia}, {Tsvetkova}, {Ulanov}, {Cline},
  {Mitrofanov}, {Golovin}, {Kozyrev}, {Litvak}, {Sanin}, {Goldstein}, {Briggs},
  {Wilson-Hodge}, {von Kienlin}, {Zhang}, {Rau}, {Savchenko}, {Bozzo},
  {Ferrigno}, {Ubertini}, {Bazzano}, {Rodi}, {Barthelmy}, {Cummings}, {Krimm},
  {Palmer}, {Boynton}, {Fellows}, {Harshman}, {Enos}, \&
  {Starr}}]{2021SvinkinGiantFlare}
{Svinkin}, D., {Frederiks}, D., {Hurley}, K., {et~al.} 2021, \nat, 589, 211

\bibitem[{{Thompson} \& {Duncan}(1993)}]{1993ThompsonPNS}
{Thompson}, C. \& {Duncan}, R.~C. 1993, \apj, 408, 194

\bibitem[{{Tiengo} {et~al.}(2013){Tiengo}, {Esposito}, {Mereghetti}, {Turolla},
  {Nobili}, {Gastaldello}, {G{\"o}tz}, {Israel}, {Rea}, {Stella}, {Zane}, \&
  {Bignami}}]{2013Tiengo}
{Tiengo}, A., {Esposito}, P., {Mereghetti}, S., {et~al.} 2013, \nat, 500, 312

\bibitem[{{Tilgner} \& {Busse}(1997)}]{1997TilgnerJFM}
{Tilgner}, A. \& {Busse}, F.~H. 1997, J. Fluid Mech., 332, 359

\bibitem[{{Turolla} {et~al.}(2015){Turolla}, {Zane}, \&
  {Watts}}]{2015TurollaMagnetarObsReview}
{Turolla}, R., {Zane}, S., \& {Watts}, A.~L. 2015, Reports on Progress in
  Physics, 78, 116901

\bibitem[{{Wicht}(2002)}]{2002WichtMagic}
{Wicht}, J. 2002, Phys. Earth Planet. Inter., 132, 281

\bibitem[{{Winteler} {et~al.}(2012){Winteler}, {K{\"a}ppeli}, {Perego},
  {Arcones}, {Vasset}, {Nishimura}, {Liebend{\"o}rfer}, \&
  {Thielemann}}]{2012WintelerHyper}
{Winteler}, C., {K{\"a}ppeli}, R., {Perego}, A., {et~al.} 2012, \apj, 750, L22

\bibitem[{{Woosley} {et~al.}(2002){Woosley}, {Heger}, \&
  {Weaver}}]{2002WoosleyProgenitor}
{Woosley}, S.~E., {Heger}, A., \& {Weaver}, T.~A. 2002, Reviews of Modern
  Physics, 74, 1015

\bibitem[{{Xue} {et~al.}(2019){Xue}, {Zheng}, {Li}, {Brandt}, {Zhang}, {Luo},
  {Zhang}, {Bauer}, {Sun}, {Lehmer}, {Wu}, {Yang}, {Kong}, {Li}, {Sun}, {Wang},
  \& {Vito}}]{2019XueXBNS}
{Xue}, Y.~Q., {Zheng}, X.~C., {Li}, Y., {et~al.} 2019, \nat, 568, 198

\bibitem[{{Yoshimura}(1975)}]{1975YoshimuraDynamo}
{Yoshimura}, H. 1975, \apj, 201, 740

\end{thebibliography}

\begin{appendix} 

\onecolumn

\section{Reference state of the anelastic model}
\label{comp_anel_state}

The comparison of the reference model in the simulation to the mixing-length theory model of 1D CCSN simulations is presented in this appendix.

\begin{figure*}[!h]
\centering
\begin{subfigure}{0.45 \textwidth}
\includegraphics[width=1.0\textwidth]{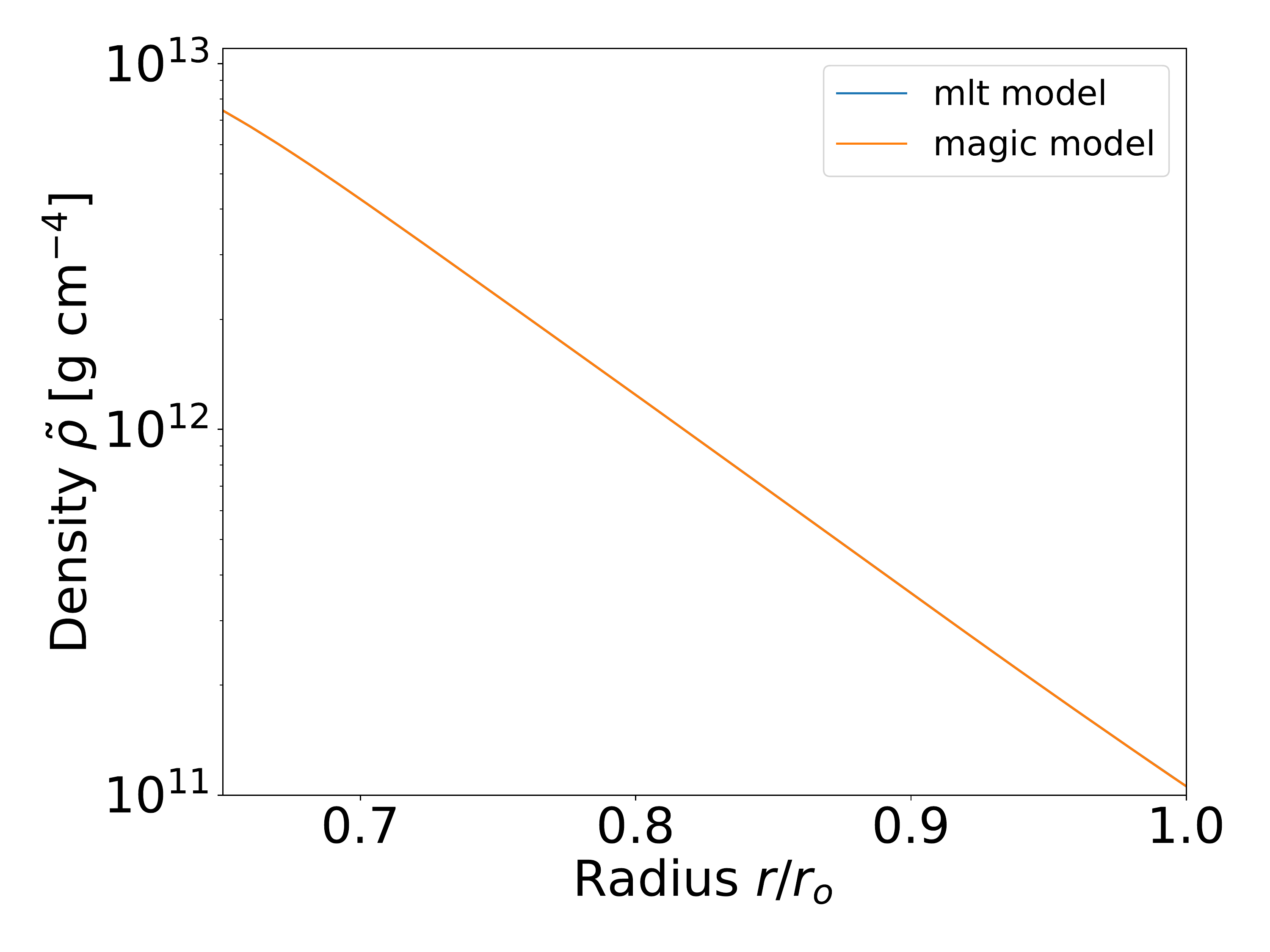}
\caption{}
\end{subfigure}
\begin{subfigure}{0.45 \textwidth}
\includegraphics[width=1.0\textwidth]{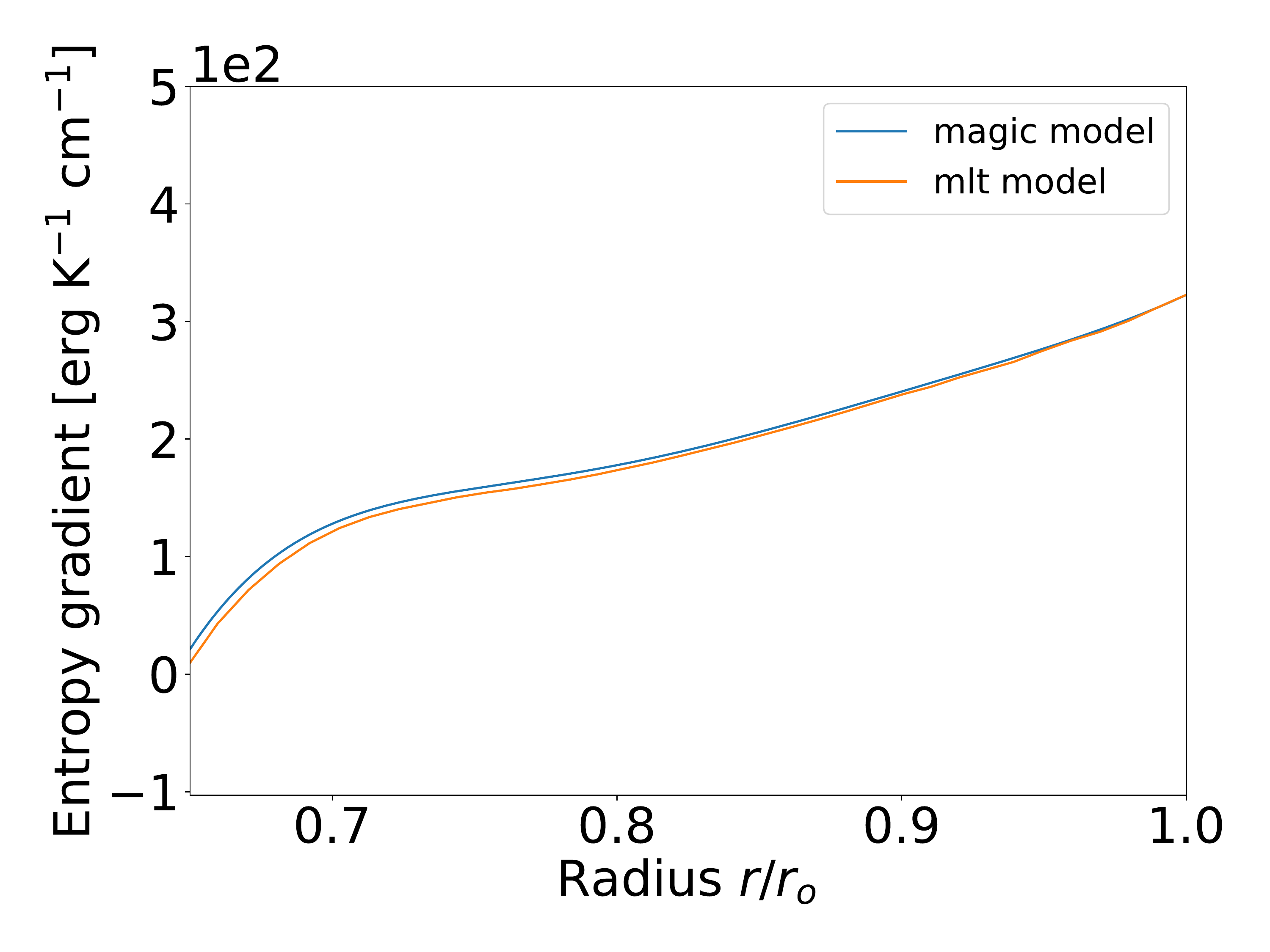} 
\caption{}
\end{subfigure}
\begin{subfigure}{0.45 \textwidth}
\includegraphics[width=1.0\textwidth]{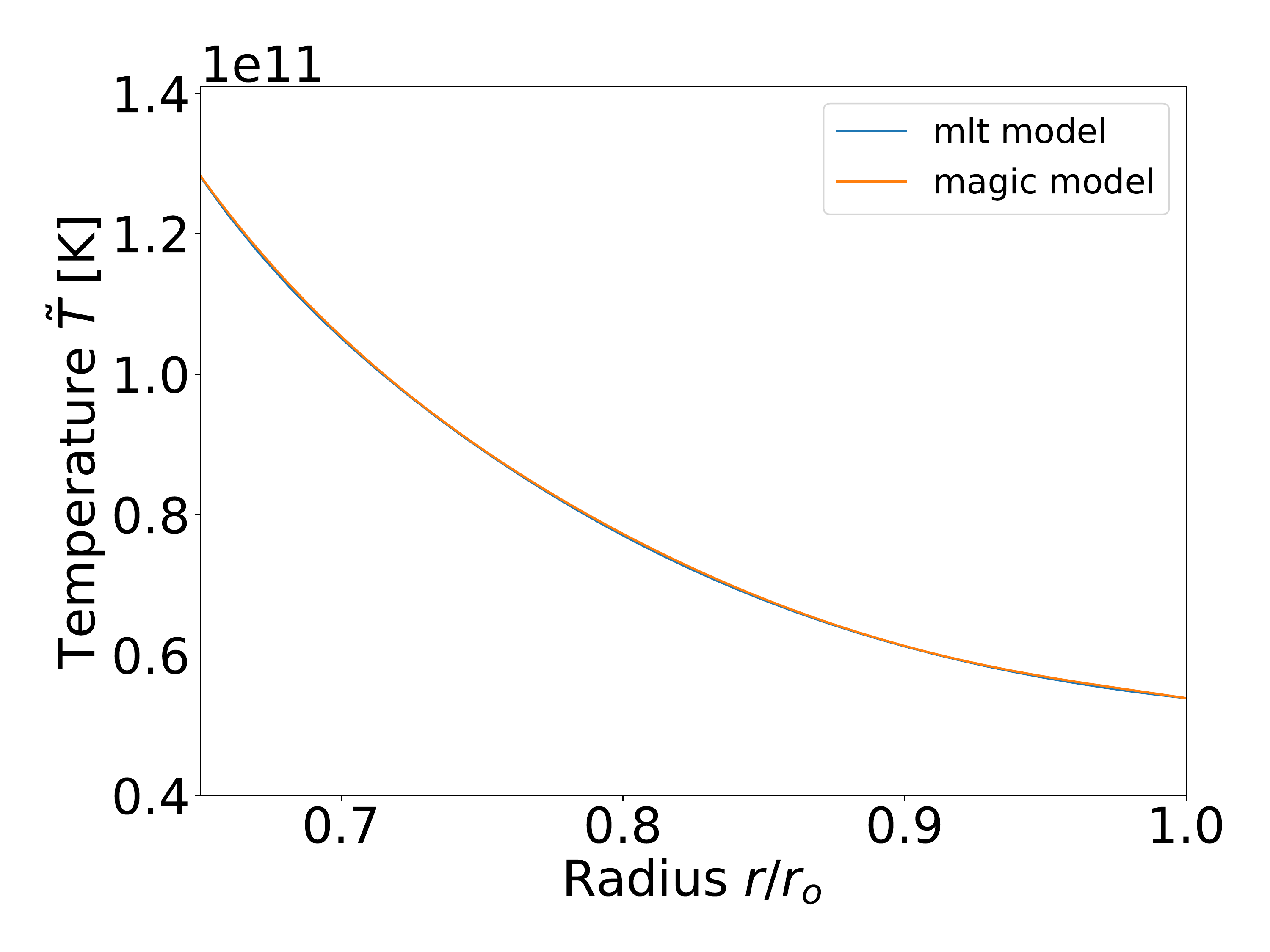}
\caption{}
\end{subfigure}
\begin{subfigure}{0.45 \textwidth}
\includegraphics[width=1.0\textwidth]{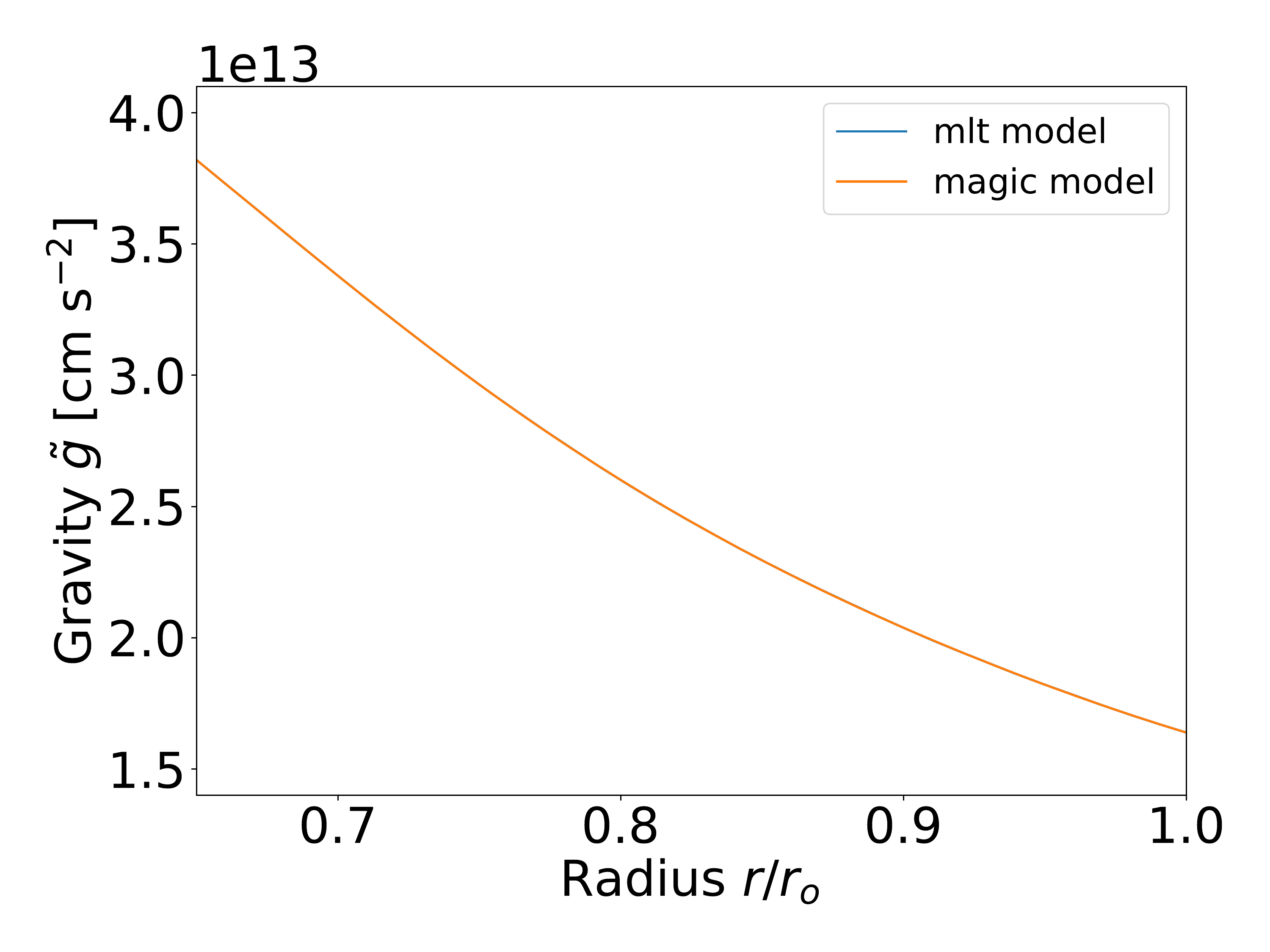}
\caption{}
\end{subfigure}
\begin{subfigure}{0.45 \textwidth}
\includegraphics[width=1.0\textwidth]{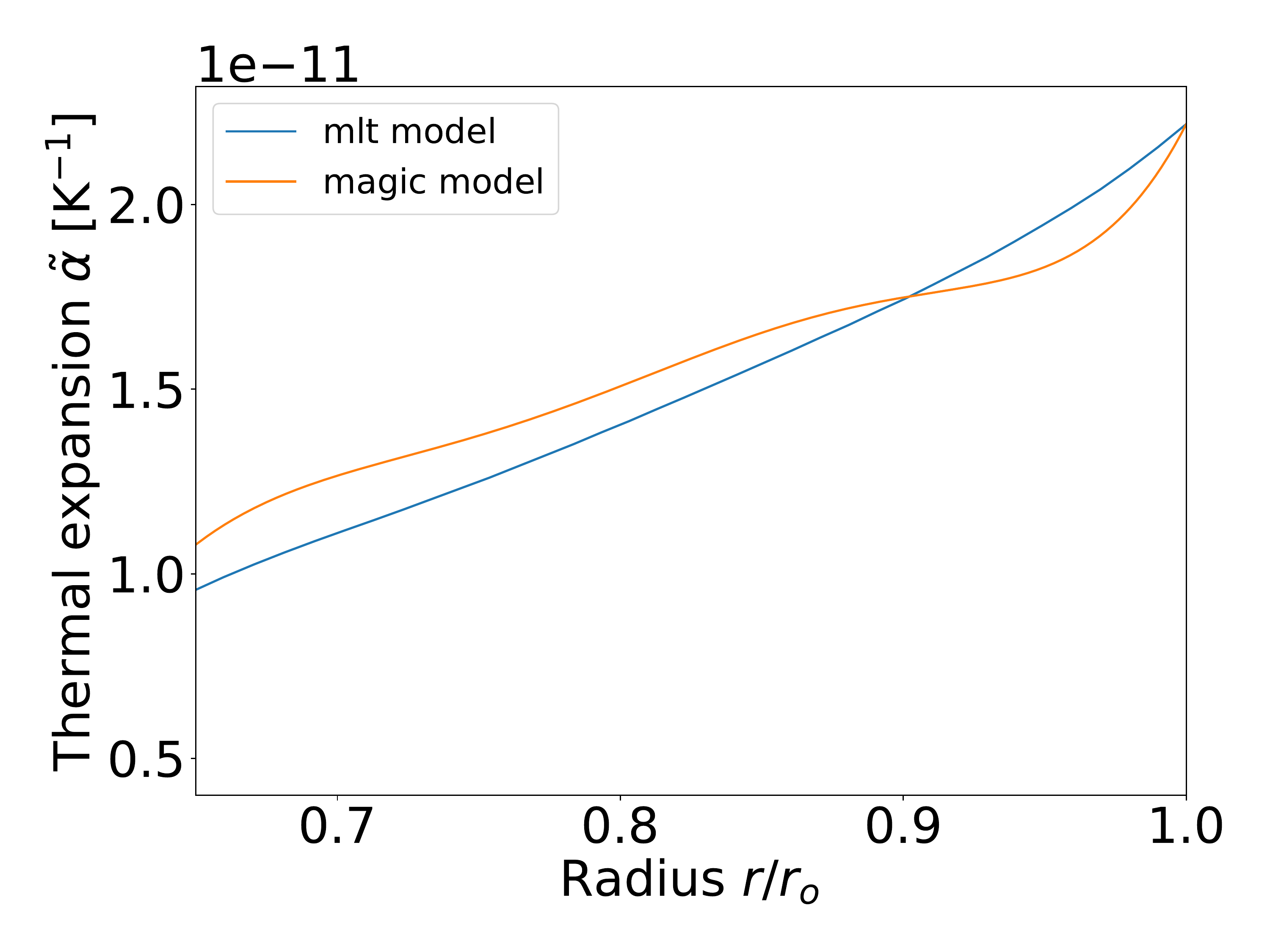}
\caption{}
\end{subfigure}
\caption{Comparison of the reference state considered in this study (\textup{}magic model) with the mlt model of the 1D CCSN simulation by \citet{2014Hudepol}.  
(a) Background density profile, (b) effective entropy profile, 
(c) temperature profile,  (d) gravity profile, and (e) thermal expansion profile as a function of the normalized radius. Dimensional and dimensionless units are converted by simple multiplication by the reference values at the outer boundary.}
\end{figure*}

\clearpage

\section{Parameters of our global simulations}

\begin{table}[h!]
\renewcommand{\arraystretch}{1.2}
\centering
\caption{Overview of the global models. All simulations use $Pm=16$ and insulating magnetic boundary conditions. The $(n_r,n_{\theta},n_{\phi})=(257,512,1024)$ resolution is called ``Medium'' and the $(n_r,n_{\theta},n_{\phi})=(385,768,1536)$ resolution is called ``High''. $Ra$ corresponds to the Rayleigh number linked to the Brunt-V\"ais\"al\"a frequency (see Eq. \ref{eq:ra_N}), $Pr$ to the thermal Prandtl number, $B_\mathrm{o}$ to the initial magnetic field strength at the outer boundary.
Magnetic field and dipole field strengths are time and space averaged. 
$P_\mathrm{sim}$ corresponds to the estimated period of the mean-field dynamo patterns at $r=0.92\,r_o$, $k_z$ to the vertical wavenumber and $q\Omega$ to the value in the turbulent region (at $r=0.92\,r_o$).
$k_z$, $q\Omega$ and the alpha component $\alpha_{\phi\phi}$ are used to compute the theoretical $\alpha\Omega$ period  $P_{\alpha\Omega}$.}
\label{table:annex}
\setlength{\tabcolsep}{5pt}
\begin{tabular}{l c c c c c c c c c c c}
\hline \hline
Model&Resolution&$Ra$&$Pr$ &$B_\mathrm{o}$ &Magnetic field &Dipole & $P_\mathrm{sim}$ & $k_z$ & $q\Omega$& $\alpha_{\phi\phi}$ & $P_\mathrm{\alpha\Omega}$  \\
&-&- &-&[$\SI{e14}{G}$]&[$\SI{e14}{G}$]&[$10^{14}$ G]&[ms]& [$\SI{}{cm^{-1}}$]&[$\SI{}{rad.s^{-1}}$]&[$\SI{}{cm \ s^{-1}}$ ]&[ms]\\ \hline 
\texttt{Standard} &Medium &-4.73e5&0.005&1.5&1.40&0.063&401&2.3e-6&385&7.7e5&342\\ 
\texttt{High Res} &High &-4.73e5&0.005&3.3&1.36&0.053&351&2.3e-6&454&7.0e5&329\\ 
\texttt{Incompressible} &High &0.0&-&1.5&2.36&0.085&NO&-&-&-&-\\
\texttt{Boussinesq} &High &-4.73e5&0.005&1.5&1.85&0.0548&NO&-&-&-&- \\
\texttt{Anel Ra0} &Medium &0.0&0.005&1.5&1.41&0.051&461&2.3e-6&369&7.8e6&349\\ 
\texttt{Anel Pr0 02} &Medium&-1.892e6&0.02&1.5&1.59&0.074&363&2.3e-6&479&8.0e5&300\\ 
\texttt{Anel Pr0 1} &Medium &-9.46e6&0.1&1.5&1.36&0.062&493&2.7e-6&480&4.5e5&366\\ 
 \hline \hline
\end{tabular}
\end{table}

\section{Correlation coefficients between the EMF and the magnetic field}
\label{an:corre}

\begin{figure*}[!h]
\centering
\begin{subfigure}{0.43 \textwidth}
\includegraphics[width=1.0\textwidth]{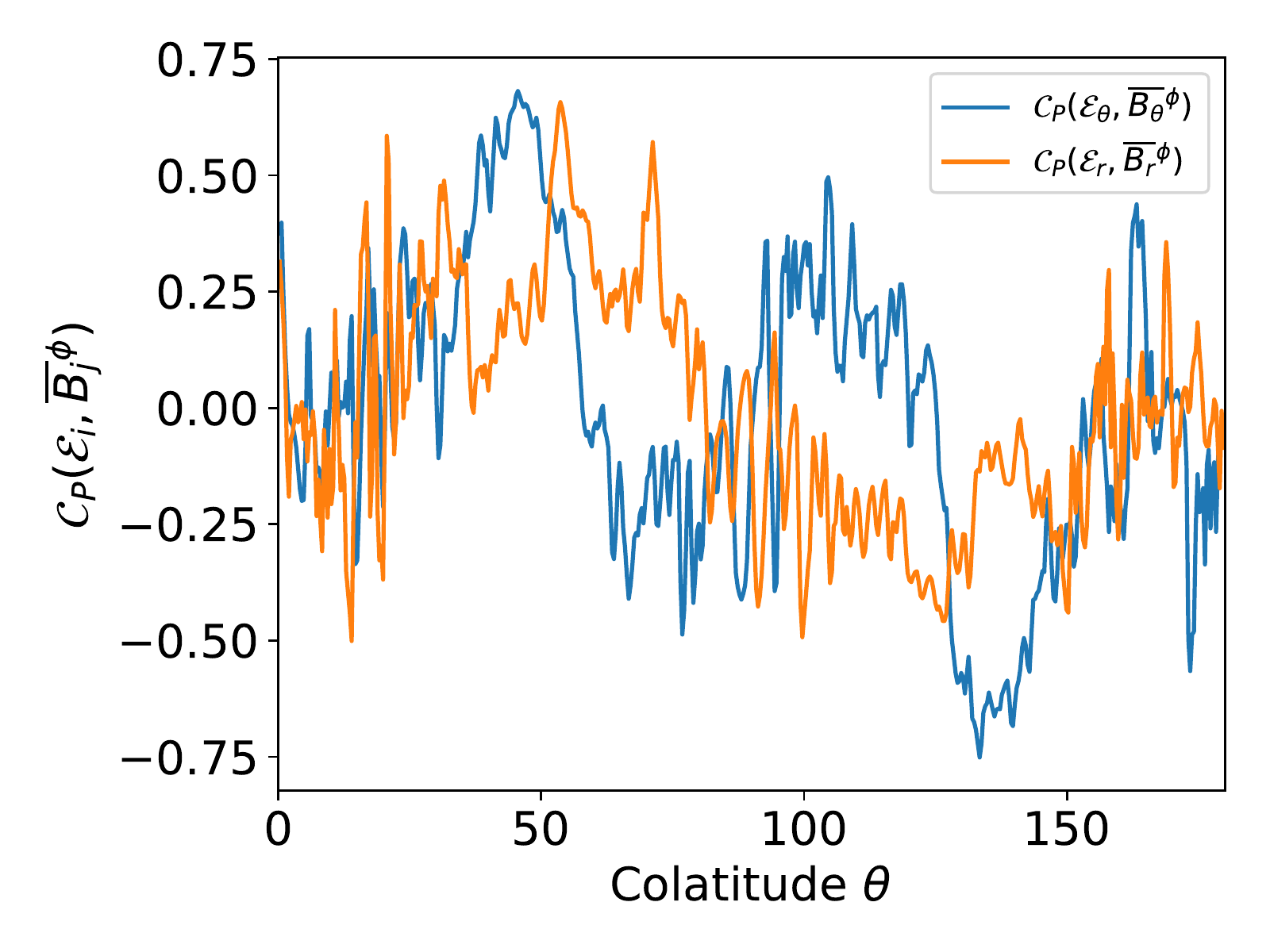}
\caption{}
\end{subfigure}
\begin{subfigure}{0.43 \textwidth}
\includegraphics[width=1.0\textwidth]{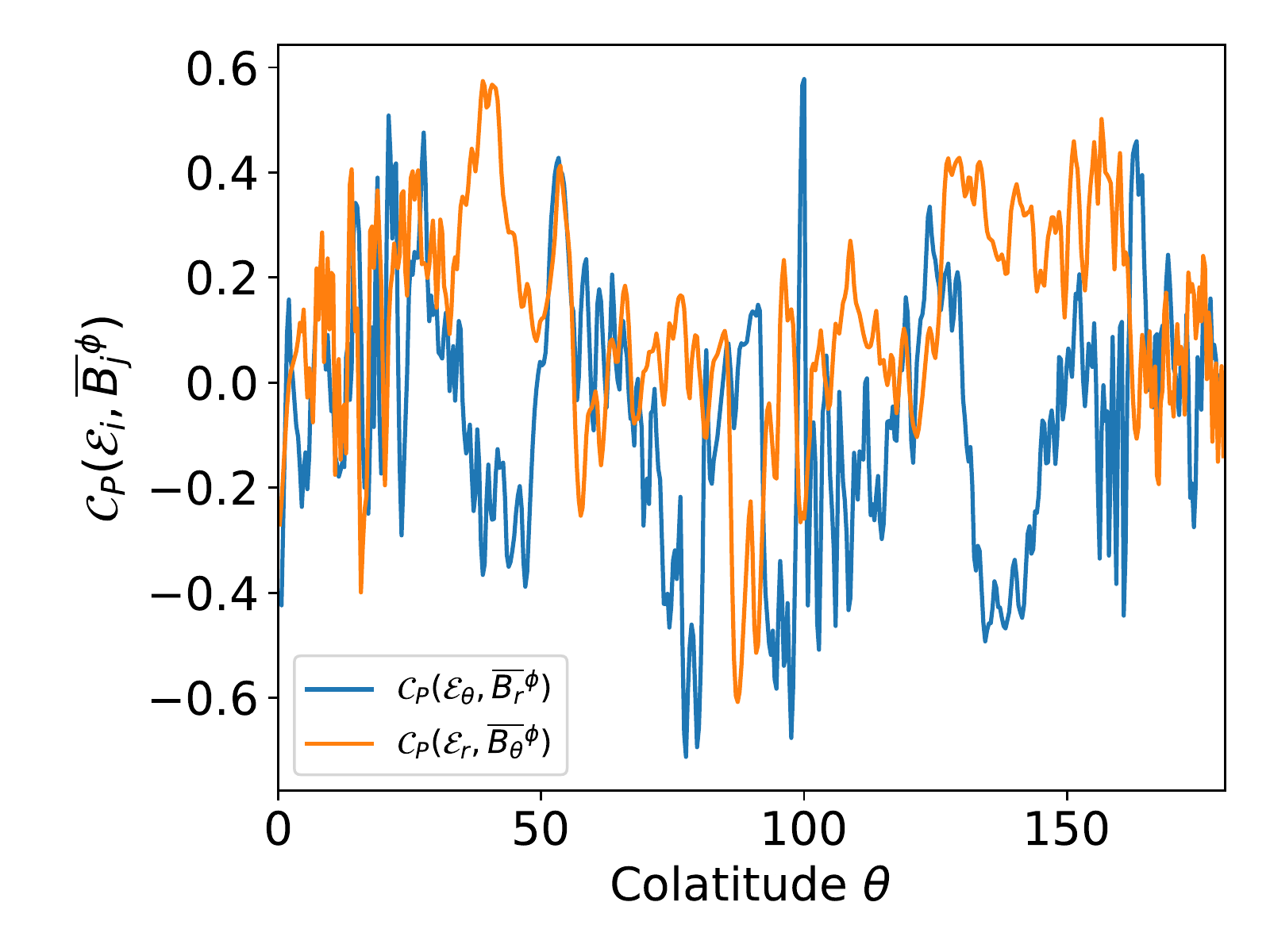} 
\caption{}
\end{subfigure}
\begin{subfigure}{0.43 \textwidth}
\includegraphics[width=1.0\textwidth]{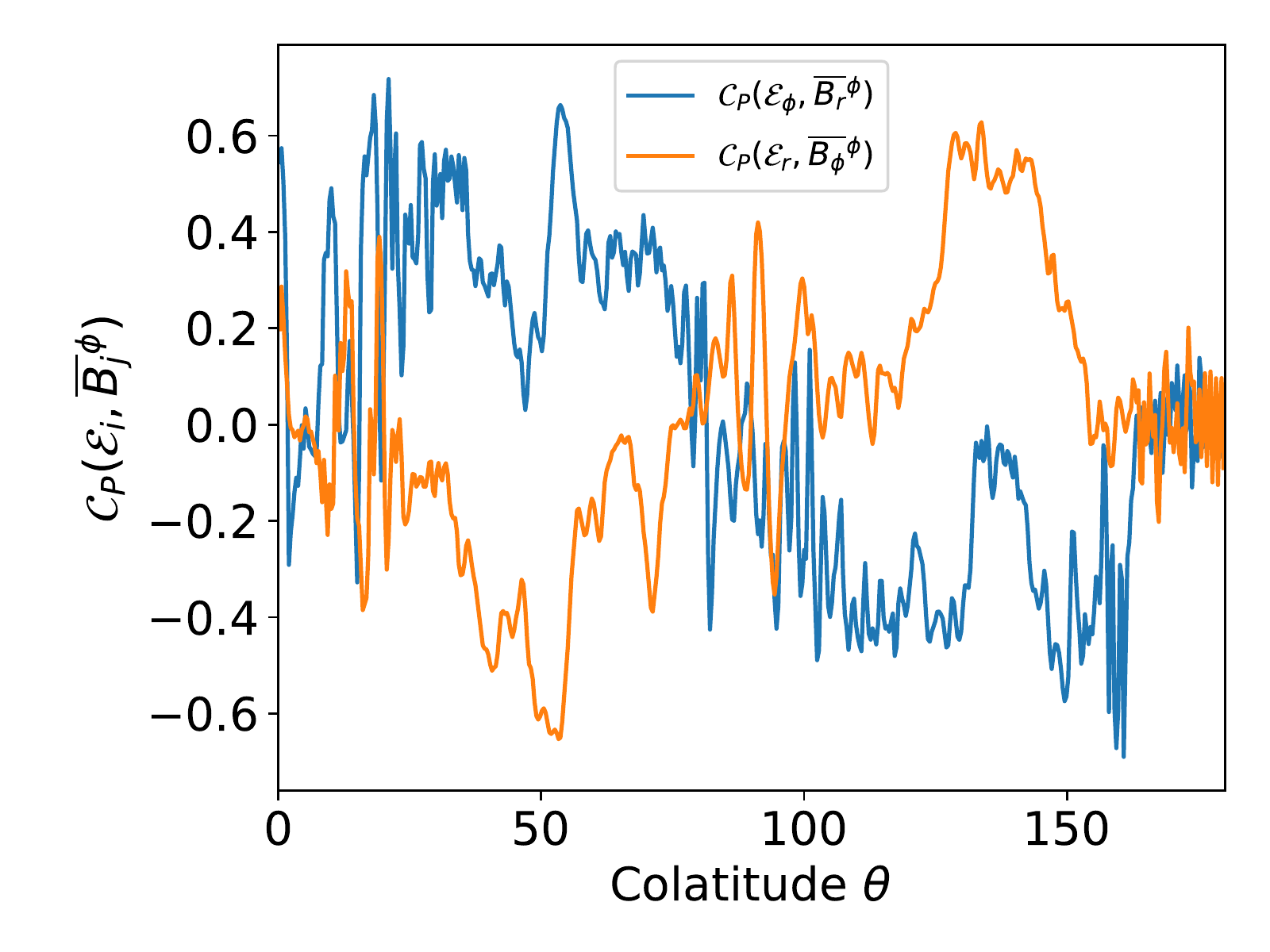}
\caption{}
\end{subfigure}
\begin{subfigure}{0.43 \textwidth}
\includegraphics[width=1.0\textwidth]{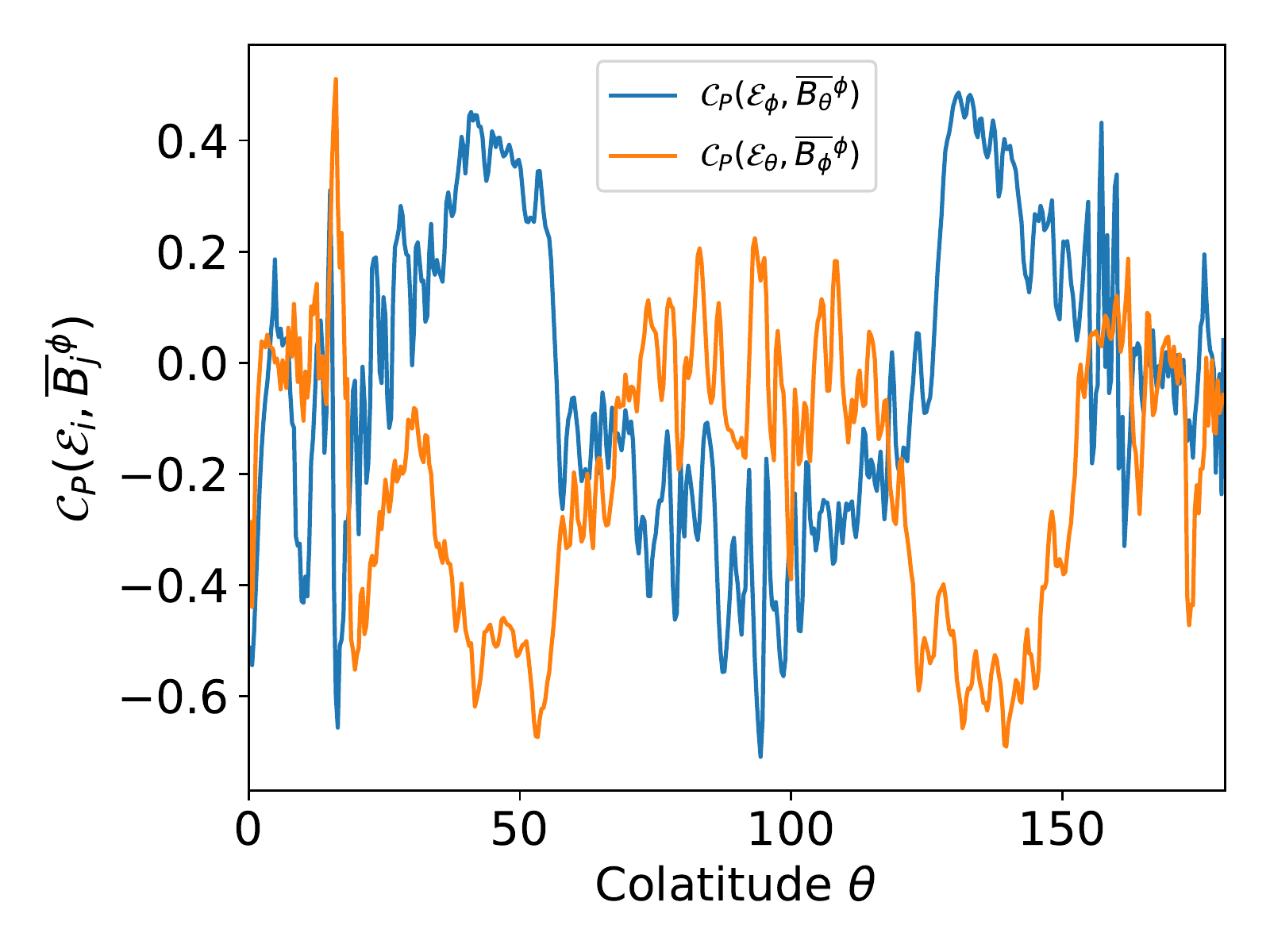}
\caption{}
\end{subfigure}
\caption{Time-averaged values of the correlation coefficient between the EMF and the magnetic field taken at $r=0.86\,r_o$.}
\end{figure*}

\section{Estimation of the components of the $\alpha$ tensor}
\label{an:alpha}

\begin{figure*}[!h]
\centering
\begin{subfigure}{0.43 \textwidth}
\includegraphics[width=1.0\textwidth]{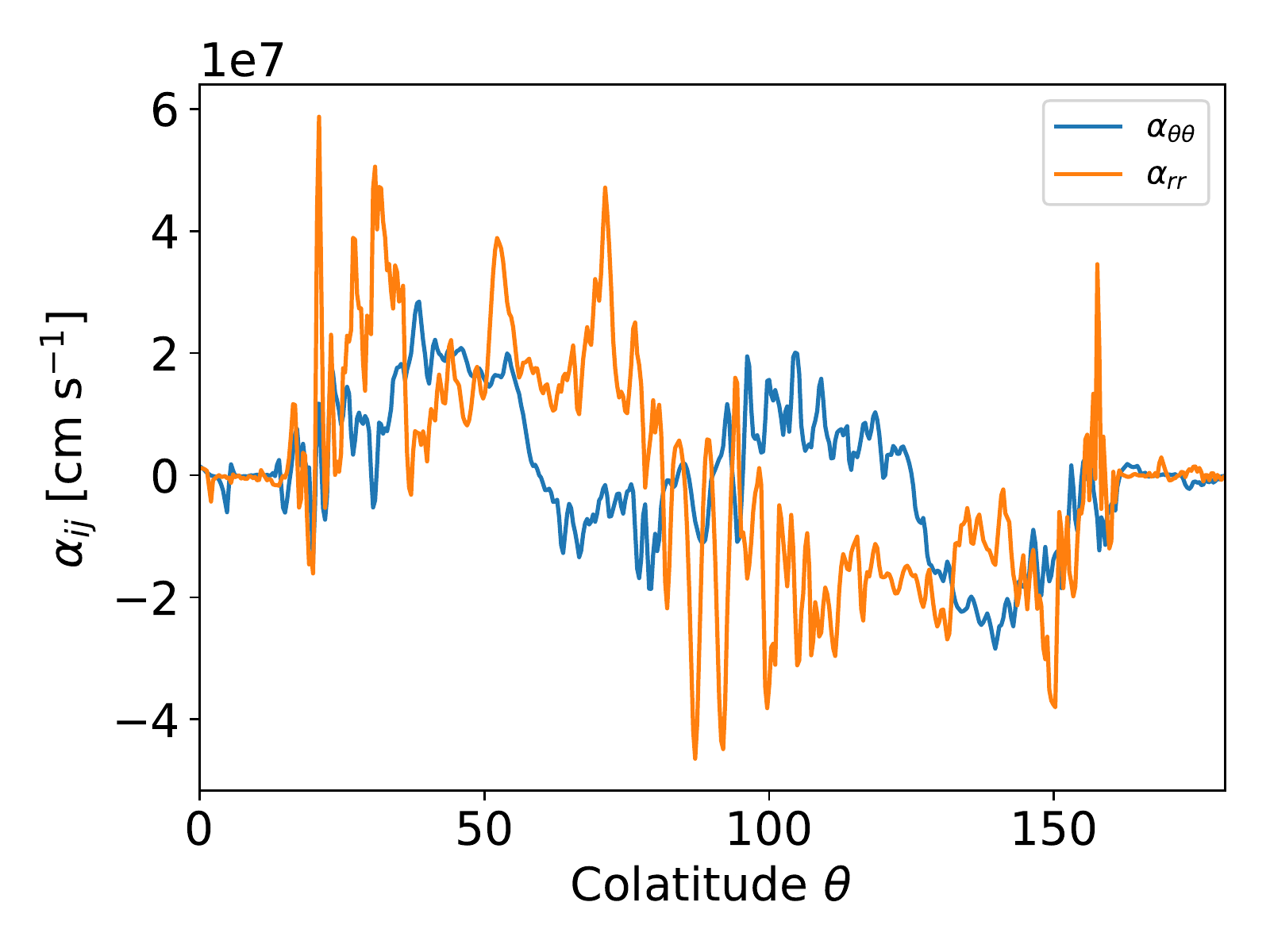}
\caption{}
\end{subfigure}
\begin{subfigure}{0.43 \textwidth}
\includegraphics[width=1.0\textwidth]{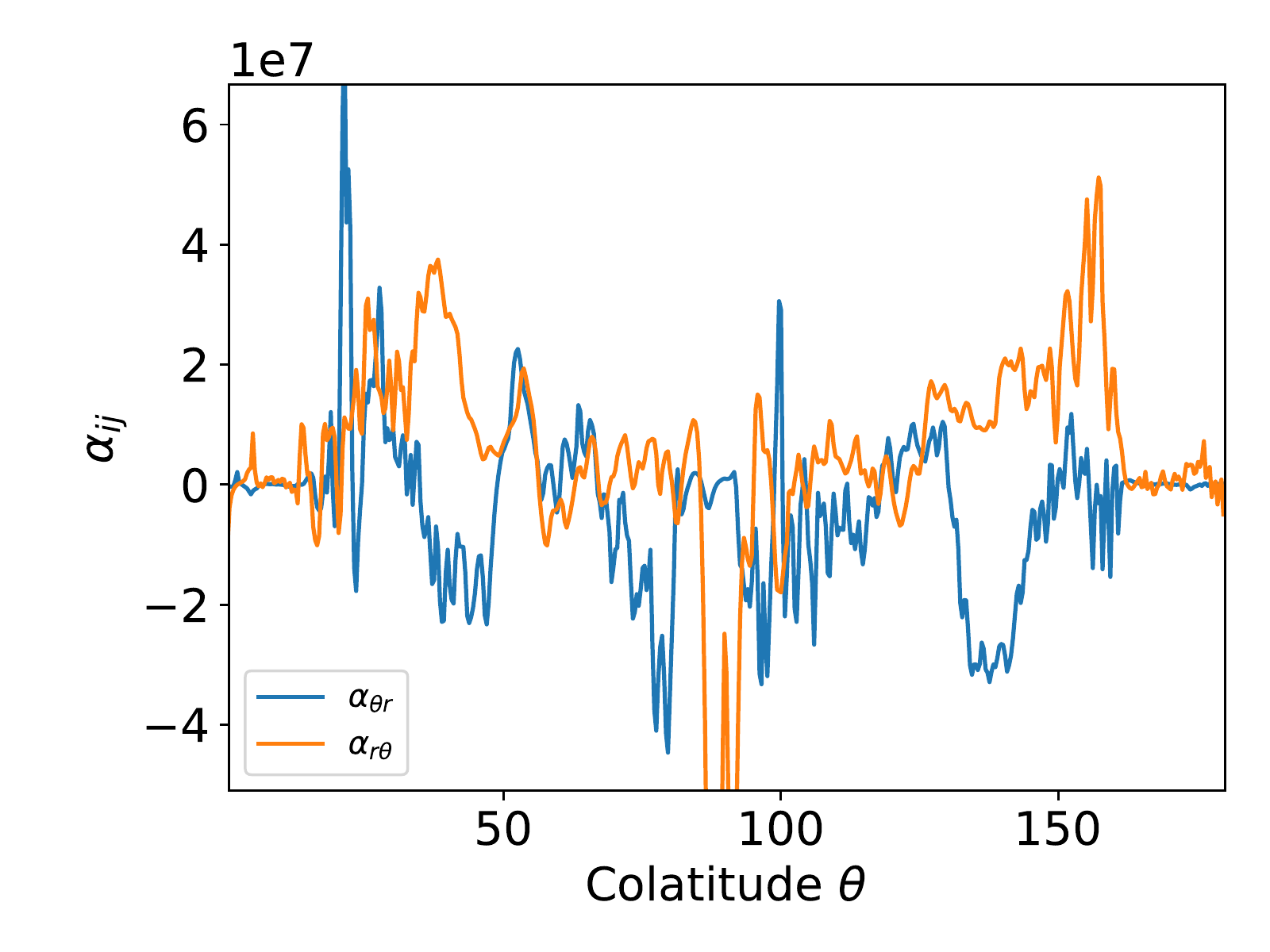} 
\caption{}
\end{subfigure}
\begin{subfigure}{0.43 \textwidth}
\includegraphics[width=1.0\textwidth]{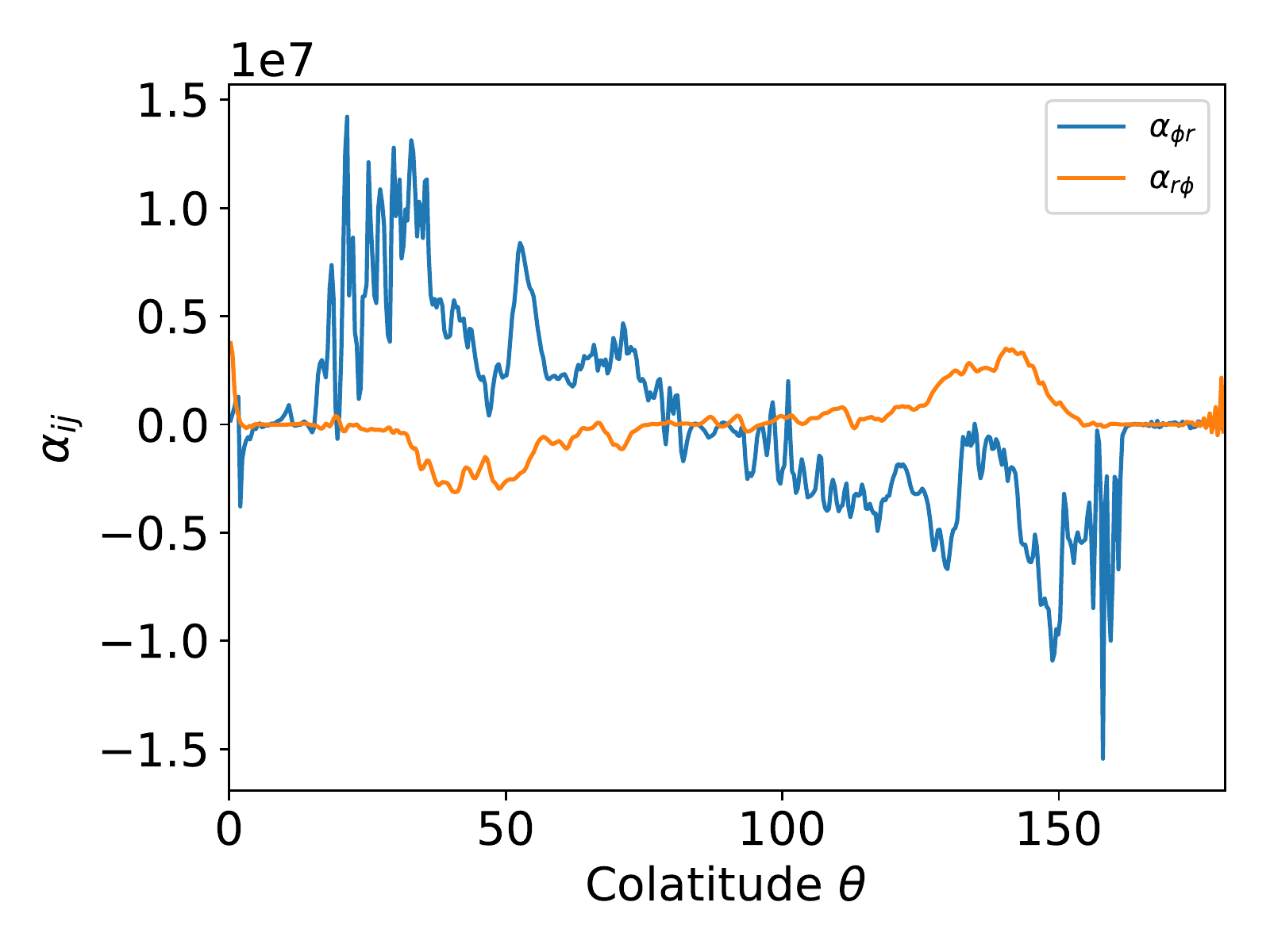}
\caption{}
\end{subfigure}
\begin{subfigure}{0.43 \textwidth}
\includegraphics[width=1.0\textwidth]{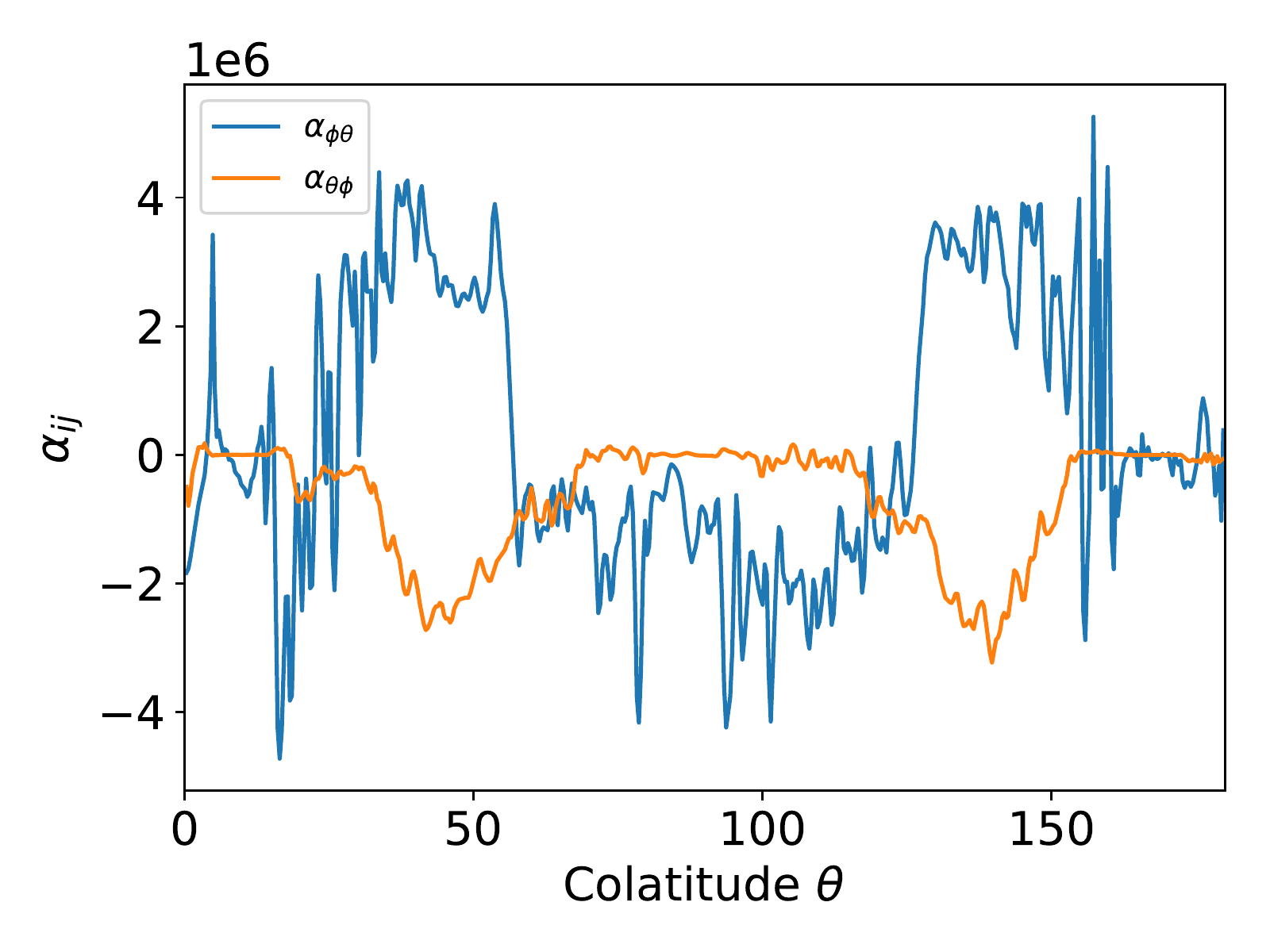}
\caption{}
\end{subfigure}
\caption{Time-averaged values of the $\alpha$ tensor components estimated by the correlation between the EMF and the magnetic field taken at $r=0.86\,r_o$.}
\end{figure*}
\newpage

\section{Azimuthal average of magnetic and velocity fields at quasi-stationary state}

\subsection{Simulation \texttt{Standard}}

\begin{figure*}[h]
     \includegraphics[width=0.24\textwidth]{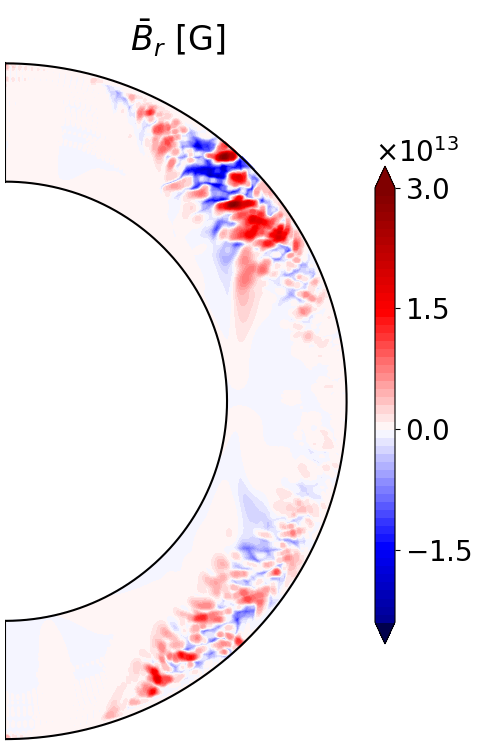}
      \includegraphics[width=0.24\textwidth]{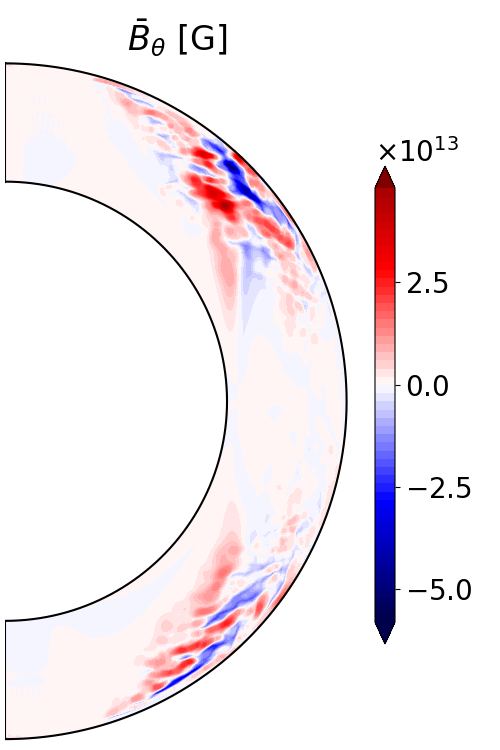}
      \includegraphics[width=0.24\textwidth]{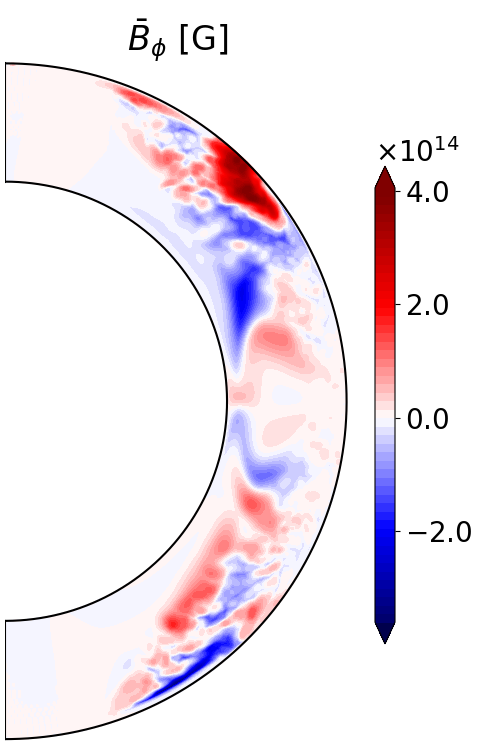}
      
     \includegraphics[width=0.24\textwidth]{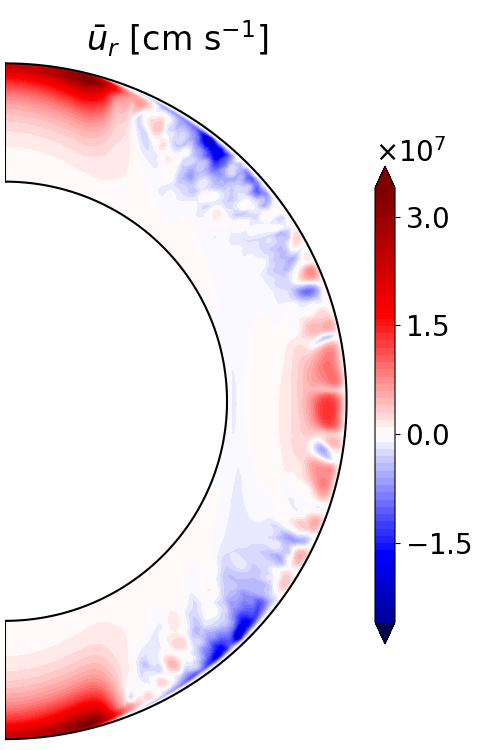}
      \includegraphics[width=0.24\textwidth]{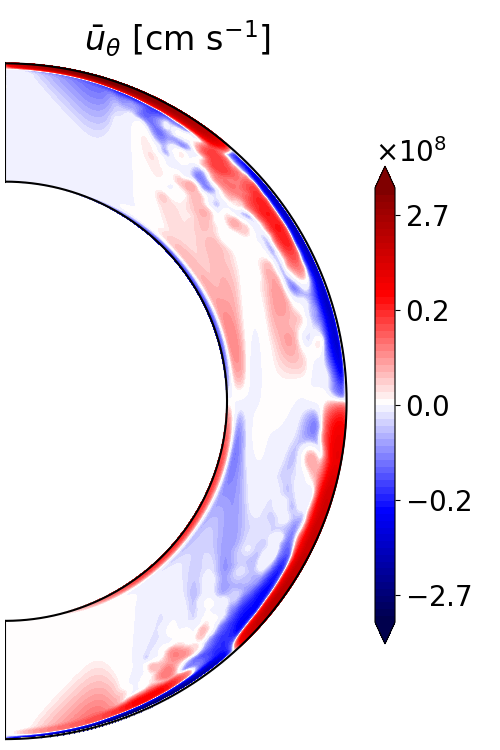}
      \includegraphics[width=0.24\textwidth]{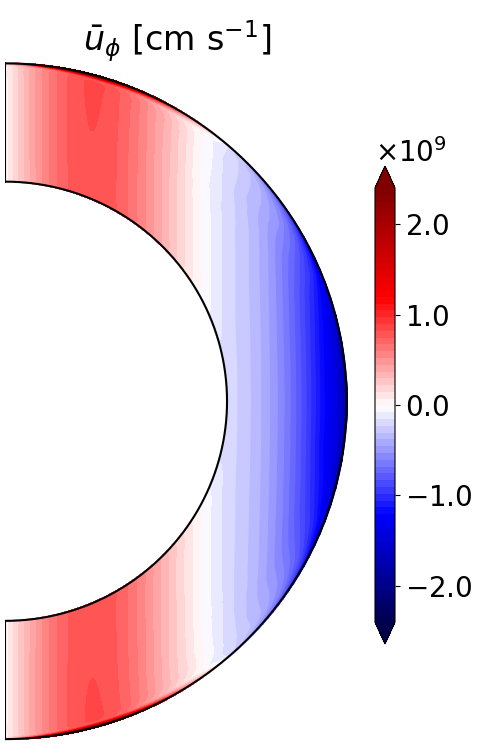}
      \includegraphics[width=0.24\textwidth]{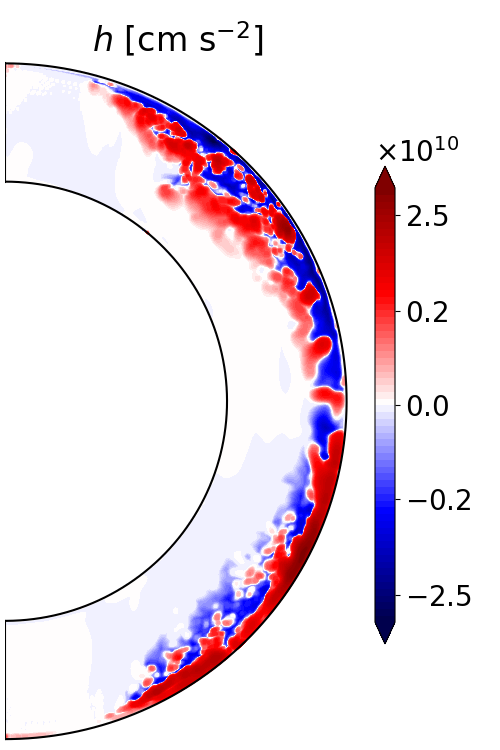}
      \caption{Snapshots of the azimuthal average of magnetic field components and velocity field components for the simulation \texttt{Standard}. 
      Top: Snapshots  of the azimuthal average of the magnetic field components $\bar{B_\phi}$, $\bar{B_r}$ , and $\bar{B_\theta}$ at $t=773$ ms for the model \texttt{Standard}. Bottom: Snapshots of the azimuthal average of the velocity field components $\bar{u_r}$, $\bar{u_\theta}$, $\bar{u_\phi}$ , and the turbulent kinetic helicity $h$. The angular velocity component can be found in Fig.~\ref{Snapshots_approx}. }
         \label{Annex_standard}
\end{figure*}
\newpage
\subsection{Simulation \texttt{Boussinesq}}
\begin{figure*}[h]


      \includegraphics[width=0.24\textwidth]{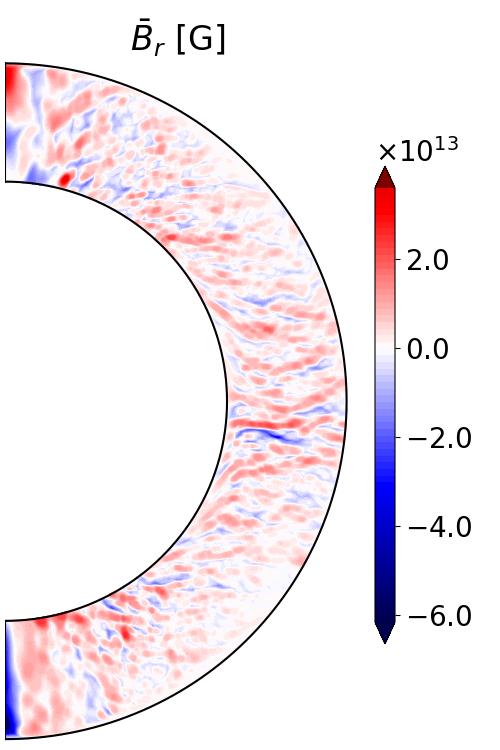}
      \includegraphics[width=0.24\textwidth]{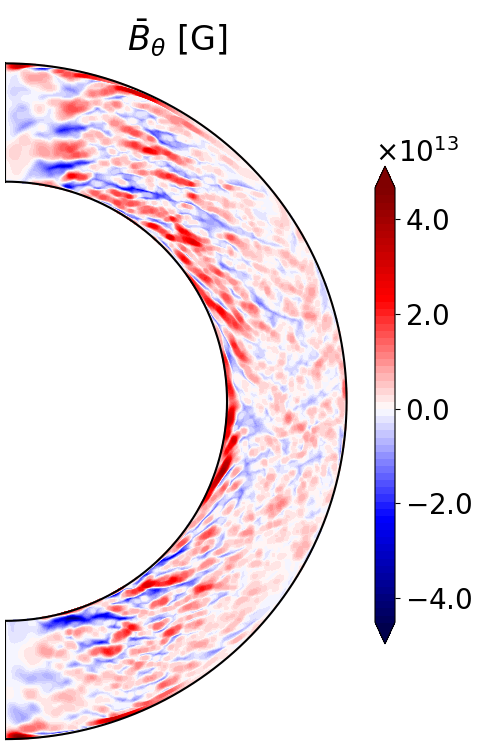}
      \includegraphics[width=0.24\textwidth]{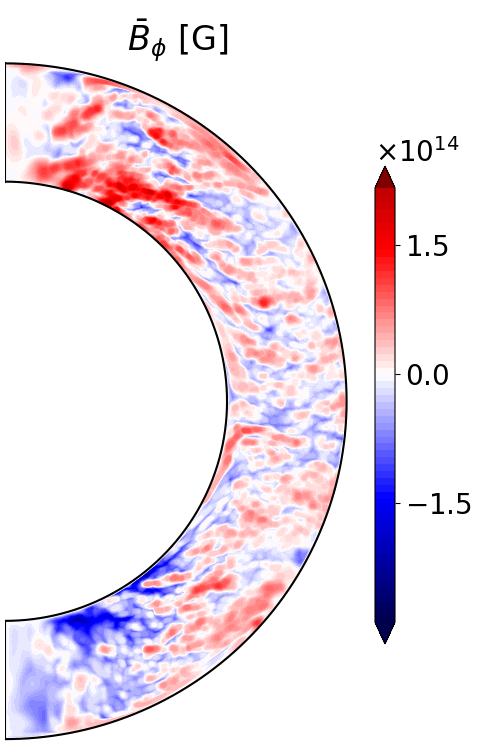}
      
     \includegraphics[width=0.24\textwidth]{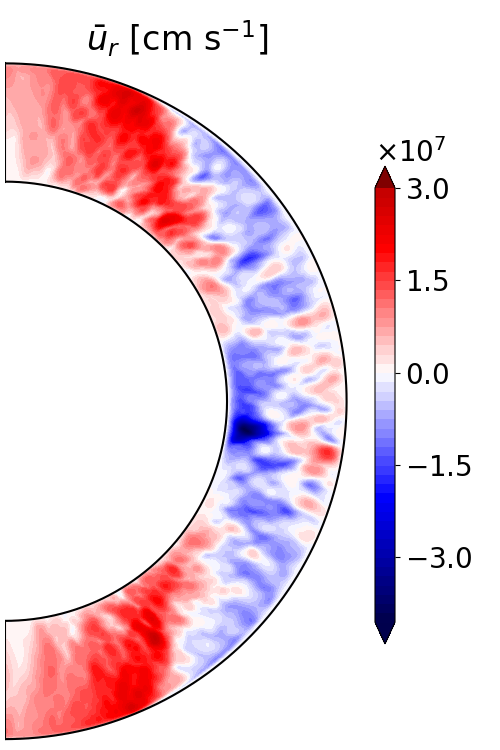}
      \includegraphics[width=0.24\textwidth]{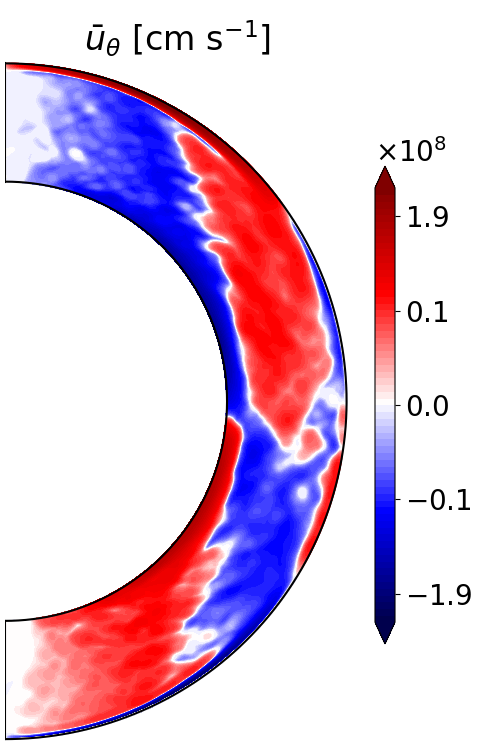}
      \includegraphics[width=0.24\textwidth]{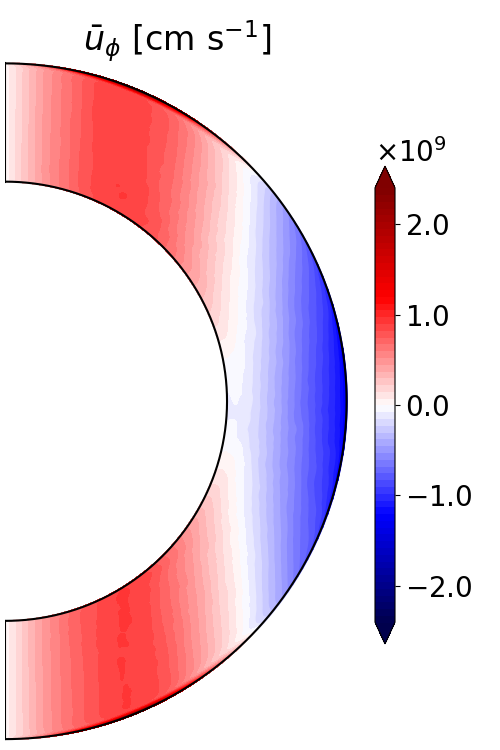}
      \includegraphics[width=0.24\textwidth]{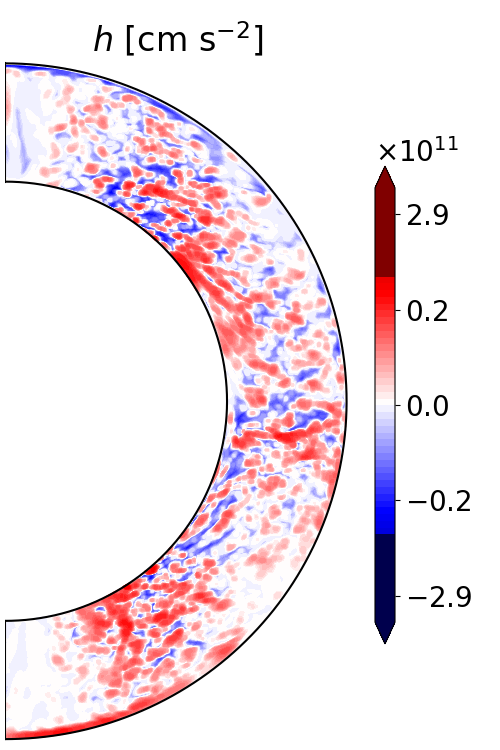}
      \caption{Snapshots of the azimuthal average of magnetic field components and velocity field components for the simulation \texttt{Boussinesq}. Top: Snapshots of the azimuthal average of the magnetic field components $\bar{B_\phi}$, $\bar{B_r}$ , and $\bar{B_\theta}$ at $t=511$ ms for model \texttt{Boussinesq}. Bottom: Snapshots of the azimuthal average of the velocity field components $\bar{u_r}$, $\bar{u_\theta}$, $\bar{u_\phi}$ , and the turbulent kinetic helicity $h$. The angular velocity can be found in Fig.~\ref{Snapshots_approx}.}
         \label{Annex_Bous}
\end{figure*}

\newpage

\subsection{Simulation \texttt{Pr=0.1}}

\begin{figure*}[h]

     \includegraphics[width=0.24\textwidth]{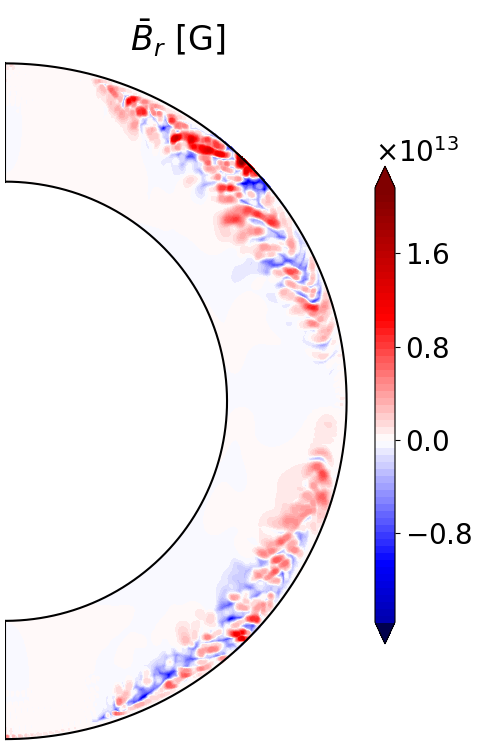}
      \includegraphics[width=0.24\textwidth]{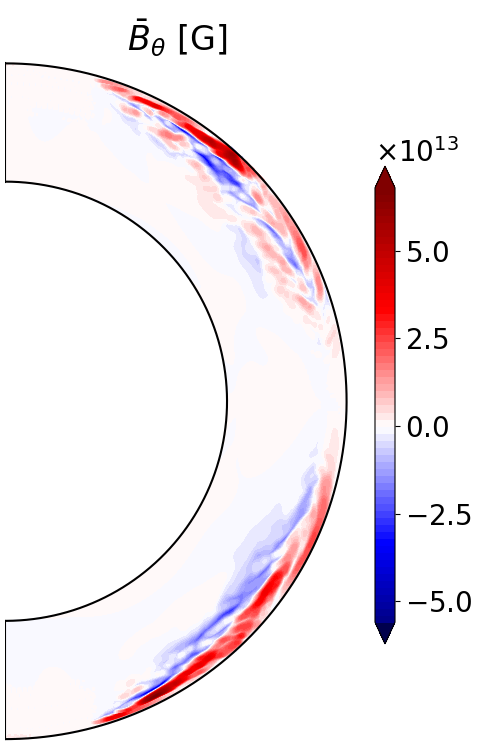}
      \includegraphics[width=0.24\textwidth]{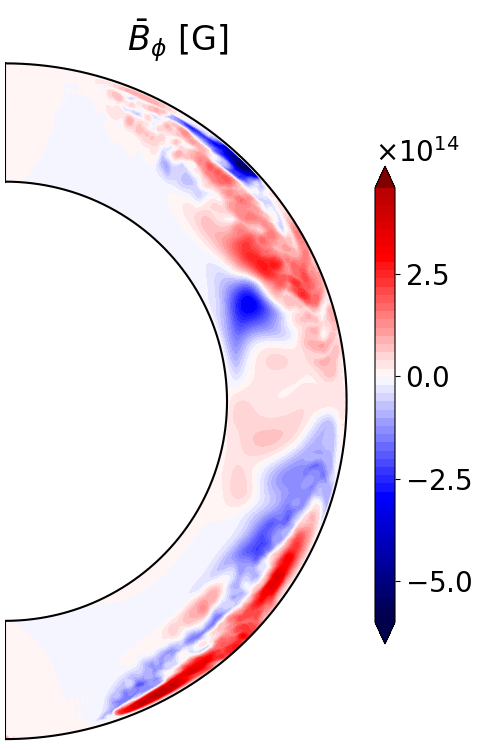}

     \includegraphics[width=0.24\textwidth]{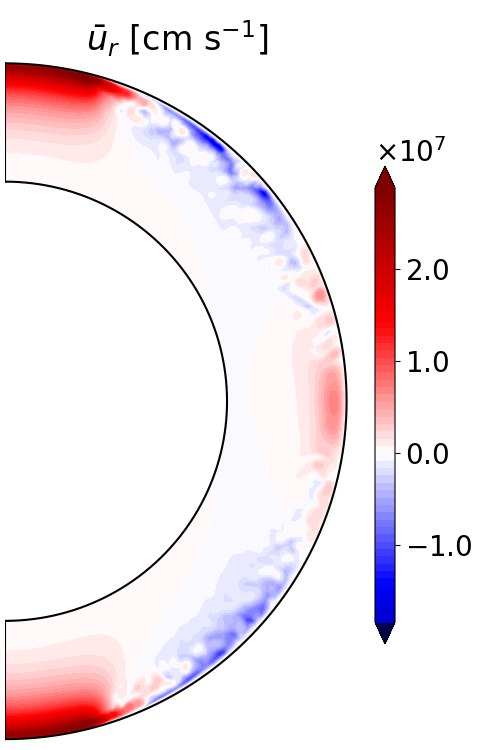}
      \includegraphics[width=0.24\textwidth]{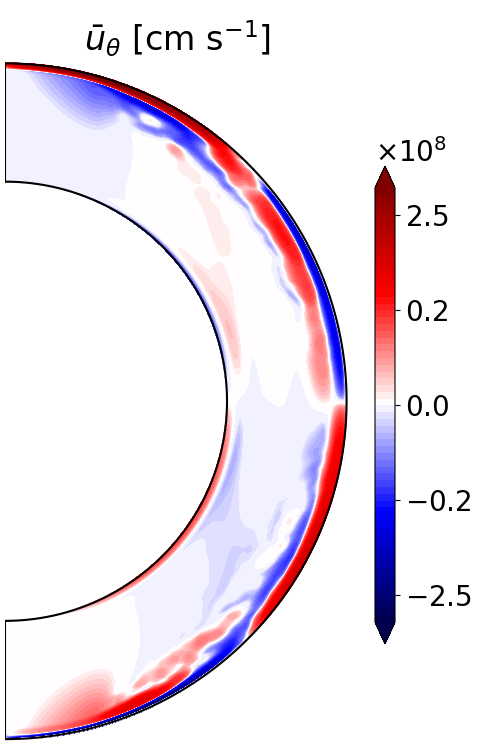}
      \includegraphics[width=0.24\textwidth]{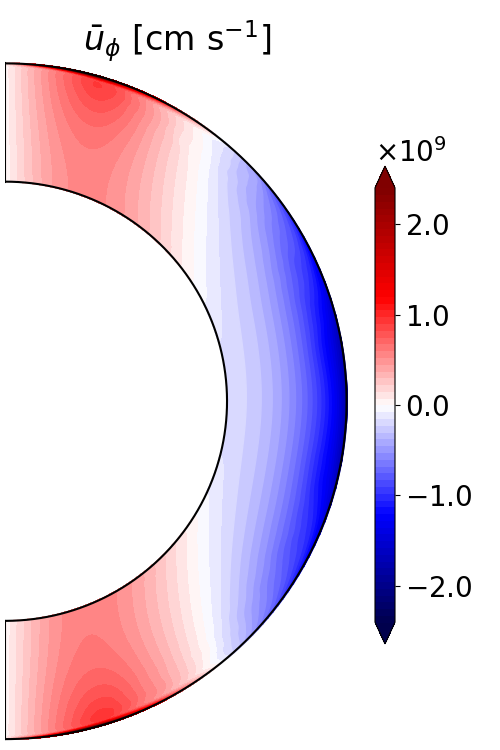}
      \includegraphics[width=0.24\textwidth]{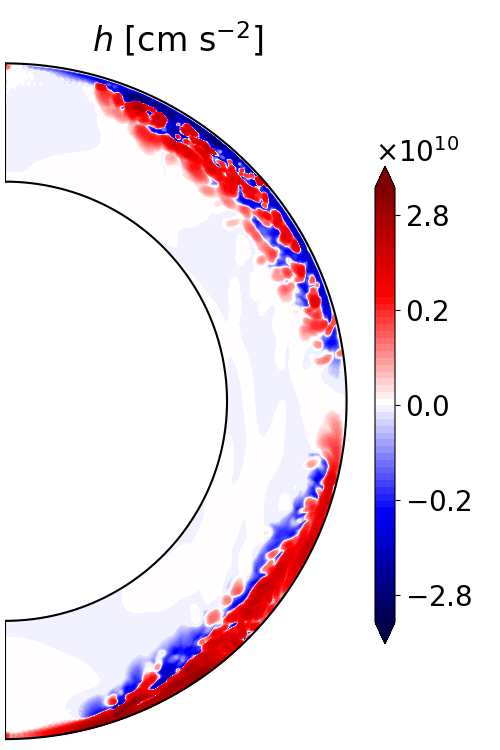}
      \caption{Snapshots of the azimuthal average of magnetic field components and velocity field components for the simulation \texttt{Pr=0.1}. Top: Snapshots of the azimuthal average of the magnetic field components $\bar{B_\phi}$, $\bar{B_r}$ , and $\bar{B_\theta}$ at $t=1128$ ms for model \texttt{Anel Pr0 1}. Bottom: Snapshots of the azimuthal average of the velocity field components $\bar{u_r}$, $\bar{u_\theta}$, $\bar{u_\phi}$ , and the turbulent kinetic helicity $h$.}\label{Annex_Pr01}
\end{figure*}

\end{appendix}

\end{document}